\documentclass[twocolumn,epjc3,nopacs]{svjour3}
\usepackage[numbers]{natbib}
\journalname{Submitted to Eur. Phys. J. A -- JLAB-PHY-26-4699}
\RequirePackage{graphicx}
\RequirePackage{enumitem}

\usepackage{graphics}
\usepackage{natbib}
\usepackage[colorlinks=true,linkcolor=blue,citecolor=blue, urlcolor=blue]{hyperref}

\usepackage{amsmath}
\usepackage{xcolor}
\usepackage{multirow}
\usepackage{float}
\usepackage{url}
\usepackage{rotating}
\usepackage{soul}
\usepackage{comment}
\usepackage{cuted}
\usepackage{tabularx}
\usepackage{subcaption}

\usepackage[switch]{lineno}

\def \rarr {\rightarrow}

\definecolor{GREEN}{rgb}{0.,0.8,0}
\definecolor{RED}{rgb}{1,0,0}
\definecolor{ORANGE}{rgb}{1,0.5,0}

\newcommand{\uRWELL}{$\mu$RWELL}
\newcommand{\uCLAS}{$\mathrm{\mu CLAS12}$}
\newcommand{\jpsi}{$J/\psi$}

\begin{document}

\title{Electro- and photoproduction of muon pairs with $\mu$CLAS12:
Double Deeply Virtual Compton Scattering, Timelike Compton Scattering, and $J/\psi$ production}
\titlerunning{The $\mu$CLAS12 experiment at JLab}
\authorrunning{S.Stepanyan \textit{et al.} (CLAS Collaboration)}  

\author{
\begin{center}
\parbox{\textwidth}{
J.~S.~Alvarado\thanksref{addr2} \and
N.~Baltzell\thanksref{addr1}$^,$\thanksref{cospok} \and
M.~Bondi\thanksref{addr4}$^,$\thanksref{cospok} \and
P.~Chatagnon\thanksref{addr3}$^,$\thanksref{cospok} \and
R.~De~Vita\thanksref{addr1}$^,$\thanksref{addr16}$^,$\thanksref{cospok} \and
M.~Hoballah\thanksref{addr2}$^,$\thanksref{cospok} \and
V.~Kubarovsky\thanksref{addr1}$^,$\thanksref{cospok} \and
R.~Paremuzyan\thanksref{addr1}$^,$\thanksref{cospok} \and
S.~Stepanyan\thanksref{addr1}$^,$\thanksref{contact} 
et al.\thanksref{al}
}
\end{center}
}

\thankstext{contact}{Contact author: stepanya@jlab.org}
\thankstext{cospok}{Co-spokesperson}
\thankstext{al}{Full list of authors provided at the end of the manuscript}

\institute{
\label{addr4} INFN, Sezione di Catania, 95123 Catania, Italy
\and
\label{addr16} INFN, Sezione di Genova, 16146 Genova, Italy
\and
\label{addr2} IJCLab, Université Paris-Saclay, CNRS–IN2P3, 91405 Orsay, France
\and
\label{addr3} IRFU, CEA, Université Paris-Saclay, 91191 Gif-sur-Yvette, France
\and
\label{addr1} Thomas Jefferson National Accelerator Facility, Newport News, VA 23606, USA
}

\date{Received: \today / Accepted: -}
\maketitle

\begin{abstract}
The CEBAF Large Acceptance Spectrometer for operation at 12~GeV (CLAS12) at the Thomas Jefferson National Accelerator Facility has played a central role in advancing the understanding of nucleon and nuclear structure. 
As increasingly precise data become available, new physics opportunities emerge that extend beyond the current capabilities of CLAS12. In this article, a program to explore the quark and gluon structure of the nucleon through di-muon electro- and photoproduction is presented. Its primary focus is the measurement of beam-spin asymmetries in Double Deeply Virtual Compton Scattering, $ep \rightarrow e^\prime \mu^+ \mu^-p^\prime $. 
By independently varying the incoming and outgoing photon virtualities and momentum transfer, the DDVCS measurement provides access to the Generalized Parton Distributions over their full three-dimensional phase space, extending beyond the kinematic constraints of Deeply Virtual Compton Scattering and Timelike Compton Scattering. In addition, the large acceptance and high luminosity of the $\mu$CLAS12 experiment will enable precision measurements of near-threshold $J/\psi$ production and high-statistics studies of Timelike Compton Scattering.
\end{abstract}

\tableofcontents

\section{Introduction}

A key goal of the 12 GeV science program at the Thomas Jefferson National Accelerator Facility (JLab) is to explore the internal structure of hadrons, facilitated by the framework of Generalized Parton Distributions (GPDs)~\cite{mueller1994wave, Ji:1996nm, Radyushkin:1997, collins:1997}. 
GPDs are universal, non-perturbative functions that enter the description of hard exclusive processes and unify the information encoded in Elastic Form Factors (EFFs) and Parton Distribution Functions (PDFs) measured in Deep-Inelastic Scattering (DIS). 
The JLab GPD program~\cite{ARRINGTON2022103985} includes measurements of beam- and target-spin asymmetries as well as cross sections in deeply virtual exclusive processes.  
Among these reactions, Deeply Virtual Compton Scattering (DVCS), in which a highly virtual photon emitted by an incoming lepton interacts with a parton in the nucleon and emerges as a real photon~\cite{mueller1994wave,Ji:1996nm,Radyushkin:1996ru}, provides the most direct and theoretically clean access to GPDs.
A large body of DVCS data has been collected since the early 2000s with 6 GeV~\cite{clasdvcs1, carlos, fx, Maxime, Jo:2015ema, erin, Pisano:2015iqa} and 12 GeV experiments~\cite{halladvcs:1,clas12dvcs:1, CLAS:2024qhy} at JLab, the H1~\cite{H1:2001nez,H1:2005gdw}, ZEUS~\cite{ZEUS:2003pwh} and HERMES~\cite{HERMES:2001bob,HERMES:2006pre,HERMES:2008abz} experiments at HERA, and the COMPASS experiment~\cite{Joerg:2016hhs,COMPASS:2018pup} at CERN. In parallel, experimental studies of Timelike Compton Scattering (TCS) ~\cite{Berger:2001xd,Goritschnig,Boer:2015hma,Boer:2015gv}, in which an incoming real photon produces a highly virtual photon that decays into a lepton pair, have been initiated. The first measurements of angular and beam-helicity asymmetries, with the CLAS12 experiment, have been published in Ref.~\cite{clas12tcs}. 

A fundamental limitation of DVCS and TCS is that they provide access to only two of the three variables ($x$, $\xi$, $t$) upon which GPDs depend.\footnote{The GPDs also depend on a renormalization scale $\mu^2$, which is often set to $Q^2$, the virtuality of the incoming photon in DVCS; or $Q'^2$, the virtuality of the outgoing photon in TCS.} The variable $x$ denotes the longitudinal momentum fraction of the active quark, $\xi$ is the longitudinal momentum transfer, also known as the skewness parameter, and $t$ is the squared four-momentum transfer to the nucleon.
DVCS and TCS experimental observables are expressed in terms of complex‑valued Compton Form Factors (CFFs). At Leading Order (LO) and Leading Twist (LT), their real parts are given by convolution integrals of GPDs over $x$, while their imaginary parts are proportional to the GPDs evaluated at the specific kinematical points $x$=$\pm\xi$.
This leads to a well-known deconvolution problem in the extraction of GPDs from experimental data~\cite{herve,partons}. As shown in Ref.~\cite{bertone}, the extraction of GPDs from CFFs is intrinsically ambiguous due to the existence of so-called shadow GPDs (SGPDs). These functions do not contribute to the CFFs and vanish in the forward limit at a given scale, yet remain allowed solutions of the inverse problem. Although QCD evolution constrains the functional space of such contributions~\cite{moffat}, resolving this ambiguity requires additional experimental constraints sensitive to the full kinematic dependence of GPDs.

Double Deeply Virtual Compton Scattering~(DDVCS) \cite{Guidal:2002kt,ddvcs_bm1,ddvcs_bm2,deja:2023,Martinez-Fernandez:2025gub}, characterized by the large virtuality of either the incoming or outgoing photons, provides access to GPDs in a broader kinematic domain. At LO and LT, DDVCS is sensitive to the $x$-dependence of GPDs in the $-\xi<x<\xi$ region, which is otherwise inaccessible in DVCS or TCS alone. As a matter of fact, the imaginary part of the DDVCS amplitude features GPDs evaluated on the line $x$ = $\xi'$, where $\xi'$ is known as the generalized Bjorken variable. However, the cross section for DDVCS is several orders of magnitude smaller than that of DVCS, making its measurement experimentally challenging. Furthermore, to avoid ambiguities associated with identical leptons in the final state, beam-decay correlations, and anti-symmetrization effects, the outgoing timelike photon must be reconstructed through its di-muon decay channel. 

JLab at the luminosity frontier, with large acceptance detectors, is the only place where DDVCS can be measured in the valence region. The CLAS12 detector~\cite{clas12} in Hall~B is particularly suited for such measurements. The upgraded detector presented in this article, $\mu$CLAS12, will be able to study the electroproduction of muon pairs in the reaction $ep\to e^{\prime}\mu^+\mu^-p^\prime$ over a wide range of spacelike and timelike virtualities, covering the skewness range $0.08 \le \xi \le 0.4$ and the generalized Bjorken variable range $-0.2 \le \xi' \le 0.1$. This upgrade effectively turns the Forward Detector (FD) of CLAS12 into a muon spectrometer. This involves installing heavy shielding in front of the FD, replacing the High Threshold \v{C}erenkov Counter (HTCC)~\cite{Sharabian:2020whm}, to suppress electromagnetic and hadronic backgrounds during operation at luminosities close to $10^{37}\,\mathrm{cm^{-2}\,s^{-1}}$.
The shielding assembly will incorporate a new PbWO$_4$ electromagnetic calorimeter for the detection of scattered electrons. In addition, new tracking systems are foreseen to provide improved forward vertex reconstruction and recoil proton detection.

In addition to DDVCS, $\mu$CLAS12 will allow studies of TCS and near-threshold $J/\psi$ production, as the di-muon final state provides a particularly clean environment for these measurements. While both TCS and near-threshold $J/\psi$ production have already been investigated experimentally at JLab, current measurements remain statistically limited. The high-luminosity data expected from $\mu$CLAS12 will enable precision extractions of cross sections and asymmetries, providing new constraints on GPDs and related quantities such as Gravitational Form Factors (GFFs)~\cite{Kobzarev:1962wt, Pagels:1966zza, Polyakov:2018zvc, Burkert2018, Lorce:2021xku, Burkert2023}. Finally, the $J/\psi$ channel will enable searches for hidden-charm pentaquark states previously reported by the LHCb collaboration, where high-statistics measurements are essential for clarification of their nature.

\section{Physics Motivation}

Quantum Chromodynamics (QCD), the non-abelian gauge theory of the strong interaction between quarks and gluons, accounts for the vast majority of the visible mass of the universe through dynamical mass generation. Despite substantial progress, a quantitative description of how nucleon properties emerge from their quark and gluon degrees of freedom remains a central challenge.

Electron scattering has played a crucial role in probing the internal structure of the nucleon. Measurements of EFFs and PDFs have provided complementary information on its spatial and longitudinal momentum structure, respectively. However, these observables access different expressions of the nucleon structure and do not yield a correlated multidimensional picture. The GPD formalism unifies these descriptions by encoding correlations between longitudinal momentum and transverse spatial degrees of freedom. In this framework, the nucleon can be characterized in terms of a multidimensional partonic structure, enabling access to quantities such as the total angular momentum carried by quarks and gluons and offering insight into the origin of nucleon mass and spin.

\subsection{Generalized Parton Distributions}

\subsubsection{Properties of the GPDs}

The GPDs encode off-forward partonic correlations inside the nucleon and interpolate between PDFs and EFFs. In appropriate kinematic limits, they provide access to the longitudinal momentum distribution of partons and to their transverse spatial distribution through Fourier transformation with respect to the transverse momentum transfer~\cite{Burkardt:2000za,Belitsky:2003nz}. At LT, there are four chiral-even GPDs per quark flavor $q$, denoted $H^q$, $E^q$, ${\tilde H}^q$, and ${\tilde E}^q$. The GPDs $H^q$ and ${\tilde H}^q$ conserve the nucleon helicity, while $E^q$ and ${\tilde E}^q$ flip it. GPDs depend on three variables: $x$, $\xi$ and $t$. The variable $x$ is the average of the initial and final parton momentum fractions, defined with respect to the average nucleon momentum $P$=$(p + p')/2$, where $p$ and $p'$ are the incoming and outgoing proton four-momenta, as defined in Fig.~\ref{subfig:DVCS}. The skewness parameter $\xi$ characterizes the fractional longitudinal momentum transfer between the initial and final nucleon states and can be expressed as $\xi $=$ -{(p' - p)^+}/{(p' + p)^+}$ where the $+$ indices refer to the corresponding light-cone coordinate. Finally, the four-momentum transfer squared is defined as $t $=$ (p' - p)^2$. The first moments of the GPDs are respectively related to the Dirac, Pauli, axial-vector and pseudoscalar form factors. In the forward limit ($t\to 0$, $\xi\to 0$), the GPDs $H^q$ and $\tilde H^q$ reduce to unpolarized and polarized PDFs, respectively. A remarkable property of GPDs is their connection to the GFFs which parameterize the QCD energy-momentum tensor (EMT), which are denoted $A_{q}(t)$, $B_{q}(t)$, $C_{q}(t)$, and $\bar{C}_{q}(t)$ using the notations of Ref.~\cite{revdiehl}. For quark GFFs, this connection appears in the second Mellin moments of the GPDs as
\begin{equation}
\begin{split}
\label{gpd_pdf2}
\int_{-1}^1dx~xH^q(x,\xi,t)&=A_{q}(t)+4\xi^2C_{q}(t), \\ 
\int_{-1}^1dx~xE^q(x,\xi,t)&=B_{q}(t)-4\xi^2C_{q}(t).
\end{split}
\end{equation}
On the one hand, the GFFs $A_{q}(t)$ and $B_{q}(t)$ offer crucial input on the proton spin carried by quarks through Ji's sum rule \cite{Ji:1996ek}. On the other hand, the GFFs $C_{q}(t)$ and $\bar{C}_{q}(t)$ can be interpreted in terms of the quark pressure and shear force distributions inside the nucleon~\cite{Polyakov:2018zvc, Burkert2018, Burkert2023}. 

The QCD factorization theorem~\cite{mueller1994wave,Ji:1996nm,Radyushkin:1997,collins:1997} provides the theoretical foundation for accessing GPDs through deeply virtual exclusive processes, including Compton scattering reactions such as DVCS illustrated in Fig.~\ref{subfig:DVCS}, and TCS illustrated in Fig.~\ref{subfig:TCS}. In these reactions, the scattering amplitude factorizes at LT into a convolution of perturbatively calculable hard-scattering coefficient functions and non-perturbative GPDs as

\begin{equation}
\mathcal{A} \sim \sum_{q} \int_{-1}^{1} dx \,
\mathcal{C}_q(x,\xi;\mu^2)\, F^q(x,\xi,t;\mu^2),
\label{eq:factorization}
\end{equation}
\noindent
where $\mathcal{C}_q$ denote the hard-scattering coefficients, $F^q(x,\xi,t)$ the quark GPDs, and $\mu^2$ the renormalization scale. The coefficient functions are complex-valued, while the GPDs are real-valued. Compton scattering with large spacelike or timelike virtuality, therefore, plays a central role among deep exclusive processes in the experimental investigation of GPDs.

\subsubsection{Compton Scattering and GPDs}

\begin{figure*}[htbp]
\centering

\begin{subfigure}{0.4\linewidth}
\centering
\includegraphics[width=\linewidth]{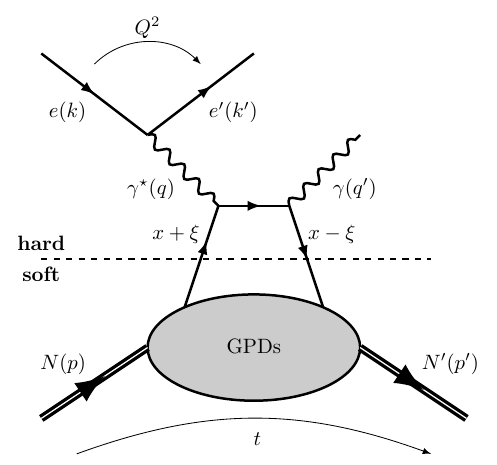}
\caption{}
\label{subfig:DVCS}
\end{subfigure}
\hspace{0.02\linewidth}
\begin{subfigure}{0.4\linewidth}
\centering
\includegraphics[width=\linewidth]{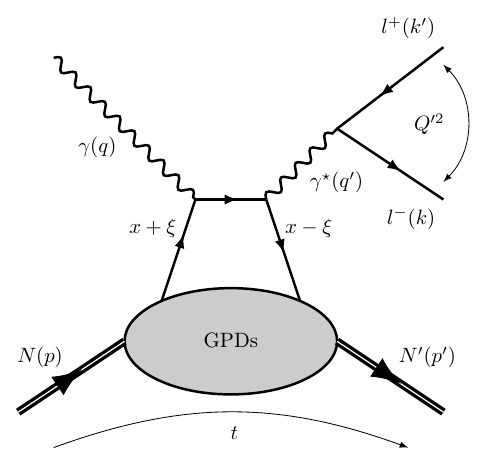}
\caption{}
\label{subfig:TCS}
\end{subfigure}
\\
\begin{subfigure}{0.4\linewidth}
\centering
\includegraphics[width=\linewidth]{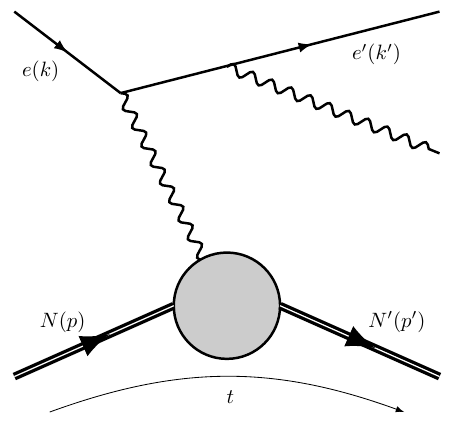}
\caption{}
\label{subfig:BH_DVCS}
\end{subfigure}
\hspace{0.02\linewidth}
\begin{subfigure}{0.4\linewidth}
\centering
\includegraphics[width=\linewidth]{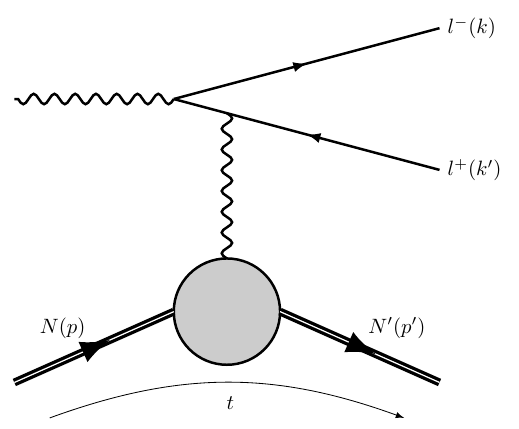}
\caption{}
\label{subfig:BH_TCS}
\end{subfigure}

     \caption{Feynman diagrams for Compton processes. Panel~\ref{subfig:DVCS}: Diagram for DVCS on the nucleon. The dashed line indicates the separation between the perturbatively calculable hard scattering subprocess and the non-perturbative part encoded in the GPDs. The four-momenta of the incoming and scattered electrons are denoted by $k$ and $k'$, respectively; those of the virtual and real photons by $q$ and $q'$; and those of the initial and final nucleons by $p$ and $p'$. Panel~\ref{subfig:TCS}: Diagram for TCS. In this case, $k$ and $k'$ denote the momenta of the produced leptons. Panels \ref{subfig:BH_DVCS} and \ref{subfig:BH_TCS} show one of the corresponding BH contributions to the same final states.}
\label{fig:all_compton_process}
 \end{figure*}

DVCS, the hard exclusive electroproduction of a real photon \mbox{$ep\rightarrow e^{\prime}p^{\prime}\gamma$} (see Fig.~\ref{subfig:DVCS} for the definition of the associated momenta), first proposed in Refs.~\cite{Ji:1996nm,  collins:1997, Radyushkin:1996ru}, stands out as the primary avenue for probing GPDs. Experimental observables of DVCS are parameterized by CFFs, where the GPDs appear in convolution integrals over the longitudinal momentum fraction $x$ of the active parton. At lowest order in $\alpha_s$, the CFFs of DVCS read 
\begin{equation}
\label{eq:CFF}
{\cal F}^\pm(\xi, t) = \int dx \, F(x, \xi, t) \left( \frac{1}{\xi-x-i\epsilon} \mp \frac{1}{\xi+ x-i\epsilon} \right),
\end{equation}
where $F$ stands for a generic GPD, and the top and bottom signs apply to the quark-helicity independent~($H^q$, $E^q$) and the quark-helicity dependent~(${\tilde H}^q$, ${\tilde E}^q$) GPDs respectively, and \mbox{$Q^2$=$-(k-k')^2$}.

At LT and LO in $\alpha_s$, there are four complex-valued CFFs related to the four relevant GPDs. The imaginary part of CFFs contains GPDs evaluated at a specific point $x$=$\pm\xi$ as 
\begin{equation}
\text{Im}[\mathcal{F}^\pm] = i\pi \sum_q e_q^2[F^q(\xi, \xi, t)\mp F^q(-\xi,\xi,t)],
\label{eq:cff_im}
\end{equation}
where $e_q$ is the electric charge of the quark flavor $q$, and are accessible in single spin asymmetry measurements.  
The real part of CFFs, accessible in cross section, double spin and beam charge asymmetries measurements, is defined as the Cauchy principal value integrals of GPDs over $x$,
\begin{equation}
\text{Re}[\mathcal{F}^\pm] = \mathcal{P} \int_{-1}^1 dx \, \left (\frac{1}{\xi-x}\mp \frac{1}{\xi+x}\right )\sum_q e_q^2{F^q(x, \xi, t)}.
\end{equation}

An analysis at LO and LT, is however far from capable of describing data at relatively low scales. It is now recognized that more physics can be extracted from data by accounting for higher-order contributions~\cite{Pire:2011st, Braun:2025xlp, Martinez-Fernandez:2025jvk}. In particular, it has been shown in Refs. \cite{Martinez-Fernandez:2025jvk,Martinez-Fernandez:2025rcg} that extracting both real and imaginary parts of CFFs at higher orders in the kinematic twist expansion allows access to the GFFs $A_{q}(t)$, $B_{q}(t)$, and $C_{q}(t)$. This further motivates the study and measurement of the observables described later in this manuscript. 


Experimentally, DVCS (see Fig.~\ref{subfig:DVCS}) is measured together with the BH process, where the photon emission is mediated by the electron, as shown in Fig.~\ref{subfig:BH_DVCS}. The measured cross section $\sigma_{ep\rightarrow e^{\prime}p^{\prime}\gamma}$ is the coherent sum of two amplitudes, $\mathcal{T}_{DVCS}$ and $\mathcal{T}_{BH}$,
\begin{equation}
\sigma_{ep\rightarrow e^{\prime}p^{\prime}\gamma} \propto |\mathcal{T}_{\rm BH}|^2 + |\mathcal{T}_{\rm DVCS}|^2 + \mathcal I,
\end{equation}
where the interference term is defined as
\begin{equation}
\mathcal I=\mathcal{T}_{\rm BH}^*\mathcal{T}_{\rm DVCS} + \mathcal{T}_{\rm BH}\mathcal{T}_{\rm DVCS}^*.
\end{equation}

The amplitude $\mathcal{T}_{BH}$ depends on the nucleon EFFs and is fully calculable in QED, whereas the DVCS amplitude $\mathcal{T}_{DVCS}$ is expressed in terms of CFFs. In much of the JLab kinematics, the BH contribution to the cross section dominates the pure DVCS contribution. However, the interference term $\mathcal I$ provides a powerful tool to extract CFFs. Its azimuthal angular dependencies, accessible via single and double spin asymmetries, as well as lepton-charge asymmetries, give access to linear combinations of imaginary and real parts of the CFFs. This azimuthal angle is defined according to the Trento convention~\cite{PhysRevD.70.117504}, as the angle formed by the leptonic and hadronic planes, defined in the lab frame, by the initial and outgoing electron momenta, and the outgoing photon and scattered proton momenta, respectively.

The TCS process mirrors the DVCS process, featuring a real incoming photon and an outgoing photon with large timelike virtuality, \mbox{$\gamma p\rightarrow \gamma^*p^\prime\rightarrow l^+l^-p^\prime $}, as shown in Fig.~\ref{subfig:TCS}. 
In TCS, the hard scale is set by the virtuality of the outgoing photon \mbox{$Q'^2$}, which is also the invariant mass squared of the produced lepton pair. As in DVCS, the BH process shown in Fig.~\ref{subfig:BH_TCS}, where the incoming photon couples to the lepton line, contributes to the same final state and typically dominates the cross section for exclusive lepton-pair photoproduction.
A remarkable property of TCS is that the amplitudes for the Compton and BH processes transform with opposite signs under the reversal of the lepton charge. Consequently, the interference term between TCS and BH in the unpolarized cross section is odd under the exchange of the $l^+$ and $l^-$ momenta, while the individual contributions of the two are even. In the handbag approximation, at LT, and using the notations in Ref.~\cite{Berger:2001xd}, the TCS interference term reads
\begin{equation}
d\sigma_{\rm INT} \propto \cos\phi\frac{1+\cos^2\theta}{\sin\theta}\mathrm{Re}\tilde{M}^{--} 
+ O\Big( \frac{1}{Q'} \Big),
\label{eq:tcsint}
\end{equation}
where $\tilde{M}^{--}$ is a combination of CFFs defined in Ref.~\cite{Berger:2001xd}. The TCS angles $\theta$ and $\phi$ are defined in the lepton–pair center-of-mass frame. The polar angle $\theta$ is the angle between the direction of the recoiling proton in this frame and the momentum of the negatively charged lepton $l^-$. The azimuthal angle $\phi$ is the angle between the leptonic plane, defined by $k$ and $k'$, and the hadronic plane, defined by the incoming and outgoing proton momenta $p$ and $p'$. From Eq.~(\ref{eq:tcsint}), it follows that
\begin{equation}
\begin{aligned}
d\sigma_{\mathrm{INT}}(\theta,\phi)
=
-\,d\sigma_{\mathrm{INT}}(\pi-\theta,\pi+\phi),
\end{aligned}
\end{equation}
reflecting the antisymmetry of the interference term under the transformation ($\theta \to \pi-\theta$,  $\phi \to \pi+\phi$). This property provides a direct access to the real part of the Compton Form Factors through appropriate angular projections, as shown in Ref.~\cite{clas12tcs}.

By measuring the real part of the Compton amplitude, one can access the $D$-term~\cite{Polyakov:2018zvc, Polyakov:2002yz, Pasquini:2014vua} in the parameterization of GPDs. On the other hand, the photon beam-polarization asymmetry projects out the imaginary part of the Compton amplitude, similar to the beam-spin asymmetry in DVCS, and allows testing of the universality of GPDs~\cite{Grocholski:2019pqj}. In 2021, the CLAS Collaboration published the first-ever experimental results on both the photon beam-polarization asymmetry and the decay lepton angular asymmetries of TCS~\cite{clas12tcs} using data obtained with the CLAS12 detector, where a $10.6$~GeV electron beam scattered off a hydrogen target.

It is worth noting that NLO corrections to the TCS cross section and asymmetries have been shown to be sizable~\cite{Pire:2011st}. They are particularly sensitive to the gluon GPDs that do not enter the LO calculation. While the TCS amplitude is almost identical to that of DVCS at LO, the calculation of the NLO corrections~\cite{Pire:2011st,Mueller:2012sma} has demonstrated that they develop a sizable difference that is proportional to the derivative of GPDs with respect to the renormalization scale.

\subsubsection{Extraction of GPDs from Experimental Observables}
\label{sec:GDP_exp}

Extracting CFFs from experimental observables, such as asymmetries and cross sections, is a crucial first step in accessing GPDs. Several methodologies have been developed to extract CFFs. These methods involve analyzing observables under well-established theoretical frameworks incorporating the symmetries and kinematic dependencies specific to the scattering process. 

Model-independent CFFs extractions have been performed with local fits at the specific kinematic points at which measurements were made~\cite{fitmick, fithermes, fittsa, fitall, kum2014, jifit, mswbsfit}. The advantage of local fits is their model independence as they directly measure the CFFs. This approach avoids biases introduced by specific GPD parameterizations and focuses solely on the experimental data. However, local fits do not inherently account for correlations across data points or the global structure of GPDs, which can limit their scope for extracting comprehensive insights about the internal structure of hadrons. 

Global fits \cite{herve, kum2014, kum2008, fitmuller} aim to simultaneously describe all available data across the entire kinematic range. These approaches use parameterized models of GPDs, which implicitly define the CFFs. The parameters of the GPD models are then optimized by fitting the entire dataset. Global fits offer several advantages: they provide a consistent description of the data, constrain GPDs over a broader kinematic range, and often lead to smaller uncertainties due to the larger dataset used. However, there are challenges due to the complexity of the GPD parameterizations and the computational demands of the fitting procedure. Different theoretical models for GPDs exist, each with its own set of parameters that need to be determined through the global fit.  

A promising approach to extract CFFs using Neural Networks (NN) is presented in Ref.~\cite{kum2011}. NNs provide a flexible and model-independent way to parameterize CFFs by learning patterns directly from experimental data without imposing rigid functional forms. The works presented in Refs.~\cite{mswfit, mksfit} offer significant advancements in these techniques. They leverage global fits using NNs alongside advanced parameterizations to accurately reconstruct CFFs from the measured data. These approaches are particularly effective in reducing model dependence and ensuring compatibility with the constraints of QCD. 

The next step, inferring GPDs from CFFs, is a challenging inverse problem. As discussed earlier, DVCS and TCS observables depend only on two of the three GPD variables. Therefore, determining GPDs from CFFs is an inherently ill-posed inverse problem with no unique solution. Consequently, various GPD models can reproduce experimental data at different scales. Furthermore, current experimental uncertainties do not strongly favor specific GPD models and parameters, as for example shown for the GPD $H$ in Ref.~\cite{PhysRevD.74.054027}. Moreover, recent studies of deconvolution have revealed the existence of a class of functions, SGPDs, with vanishing CFFs and vanishing forward limit at a given scale $\mu^2$, which contribute to solutions in the GPD extraction~\cite{bertone}. Measuring the evolution of GPDs in $\xi$ and $Q^2$ could be used to exclude a large class of SGPDs \cite{bertone, moffat}. Lattice QCD may also offer a promising approach to determining the $x$‑dependence of GPDs \cite{lattice}. Nevertheless, only a process directly sensitive to the $x$‑dependence of GPDs can address this challenge experimentally.

\subsection{Double Deeply Virtual Compton Scattering}

\subsubsection{Phenomenology Overview}

As mentioned above, DVCS and TCS only probe the $\xi$- and $t$-dependencies of GPDs. In contrast, DDVCS~\cite{Guidal:2002kt,ddvcs_bm1,ddvcs_bm2,deja:2023, Martinez-Fernandez:2025gub}, where both the incoming and outgoing photons are virtual, features two independent scales via the spacelike virtuality $Q^2$ and the timelike virtuality $Q'^2$. This additional degree of freedom enables direct sensitivity to GPDs away from the $x$=$\pm\xi$ ridge, already at LO and LT.

\begin{figure}[htbp]
  \centering
     \includegraphics[width=\linewidth]{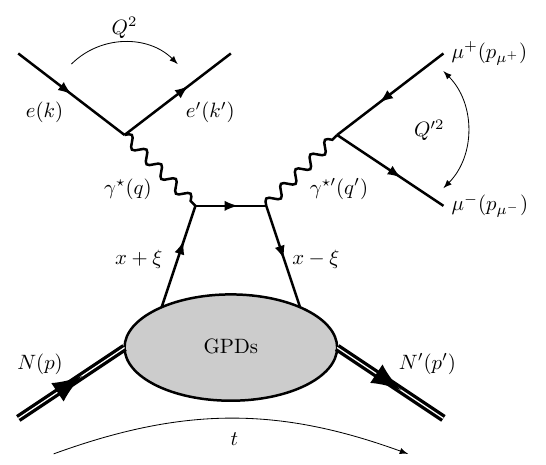}
\caption{Handbag diagram for DDVCS with a di-muon final state.}
\label{fig:ddvcs}
 \end{figure}

The DDVCS process can be measured using the exclusive electroproduction of a pair of leptons, \mbox{$ep\rightarrow e^\prime  \gamma^*p^\prime\rightarrow e^\prime  l^+l^-p^\prime$}.
At LT and LO, DDVCS can be interpreted as the absorption of a spacelike photon by a parton inside the nucleon and emission of a timelike photon, which then decays into a lepton pair, as shown in Fig.~\ref{fig:ddvcs}. 
CFFs enter the DDVCS amplitude as convolution integrals over the longitudinal momentum of the probed parton, which, at the lowest order in $\alpha_s$, reads
\begin{equation}
\begin{aligned}
{\cal F}^\pm(\xi^\prime,\xi, t) = &\mathcal{P} \int dx  \left( \frac{1}{\xi^\prime-x} \mp \frac{1}{\xi^\prime+ x} \right)\sum_q e_q^2{F^q(x, \xi, t)} \\ &+  i\pi \sum_q e_q^2 [F^q(\xi^\prime,\xi,t)\mp F^q(-\xi^\prime,\xi,t)],
\end{aligned}
\label{eq:cff_gpd}
\end{equation}
where the scaling variables, the skewness $\xi$, and the generalized Bjorken variable $\xi^\prime$, are defined in Ref~\cite{Martinez-Fernandez:2025gub} as
\begin{equation}
\begin{split}
\xi&=\frac{ \sqrt{ (Q^2+Q^{\prime 2})^2 + t^2 +2t(Q^2-Q'^2) } }{2Q^2/x_B-Q^2-Q^{\prime 2}+t}, \\
\xi^\prime&=\xi\frac{Q^2-Q'^2 +t}{Q^2+Q'^2 +t},\\ 
\end{split}
\end{equation}
where $x_B$=$Q^2/(2p\cdot q)$ is the usual Bjorken variable.
Therefore, by varying the virtualities of the incoming and outgoing photons, one can map the GPDs as a function of both $x$=$\xi^\prime$ and $\xi$. 

\begin{figure*}[htbp]
  \centering
  \begin{subfigure}{0.4\linewidth}
\centering
\includegraphics[width=\linewidth]{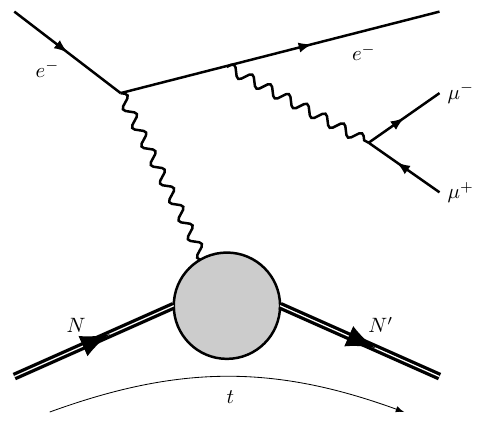}
\caption{}
\label{subfig:BH_DDVCS1}
\end{subfigure}
\hspace{0.02\linewidth}
\begin{subfigure}{0.4\linewidth}
\centering
\includegraphics[width=\linewidth]{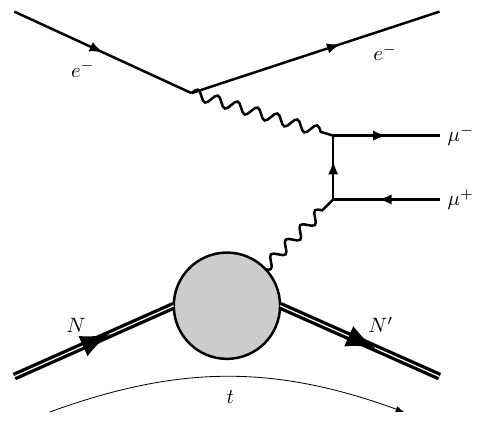}
\caption{}
\label{subfig:BH_DDVCS2}
\end{subfigure}
\caption{Feynman diagrams for the Bethe–Heitler (BH) processes associated with DDVCS. The two diagrams shown correspond to BH1 (Panel~\ref{subfig:BH_DDVCS1}) and BH2 (Panel~\ref{subfig:BH_DDVCS2}). The crossed counterparts are not displayed.}
\label{fig:ddvcs_bh}
 \end{figure*} 
\begin{figure}[htbp]
\centering
    \includegraphics[width=\linewidth]{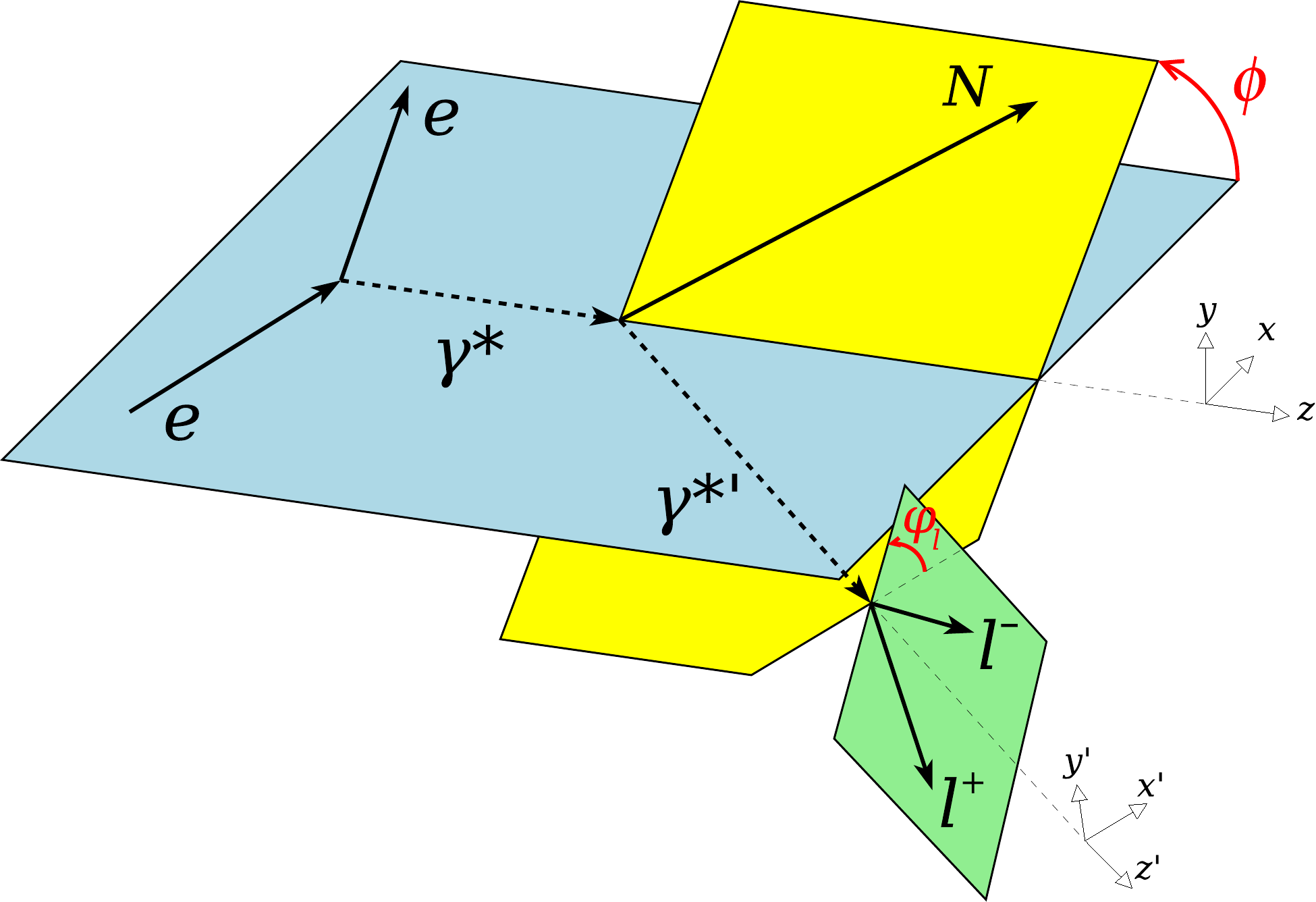}
    \caption{Scattering planes and the definition of angles in lepton-pair electroproduction.}
    \label{fig:dDVCS_planes}
\end{figure} 

As in the case of DVCS and TCS, the BH process interferes at the amplitude level. The diagrams of the two BH processes associated with DDVCS are shown in Fig.~\ref{fig:ddvcs_bh}. With these three interfering processes, the 7-fold differential cross section of electroproduction of lepton pairs can be expressed following the prescriptions in Refs.~\cite{Guidal:2002kt, ddvcs_bm1, ddvcs_bm2}, and using the notations of Ref.~\cite{ddvcs_bm1}, as
\begin{equation}
\begin{aligned}
 {d^7\sigma\over{dQ^2dtdx_Bd\phi dQ^{\prime 2}}d\Omega_l} = 
    {\alpha\over{16(2\pi)^3}}{x_By\over{Q^2}}(|\mathcal{T}_{\rm BH1}+\mathcal{T}_{\rm BH2}|^2 
    \\ +|\mathcal{T}_{\rm VCS}|^2
    +\mathcal{I}_1
    +\mathcal{I}_2),
\label{eq:ddvcs7}
\end{aligned}
\end{equation}
where the interference terms are written as
\begin{equation}
  \mathcal{I}_{1(2)}=\mathcal{T}_{\rm VCS} \mathcal{T}^*_{\rm BH1(2)}+\mathcal{T}_{\rm VCS}^* \mathcal{T}_{\rm BH1(2)}.   
\end{equation}
The solid angle of the pair of leptons is defined as $d\Omega_l$=$\sin{\vartheta_l} d\vartheta_ld\varphi_l$, $\alpha$ is the fine structure constant, and $y$=$p\cdot q/p\cdot k$. The definitions of the angles $\phi$ and $\varphi_l$ are given in Fig.~\ref{fig:dDVCS_planes}, while $\vartheta_l$ is defined as the angle between the negatively-charged lepton and the outgoing virtual photon momenta.
The most direct information on GPDs is encoded in the observables that arise from the interference of VCS and BH amplitudes. In DDVCS, isolating the interference terms is more involved than in DVCS or TCS. There are two BH processes and three interference terms (including the interference between the two BH processes). The VCS amplitude is odd under the beam lepton charge interchange and even under the exchange of the momenta of the decay leptons. The first BH amplitude, labeled BH1 in Fig.~\ref{fig:ddvcs_bh}, is even, while the second, labeled BH2, is odd with respect to the interchange of both the beam lepton charge and the momenta of the decay leptons. These symmetries allow access to DVCS-like single-spin asymmetries, such as the longitudinal beam-spin asymmetry, in the five-fold differential cross section measurement. Here, the integration over the solid angle of the decay leptons, $\Omega_l$, eliminates the contribution of the interference of BH2 with the other two amplitudes and only the interference term $\mathcal{T}_{\rm VCS}\mathcal{T}^*_{\rm BH1}$, denoted $d^5\sigma_{\rm Int_1}$, remains,
\begin{equation}
\begin{aligned}
    {d^5\sigma\over{dQ^2dtdx_Bd\phi dQ^{\prime 2}}}  
    =~& d^5\sigma_{\rm BH1}+d^5\sigma_{\rm BH2} +d^5\sigma_{\rm VCS} \\ &+d^5\sigma_{\rm Int_1} +\lambda(d^5\tilde{\sigma}_{\rm VCS}+d^5\tilde{\sigma}_{\rm Int_1}),
\label{eq:ddvcs5}
\end{aligned}
\end{equation}
where $\tilde{\sigma}_X$ are beam polarization-dependent cross sections, and $\lambda$ is the beam polarization. The polarized cross section difference then read
\begin{equation}
    \Delta\sigma_{ LU}={d^5\overrightarrow{\sigma}-d^5\overleftarrow{\sigma}}=d^5\tilde{\sigma}_{\rm VCS}+d^5\tilde{\sigma}_{\rm Int_1}.
    \label{eq:slu_ddvcs5}
\end{equation}
The $d^5\tilde{\sigma}_{VCS}\propto \text{Im}[\mathcal{T}_{VCS}\mathcal{T}^*_{VCS}]$ is expected to be negligible as it arises from twist-three CFFs~\cite{BELITSKY2002323}, and the beam-spin asymmetry is proportional to $\text{Im}[\mathcal{T}^*_{BH_1}\mathcal{T}_{VCS}]$ and depends on a linear combination of CFFs as
\begin{equation}
\begin{aligned}
    \Delta\sigma_{ LU} \propto & \operatorname{Im}[F_1\mathcal{H}(\xi^\prime,\xi,t)
    \\ + & \xi^\prime (F_1+F_2)\tilde{\mathcal{H}}(\xi^\prime,\xi,t)
    - {t\over{4M^2}}F_2\mathcal{E}(\xi^\prime,\xi,t)]\sin{\phi},
\end{aligned}
\end{equation}
where the imaginary part of the CFFs relates to GPDs at $\xi^\prime$ and $\xi$.

\subsubsection{Observables of Interest}

DDVCS has gained significant attention in theoretical and phenomenological studies due to its potential to provide detailed information on GPDs. Early theoretical frameworks laid the foundation for understanding DDVCS in terms of GPDs \cite{Guidal:2002kt, ddvcs_bm1, ddvcs_bm2} and provided predictions for cross sections and beam-helicity asymmetries, emphasizing the feasibility of experimental measurements. Subsequent studies further advanced the theoretical understanding \cite{Martinez-Fernandez:2025gub} and phenomenological modeling \cite{deja:2023} of DDVCS, while highlighting key measurements \cite{deja:2023, zhao:2021, alvarado2025} enabled by improved GPD parameterizations and expanding experimental capabilities at facilities such as JLab (including the positron program~\cite{Accardi:2020swt} and the $22$ GeV upgrade~\cite{Accardi:2023chb}) and the Electron Ion Collider (EIC)~\cite{AbdulKhalek:2021gbh}. 

These studies stressed the importance of measuring the beam-spin asymmetries, defined as 
\begin{equation}
    A_{ LU}={{d^5\overrightarrow{\sigma}-d^5\overleftarrow{\sigma}}\over{d^5\overrightarrow{\sigma}+d^5\overleftarrow{\sigma}}},
    \label{eq:alu_ddvcs5}
\end{equation}
in both spacelike ($Q^2>Q^{\prime 2}$) and timelike ($Q^2<Q^{\prime 2}$) regions. A sign change of $A_{LU}$ is expected in the transition, providing a stringent test of the non-perturbative QCD regime. Studies in Ref.~\cite{zhao:2021} show how one can combine measurements of beam-charge and spin asymmetries with polarized $e^-$ and $e^+$ beams to separate the interference ($d^5\tilde{\sigma}_{\rm Int_1}$) and the DDVCS ($d^5\tilde{\sigma}_{\rm VCS}$) terms in Eq.~(\ref{eq:slu_ddvcs5}). Moreover, combining beam-charge and decay-lepton angular asymmetries offers clean access to the real part of the DDVCS and BH interference part, $\text{Re}[\mathcal{T}_{\rm VCS}\mathcal{T}^*_{\rm BH1}]$. In Ref.~\cite{deja:2023}, numerical estimates of DDVCS observables at the kinematics of JLab and the EIC using the PARTONS software \cite{partons} are presented, comparing different GPD models and demonstrating the measurability of the DDVCS reaction at these experimental facilities. Finally, the most relevant study of DDVCS measurements at JLab has been published in Ref.~\cite{alvarado2025}, and the expected results of $\mu$CLAS12 and the planned SoLID spectrometer~\cite{JeffersonLabSoLID:2022iod} are discussed.

\begin{figure*}[htbp]
  \centering   
\begin{subfigure}{0.49\linewidth}
\centering  \includegraphics[width=\linewidth]{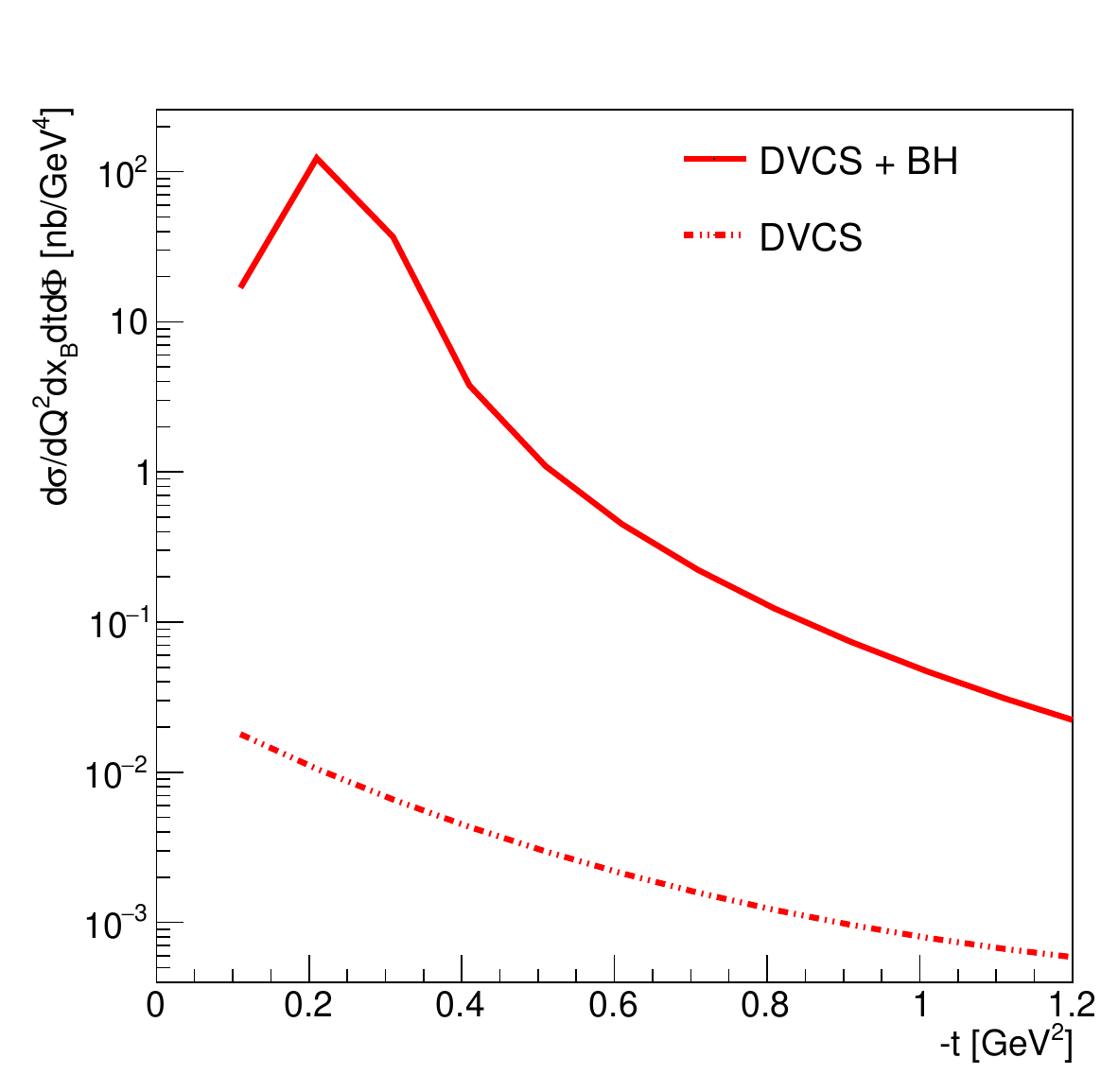}   
  \caption{}
\label{fig:xs_dDVCS_1}
\end{subfigure}
\begin{subfigure}{0.49\linewidth}
\centering  
\includegraphics[width=\linewidth]{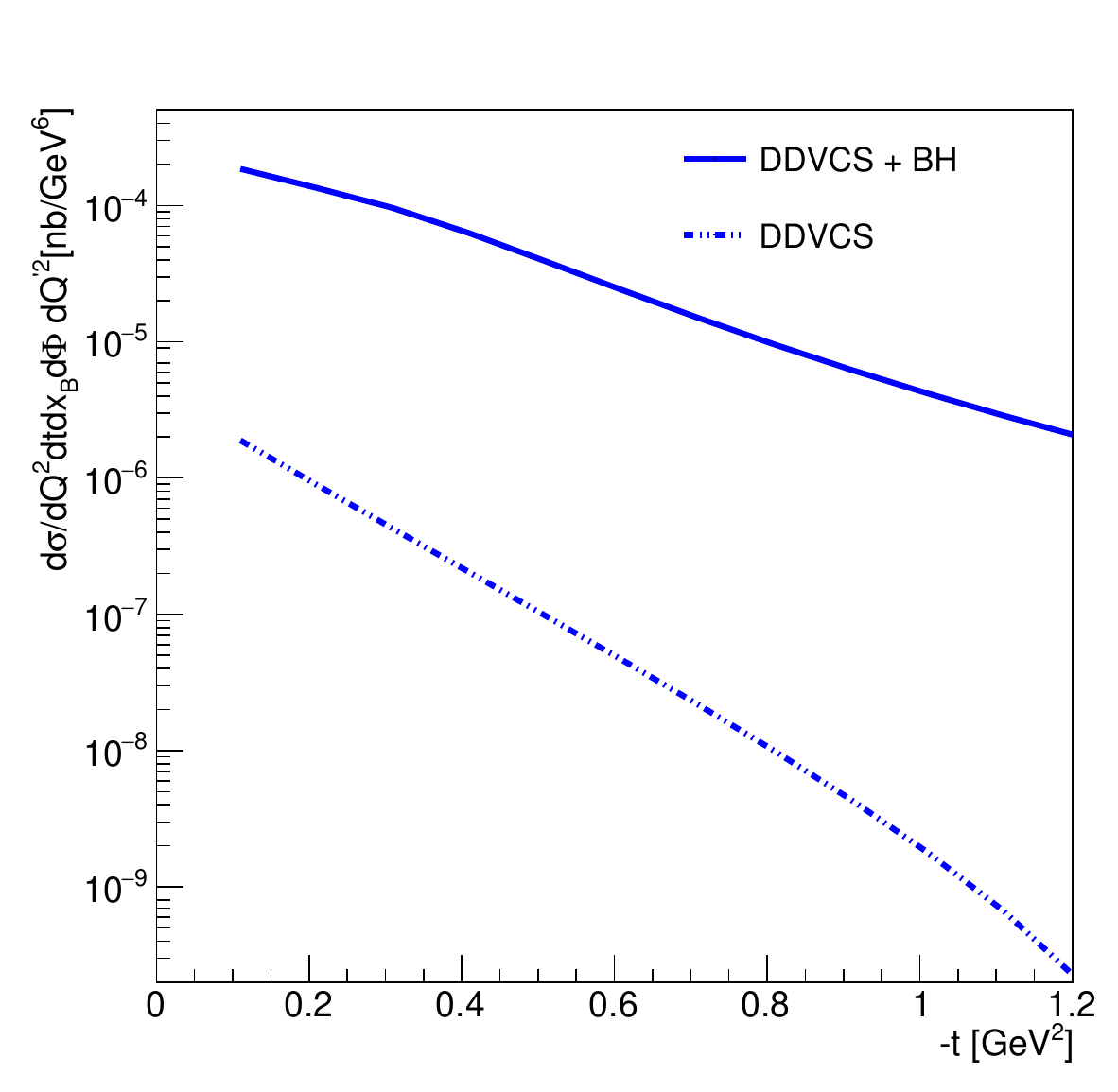}
\caption{}
\label{fig:xs_dDVCS_2}
\end{subfigure}
  
\caption{Differential cross sections for DVCS (Panel~\ref{fig:xs_dDVCS_1}) and DDVCS (Panel~\ref{fig:xs_dDVCS_2}) with a $10.6~\mathrm{GeV}$ electron beam, obtained using the PARTONS software~\cite{partons}. 
The scattered-electron kinematics are fixed at $Q^2 \,$=$\, 2.75~\mathrm{GeV}^2$ and $x_B \,$=$\, 0.15$. 
For DDVCS, the virtuality of the timelike photon is $Q'^2 \,$=$\, 1.4~\mathrm{GeV}^2$.}
\label{fig:xs_dDVCS}
 \end{figure*}

The main challenge of studying DDVCS experimentally is its cross section, which is orders of magnitude smaller than that of DVCS. In Fig.~\ref{fig:xs_dDVCS}, the differential cross sections for DVCS+BH~(Panel~\ref{fig:xs_dDVCS_1}) and DDVCS+BH~(Panel~\ref{fig:xs_dDVCS_2}) are presented for a 10.6~GeV electron scattering off a proton. The kinematics of the scattered electron is fixed at $Q^2$=$2.75$ GeV$^2$ and $x_B$=$0.15$. For DDVCS, the virtuality of the outgoing photon is $Q^{\prime 2}$=$1.4$ GeV$^2$. In the whole $t$-range of interest, cross sections for DVCS and DVCS+BH are about four orders of magnitude larger than those of DDVCS and DDVCS+BH. 
Moreover, the outgoing timelike photon must decay into a different lepton flavor than that of the initial beam, in order to avoid ambiguities and antisymmetrization effects.

To overcome these challenges and study DDVCS in the \mbox{$ep\rightarrow e^\prime \mu^+\mu^-p^\prime$} channel, a large-acceptance detector such as $\mu$CLAS12, capable of operating at very high luminosities (~\mbox{$\ge 10^{37}\,\mathrm{cm^{-2}\,s^{-1}}$}) while maintaining excellent muon detection capabilities, is required.  

\subsection{Near-Threshold $J/\psi$ Production}

In addition to the critical measurement of DDVCS, $\mu$CLAS12 will also be capable of detecting muon pairs originating from the decay of \jpsi~mesons. Although the branching ratio for \mbox{$J/\psi \rightarrow \mu^+ \mu^-$} is only 6\%, the high luminosity of the experiment, combined with efficient muon detection, will enable the collection of a large number of \jpsi~events. Such a dataset will provide the foundation for a detailed investigation of the gluon content of the proton.

\begin{figure*}[h!]
\centering
\begin{subfigure}{0.31\linewidth}
\centering
\includegraphics[width=\linewidth]{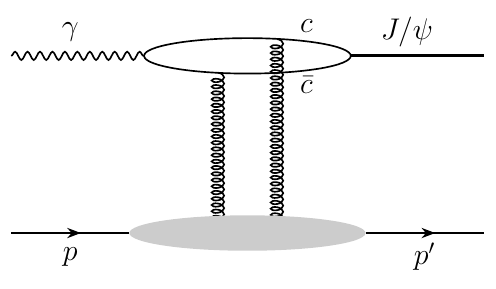}
\caption{}
\label{subfig:JPsigluon}
\end{subfigure}
\hspace{0.02\linewidth}
\begin{subfigure}{0.31\linewidth}
\centering
\includegraphics[width=\linewidth]{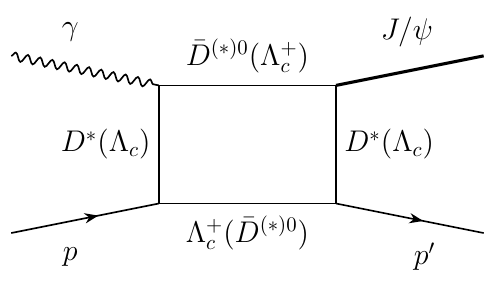}
\caption{}
\label{subfig:opencharm}
\end{subfigure}
\hspace{0.02\linewidth}
\begin{subfigure}{0.31\linewidth}
\centering
\includegraphics[width=\linewidth]{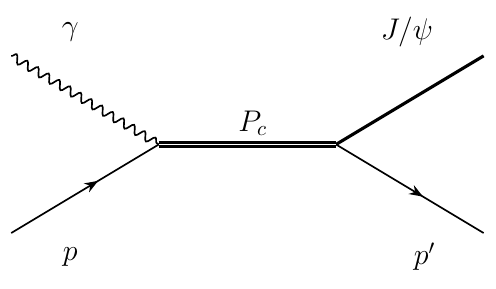}
\caption{}
\label{subfig:pentaquark}
\end{subfigure}

\caption{Representative diagrams contributing to the amplitude for $J/\psi$ photoproduction on the proton. Panel~\ref{subfig:JPsigluon}: two-gluon exchange mechanism. Panel~\ref{subfig:opencharm}: open-charm box diagrams. Panel~\ref{subfig:pentaquark}: pentaquark production.}
\label{fig:diagram}
\end{figure*}

Indeed, the exclusive photoproduction of the $J/\psi$ meson has long been recognized as a key process to probe the gluon content of the nucleon~\cite{KHARZEEV1999568}. Figure~\ref{fig:diagram} illustrates the reaction diagram, under the assumption that the produced $J/\psi$ interacts with the nucleon exclusively via 2-gluon exchange. Recent theoretical developments~\cite{PhysRevD.100.014032, PhysRevD.103.096010, PhysRevD.104.054015, Mamo:2022eui, Guo:2023pqw} suggest that the gluon GFFs of the proton can be accessed experimentally through the measurement of the $t$-dependence of the cross section. It is also worth noting that recent progress in Lattice QCD has led to reliable extractions of the gluon GFFs, as demonstrated in Refs.~\cite{Shanahan:2018pib, Pefkou:2021fni, Hackett:2023rif}.

\subsubsection{Current Experimental Status}

Although exclusive photoproduction of $J/\psi$ on a proton has already been measured in experiments at Cornell~\cite{PhysRevLett.35.1616}, SLAC~\cite{PhysRevLett.35.483}, HERA~\cite{Zeus:2002fa,H1:2005dtp,H1:2013okq}, and at the LHC in ultra-peripheral collisions~\cite{ALICE:2014eof, ALICE:2018oyo, LHCb:2018rcm}, such a measurement near the production threshold only became feasible thanks to the 12~GeV upgrade of the CEBAF accelerator at JLab~\cite{Adderley:2024czm}.

The measurement by the GlueX experiment, reported in Refs.~\cite{gluexjp:2019,gluexjp:2023}, used a tagged-photon beam incident on a hydrogen target, and the $J/\psi$ meson was reconstructed via its electron-positron decay channel. Both the total cross section as a function of the incident real photon energy $E_\gamma$ and the differential cross section as a function of $-t$ were extracted. The E12-16-007 experiment in Hall C at JLab (also referred to as \jpsi-007) used an untagged photon beam scattering off a proton target. The lepton pair (both electron-positron and di-muons) from the decay of the $J/\psi$ meson was detected in the two focusing magnetic spectrometers positioned on both side of the incoming beam. This experiment measured the differential cross sections as a function of $-t$ in Ref.~\cite{hallc:007}, and the total cross section as a function of incident photon energy in Ref.~\cite{007:2026dow}. The CLAS12 experiment in Hall B collected data on a proton target in 2018 and 2019. The results of the E12-12-001 and E12-12-001A experiments~\cite{E12_12_001, E12_12_001A}, on the measurements of the total cross section as a function of $E_\gamma$ and the $t$-dependent differential cross section have been published in Ref.~\cite{CLAS:2026lls}.

\subsubsection{Gluon GFFs of the Proton}
\label{sec:theo_models}

The GFFs of the proton have been an active topic of recent research. They appear in the matrix element of the QCD energy-momentum tensor, and the gluon GFFs can be related to the gluon GPDs via their integration of the momentum fraction $x$ as
\begin{equation}
\begin{split}
\int_0^1dx\,H_g(x,\xi,t)  =  A_g(t) + 4\xi^2C_g(t),\\
\int_0^1dx\, E_g(x,\xi,t)  =  B_g(t) - 4\xi^2C_g(t).
\end{split}
\end{equation}

Assuming vector-meson dominance, i.e. the exchange of a pair of gluons between the proton and the $J/\psi$, as depicted in Fig.~\ref{fig:diagram}, various models have been developed to relate the differential cross section of near-threshold $J/\psi$ photoproduction to the gluon GFFs of the proton. Models based on holographic QCD have been developed in Refs.~\cite{Mamo:2022eui,PhysRevD.104.066023,PhysRevD.101.086003,PhysRevD.103.094010}, while models based on GPDs have been detailed in Refs.~\cite{PhysRevD.103.096010,Guo:2023pqw}. The transverse and shear pressure distributions generated by gluons in the proton can then be inferred from the Fourier transform of the $C_g(t)$ form factor following the prescription of Refs.~\cite{Polyakov:2018zvc,Lorce:2018egm}. The mass radius $\langle r_m^2 \rangle_g$ and the scalar radius $\langle r_s^2 \rangle_g$ of the proton, defined in Ref.~\cite{Ji:2021mtz}, can also be extracted from $A_g(t)$ and $C_g(t)$.

An alternative description of $J/\psi$ photoproduction, based on Regge theory, has been proposed in Ref.~\cite{Tang:2025qqe}. This model reproduces reasonably well both the total cross section near the $J/\psi$ production threshold and the differential cross section as a function of the momentum transfer $t$. Such a
model, solely based on Pomeron exchange, offers a description of the process which is unrelated to the gluon dynamics in the proton.

\subsubsection{Open-Charm and Pentaquark Contributions}

The interpretation of the \jpsi~differential cross section in terms of gluon GFFs is valid if the process is dominated by the exchange of two gluons. However, to accurately describe the near-threshold dynamics, it is essential to consider additional contributions such as open-charm loops and potential pentaquark states, as illustrated in Figs.~\ref{subfig:opencharm} and~\ref{subfig:pentaquark}. Significant theoretical efforts have been devoted to estimating the impact of these mechanisms and identifying their potential signatures in the data. Open-charm contributions have been studied in Refs.~\cite{Du:2020bqj, PhysRevD.108.054018}, while effects of pentaquarks have been discussed in Refs.~\cite{Eides:2015dtr, Kubarovsky:2015aaa, Guo:2015umn, Blin:2016dlf, Strakovsky:2023kqu}. The results suggest that VMD might not be applicable and emphasize the necessity for further experimental data, especially as a function of the incoming photon energy in the range of the $\Lambda_c\overline{D}$ and $\Lambda_c\overline{D}^*$ energy thresholds, at 8.7 and 9.4~GeV.

\section{Detector Configuration}
\label{detector}
\vspace{-0.cm}

The $\mu$CLAS12 setup will utilize a modified CLAS12 detector for enhanced muon capabilities and operation at luminosities greater than $10^{37}\,\mathrm{cm^{-2}\,s^{-1}}$. The CLAS12 detector~\cite{clas12} has been in operation since 2018, successfully collecting data on cryogenic, solid, and polarized targets with electron beams energies up to 10.6~GeV, at close to the design luminosity of $10^{37}\,\mathrm{cm^{-2}\,s^{-1}}$. 
The success of running such an open-acceptance detector at high luminosities relies on effectively shielding sensitive elements from electromagnetic background, a significant fraction of which is caused by M{\o}ller scattering. The CLAS12 FD is shielded from this background with the help of the $5$~T field of the CLAS12 solenoid magnet~\cite{Fair:2020yfx} and a so-called M{\o}ller cone, made of tungsten, that covers forward angles up to $2.5^\circ$. A well-shielded FD is the core element of the $\mu$CLAS12 detector for the DDVCS experiment.

The performance of CLAS12 in terms of efficiencies and resolutions is well understood and supported by a validated GEANT4-based Monte Carlo (MC) model, GEMC \cite{gemc}. Since its inception, significant efforts have been made to enhance reconstruction and particle identification (PID) algorithms.
One recent advancement has been Machine Learning (ML) for forward tracking, with now over $90\%$ efficiency at luminosities exceeding design specifications~\cite{Thomadakis:2022zcd}, and future tracking detector upgrades are also expected to enable CLAS12 to operate efficiently at twice the design luminosity~\cite{Gnanvo:2024jag}.

\begin{figure}[htbp]
\centering
{\includegraphics[width=\linewidth]{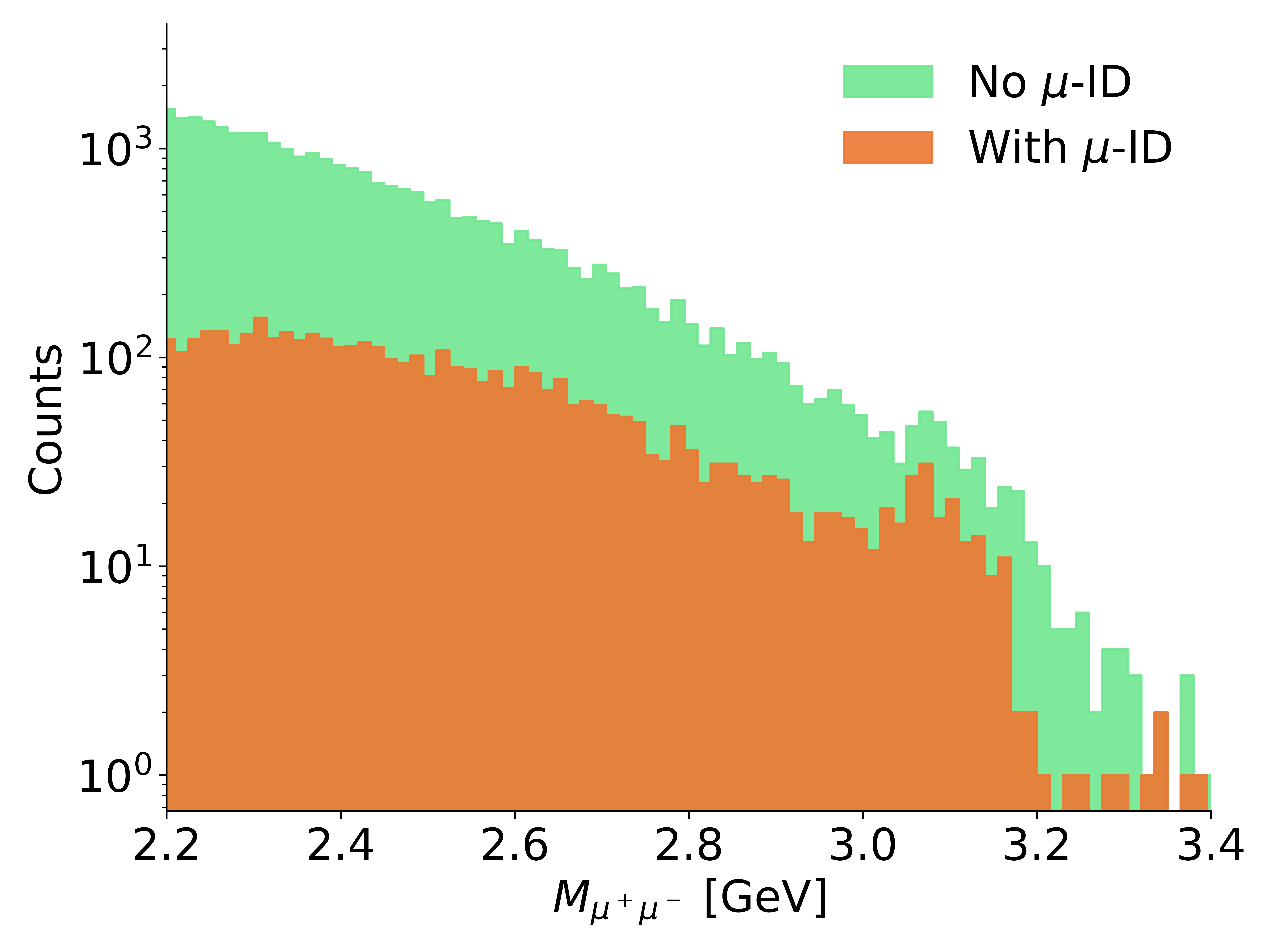}}
\caption{Invariant mass distribution of 2-MIPs events without any muon ID (green) and with the ML-based muon ID (orange). Without muon ID, the spectrum is mostly dominated by pion pairs. After applying the muon PID, the $J/\psi$ signal becomes more visible. Figure adapted from Ref.~\cite{Tyson:2023yer}.}
\label{fig:mujpsi}
\end{figure}

ML-based methods have also been applied to significantly improve particle identification in kinematic regions outside the reach of traditional methods. As demonstrated in Refs.~\cite{clas12tcs, eai:tenorio}, a dedicated ML-based algorithm has allowed a clean separation of electrons and positrons from pions with momenta above $4.7$~GeV/$c$, the pion threshold in the CLAS12 HTCC.
For the $\mu$CLAS12 program, the muon PID algorithm, based on Boosted Decision Trees (BDT) and developed for the $J/\psi$ studies with CLAS12, will be particularly valuable. This tool leverages information from the forward calorimeter to enhance the purity of muon identification. In Fig.~\ref{fig:mujpsi} adapted from Ref.~\cite{Tyson:2023yer}, the implementation of the ML muon identification is demonstrated. The invariant mass distribution of pairs of opposite-charge particles with energy deposition in the forward Electromagnetic Calorimeter~(ECal)~\cite{Asryan:2020iqj} consistent with Minimum Ionizing Particles~(MIP) is shown in green. The distribution is dominated by pairs of pions, and no clear peak is visible in the $J/\psi$ mass region. With a cut applied on the BDT output (in orange), the $J/\psi$  peak becomes visible. This classifier eliminates more than $90\%$ of the background events, reducing the single-pion contamination in the muon sample by a factor of at least three.

\subsection{The $\mu$CLAS12 Detector}

With the final states $e^\prime\mu^+\mu^-(p^\prime)$ and $\mu^+\mu^-p^\prime(e^\prime)$ in mind, the planned modifications to CLAS12 pursue several goals. These include shielding the FD from electromagnetic background for high-luminosity operation, enhancing muon identification with a charged-pion suppression factor of over 100 (to keep the pion-pair background below a few percent of the reconstructed $e'\mu^+\mu^-$ events), providing electron reconstruction and vertex determination, and detecting the recoil proton in a high-background environment.

\begin{figure*}[htbp]
\centering
\includegraphics[width=\linewidth, , page = 2]{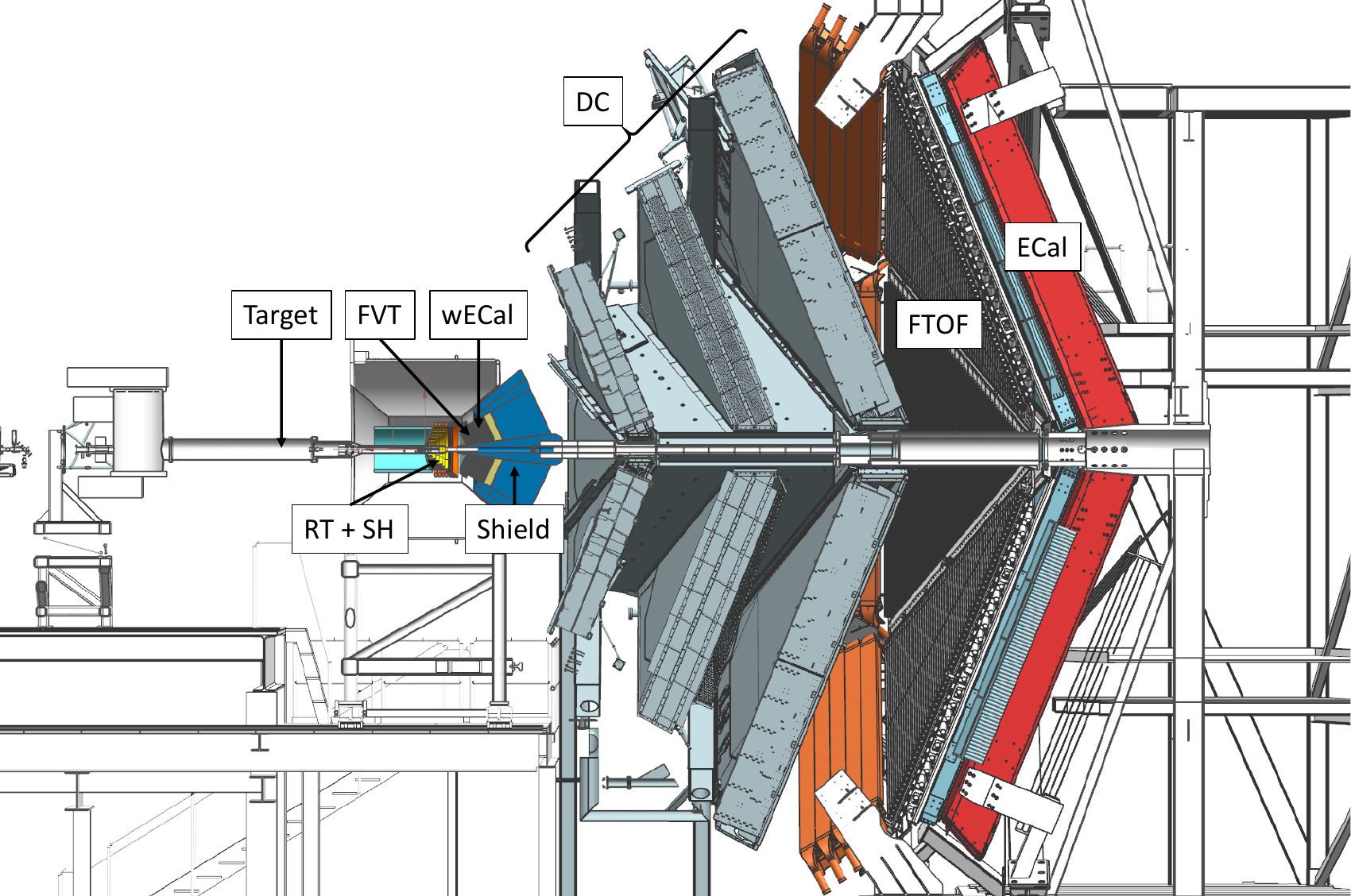}
\caption{Conceptual layout of $\mu$CLAS12. The lead shield and Electron Calorimeter (wECAL) are installed in place of the CLAS12 High Threshold \v{C}erenkov Counter (HTCC)~\cite{Sharabian:2020whm}. The Forward Vertex Tracker (FVT) is placed in front of the wECAL. The Recoil Tracker (RT) and Scintillation Hodoscope (SH) surround the target, and are located inside the existing CLAS12 solenoid magnet~\cite{Fair:2020yfx}. In the Forward Detector (FD), the existing Drift Chambers (DC)~\cite{Mestayer:2020saf}, forward Electromagnetic Calorimeter (ECal)~\cite{Asryan:2020iqj}, and Forward Time-Of-Flight (FTOF)~\cite{Carman:2020fsv} are kept in the $\mu$CLAS12 configuration.}
\label{fig:muclas12}
\end{figure*}

\begin{figure*}[htbp]
\centering
\includegraphics[width=1\textwidth, page =1]{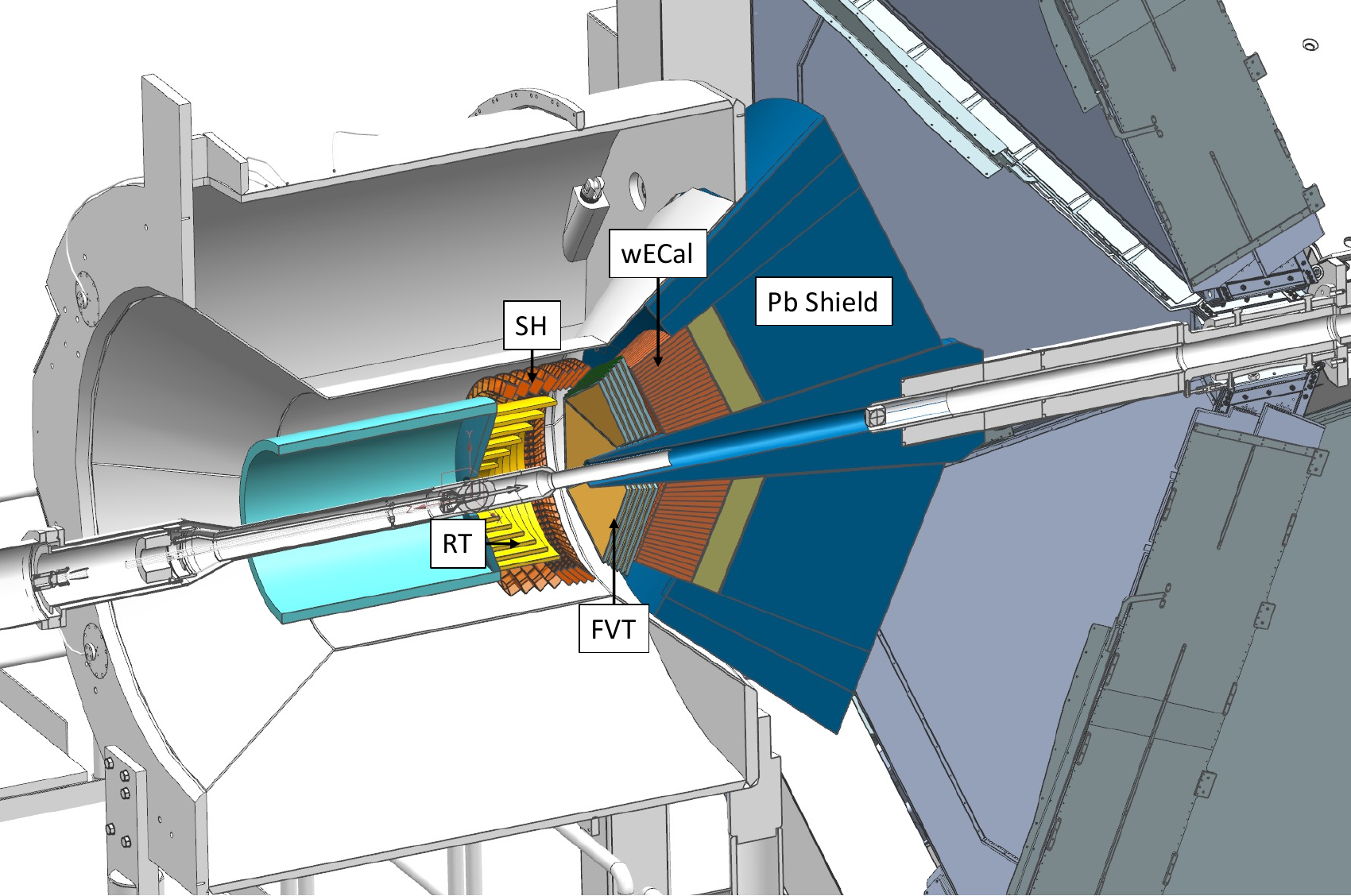}
\caption{Overview of the new detector elements in the $\mu$CLAS12 configuration. This includes a lead shield, shown in dark blue, the wECal located in front of the shield. The FVT will be installed in front of the wECal. 
A recoil detection system composed of a SH and a RT will provide precise tracking for protons recoiling in the polar angle 
range between $40^\circ$ and $70^\circ$.}
\label{fig:recoil}
\end{figure*}

These goals will be achieved by replacing the CLAS12 HTCC with a PbWO$_4$ calorimeter (wECal) surrounded by a lead shield. Before the calorimeter and shield, a high-rate Forward Vertex Tracker (FVT) will replace the existing Forward Micromegas detectors. A new detector for recoil proton detection, composed of a scintillation hodoscope and Micro-Pattern Gaseous Detector (MPGD), will replace the Central Vertex Tracker system (CVT)~\cite{Antonioli:2020ylv, Acker:2020qkv}, the Central Time-of-Flight counters (CTOF)~\cite{Carman:2020yma}, and the Central Neutron Detector (CND)~\cite{Chatagnon:2020lwt} of CLAS12. The Backward Neutron Detector (BAND)~\cite{Segarra:2020txy} and the Low Threshold Cherenkov Counter (LTCC)~\cite{Ungaro:2020hbs} will also be removed. Other changes for converting CLAS12 into $\mu$CLAS12 include removing the Forward Tagger system (FT)~\cite{ftcal} and extending the coverage of the M{\o}ller cone to $7^\circ$ in polar angle. The new PbWO$_4$ calorimeter and the $60$-cm-thick Pb-shield will cover the $7^\circ$ to $35^\circ$ polar angular range with $2\pi$ azimuthal coverage. The conceptual design of the $\mu$CLAS12 setup has been modeled in CAD and is shown in Fig.~\ref{fig:muclas12}. A detailed overview of the new detector elements is provided in Fig.~\ref{fig:recoil}. 

 The CLAS12 GEANT4 model has been modified to create a $\mu$CLAS12 model based on this design. Together with the CLAS12 event reconstruction algorithms, COATJAVA \cite{coatjava}, this set of tools was used to study backgrounds, occupancies, rates, and to validate muon identification and event reconstruction. At the time of this study, some of the new detector components did not have a full digitization or reconstruction algorithm implemented. In these cases, their responses were modeled based on the performance of comparable CLAS12 subsystems. For example, the energy resolution of the wECal was taken from the FT calorimeter (FTCal)~\cite{ftcal}, the vertex reconstruction for the FVT was assumed to match that of the CLAS12 Micromegas Vertex Tracker (MVT)~\cite{Acker:2020qkv}, and the recoil detector was expected to perform similarly to the CLAS12 CVT and CTOF. 

 The following sections describe the new detectors, the beamline, and the target system in detail.
 As demonstrated below, $\mu$CLAS12 will operate at luminosities equal or larger than $10^{37}\,\mathrm{cm^{-2}\,s^{-1}}$ and produce high-quality data with wide kinematical coverage. 

\subsubsection{Electron Calorimeter}

The electron calorimeter wECal will be mounted at $60$~cm from the center of the target and consists of about $1320$, $20$-cm long, PbWO$_4$ modules. Lead tungstate is the only viable technology for this application due to its compactness, radiation hardness, and good energy resolution in a high-rate environment. Tapered crystals will be arranged to form a ring around the beamline with a hole in the center, similar to the Inner Calorimeter of the Hall~B DVCS experiment~\cite{ic}. The central hole will extend to $7^\circ$, with the outer perimeter of the calorimeter at a polar angle of $30^\circ$. The majority of crystals will be $1.5\times 1.5$ cm$^2$ (at the front face), but $1.3\times 1.3$ cm$^2$ crystals will be used below $12^\circ$ to maintain acceptable rates. The readout will be performed with Avalanche Photodiodes (APD) from the downstream face of the crystals.

Similar calorimeters have been successfully used at JLab since the early 2000s. The first implementation of a compact, lead-tungstate calorimeter in Hall~B was for the $6$ GeV DVCS experiment~\cite{ic}, with $424$ tapered crystals from the CMS experiment and APD readout. 
Later, these modules were repurposed for the Heavy Photon Search (HPS) experiment~\cite{HPS:2016rgp,Baltzell:2022rpd}. These calorimeters operated at close to $1.5$ MHz hit rate per channel with \mbox{$\sigma/E\simeq 4.5\%/\sqrt{E}$} resolution.  Another implementation in Hall B was the PRad experiment's Hybrid Calorimeter~\cite{Xiong:2019umf, doi:10.1142/9789812701978_0014}, which achieved an energy resolution of \mbox{$\sigma/E \le 2.5\%/\sqrt{E}$} using PMT readout.

In the 12~GeV era, several additional PbWO$_4$ calorimeters have been built and operated in high-rate environments. The FTCal of CLAS12 uses 332 $1.5\times1.5\times20$~cm$^3$ crystals, read out by $1\times1$~cm$^2$ LAAPDs. It operates at $0^\circ$C and achieves an energy resolution of the order of $3.5\%/\sqrt{E}$.  More recently, two large-area PbWO$_4$ calorimeters have been built and commissioned at JLab: the NPS calorimeter~\cite{nps} in Hall~C with close to 1000 channels, and a 1600-module upgrade of the GlueX forward calorimeter in Hall~D~\cite{Somov:2025eiq}, both using PMT readout. Beam tests of a small GlueX calorimeter prototype consisting of 140 modules of size $2\times2\times20$~cm$^3$ demonstrated the expected energy resolution of $\sigma/E \le 3\%/\sqrt{E}$~\cite{sgluex}.



\subsubsection{Forward Vertex Tracker}
 
The planned measurements require the detection of three, forward-going, charged particles: an electron, identified in the calorimeter, and a pair of muons detected in the $\mu$CLAS12 FD. All three particles originate from the target and traverse the strong magnetic field of the solenoid magnet before reaching their respective detectors. In particular, the muons pass through both the calorimeter and the shielding, undergoing significant energy loss and multiple scattering prior to momentum analysis. The FVT located close to the target is therefore essential to ensure an accurate vertex reconstruction.

The FVT will consist of an MPGD tracker that will be positioned in front of the wECal calorimeter and downstream of the target.  A triple-GEM \cite{SAULI1997531} design is well-suited to this high-rate environment. The GEM technology relies on gas-avalanche multiplication in micro-scale holes ($\approx$~50~$\mu$m), with multiple cascaded GEM foils providing high gain and operational stability. GEM-based tracking detectors have been widely used in JLab experiments since the early 2000s, including the BONuS~\cite{bonus_gem}, eg6~\cite{eg6_gem}, and PRad \cite{prad_gem} experiments in Hall~B. Furthermore, GEM trackers designed for rates of approximately $ 1$~MHz/cm$^2$ have been successfully fabricated and operated in the Hall~A SBS~\cite{rd51_mw} spectrometer.

A six-station tracking system for the $\mu$CLAS12 FVT is envisioned. Each station comprises six trapezoidal modules to cover the $2\pi$ azimuthal acceptance from $7^\circ$ to $35^\circ$ in polar angle. A 2D stereo strip readout~\cite{COMPASS:2002,GNANVO:2016nim} is planned, where the $U$ and $V$ strips will be oriented parallel to the legs of the trapezoid. With an expected pitch of $0.5$~mm, each module will have approximately 1200 readout channels. This arrangement will allow the front-end readout electronics to be placed at the base of the trapezoid, outside of the detector acceptance. The first station will be approximately $40$~cm from the target, and the longest strip in this design will be less than $25$~cm. Given the relatively small detector size and short strip lengths, and based on operational experience of GEM detectors for SBS, the FVT will be able to sustain rates of 250~kHz/cm$^2$ with position resolution better than $100~\mu$m.

\subsubsection{Recoil Detector}

A Recoil Detector (RD) is essential for tagging protons in quasi-real photoproduction reactions, such as TCS, and for systematic studies of DDVCS. The detector is designed to cover scattering angles from $40^\circ$ to $70^\circ$.
For forward tracking, material budget does not significantly affect track reconstruction, since the dominant energy loss occurs in the shielding. Therefore, high-rate capable GEM detectors can be effectively employed. In contrast, minimizing the material budget is critical for detecting low-energy recoil protons.
In addition to tracking, measuring the time-of-flight and energy loss of the proton is required for its identification. Therefore, the RD consists of two sub-detectors: a Recoil Tracker (RT) for charged particle tracking, and a Scintillation Hodoscope (SH) for precise proton identification.

\paragraph{Recoil Tracker:}

The recently developed Micro-Resistive Well (\uRWELL) detectors~\cite{Bencivenni:2014exa} represent a promising technology for the RT. These detectors combine a low material budget with a relatively simple structure, making them well-suited for high-precision tracking applications. As in GEM detectors, charged particles ionize the gas in a drift region; however, \uRWELL~detectors employ a single amplification stage, reducing the amount of material traversed by the particles.

The initial \uRWELL ~prototype detectors had rate capability limitations, with a gain drop observed above 100 $\mathrm{kHz/cm^2}$. This was mainly caused by the collected charge on the resistive layer, which could not dissipate fast enough, effectively reducing the amplification field inside the wells. In recent years, new developments in this direction have allowed \uRWELL\ detectors to withstand rates larger than $\mathrm{1\;MHz/cm^2}$ without compromising their gain. This was achieved by adding more grounding lines on the resistive layer (Patterning-Etching-Plating, PEP-groove and PEP-dots), significantly speeding up the charge evacuation, as demonstrated in Ref.~\cite{Bencivenni:2024jgp}.

The concept of the RT is shown in Fig.~\ref{fig:recoil}, and consists of six concentric cylindrical \uRWELL\ layers, each comprising three sectors, featuring a 2D readout, and covering polar angles from $40^{\circ}$ to $70^{\circ}$. The innermost layer will be positioned at a radius of 8 cm from the beamline, the outermost layer at 23.5~cm, with an approximately 2~cm spacing between layers. The 2D strip layout features Z‑strips, oriented parallel to the beam direction, and C‑strips, oriented perpendicularly, allowing precise determination of azimuthal and longitudinal hit coordinates. The strip layout will closely resemble the design used in the current CLAS12 MVT system described in Ref.~\cite{Acker:2020qkv}. The Z-strips have a uniform $\mathrm{500~\mu m}$ pitch, while the C-strips have the same average pitch but compensated to maintain a uniform polar angle resolution. The total number of readout channels for the entire detector will be below $22\times 10^3$, well within the practical limits for efficient data acquisition.  

\paragraph{Scintillation Hodoscope:}
The proton identification will be provided by a scintillator-based detector capable of measuring both the time-of-flight and the energy loss of the recoil proton. The SH will consist of 540 truncated pyramidal modules arranged in concentric rings around the beamline at a radius of 25~cm, as shown in Fig.~\ref{fig:recoil}. The modules will follow a projective geometry that points toward the center of the target.

The detector layout includes four forward rings that cover the scattering angles from $40^\circ$ to $50^\circ$, populated with modules of $2 \times 2$ cm$^2$ cross section. Modules covering the $50^\circ$ to $70^\circ$ range will be larger, with $4 \times 4$ cm$^2$ dimensions. Light readout will be achieved using 2-mm diameter, green wavelength-shifting fibers glued along the outer surface of each pyramid. These fibers will extend two meters upstream, where the light will be collected by photodetectors, either silicon photomultipliers (SiPMs) or multi-anode photomultiplier tubes (MAPMTs). The decreasing height of pyramids in successive rings accommodates the routing of the 2 mm fibers. This readout configuration is based on proven designs currently in use in three Hall B detectors: the CLAS12 FT hodoscope~\cite{ftcal} (using SiPMs), the HPS scintillation hodoscope~\cite{Baltzell:2022rpd} (with 16-channel Hamamatsu MAPMTs), and the scintillation hodoscope for the CLAS12 $\mu$RWELL project~\cite{Gnanvo:2024jag} (featuring 64-channel Hamamatsu MAPMTs).

\subsubsection{Beamline and Target}

The planned luminosity of $10^{37}\,\mathrm{cm^{-2}\,s^{-1}}$ will be achieved with a 5-cm-long liquid hydrogen (LH$_2$) target and a beam current of 7.5~$\mu$A. The Hall~B beamline will support high‑current operation with minimal modifications, the primary requirement being an increase in beam‑dump power. In 2021, a two-phase upgrade of the Hall~B beam dump was initiated. The first phase was completed, allowing the dump to handle a 17~kW beam on a water-cooled movable beam blocker and up to 1~kW on the Faraday cup. The second phase is planned for the next few years. Several options are being considered and will increase the beam dump capacity to up to 100~kW, allowing operation at the required beam currents for this experiment. 

The LH$_2$ target of $\mu$CLAS12 will be positioned at the center of the solenoid magnet. Unlike the current design, which uses a Kapton cell with aluminum windows, the new target cell will be constructed entirely from aluminum to enhance thermal performance and improve heat dissipation, ensuring stable operation under high beam intensity conditions. While beam heating will not be an issue regarding maintaining LH$_2$ in the cell, it is expected to incur some local boiling and density fluctuations. This effect is not uncommon for high-power targets, and a typical mitigation is to use luminosity scans to track density changes as a function of beam current.

\subsection{Rates and Occupancies}

GEMC, the CLAS12 GEANT4 simulation software, was modified to allow background, detector occupancies, event reconstruction, and experimental reach to be studied for the new configuration. The $\mu$CLAS12 CAD model for the new detector components was integrated into GEMC and the existing CLAS12 software framework. This approach enabled the use of COATJAVA for realistic event reconstruction and provided a foundation to conduct physics analyses.

\begin{figure*}[htbp]
\centering
\includegraphics[width=1\textwidth]{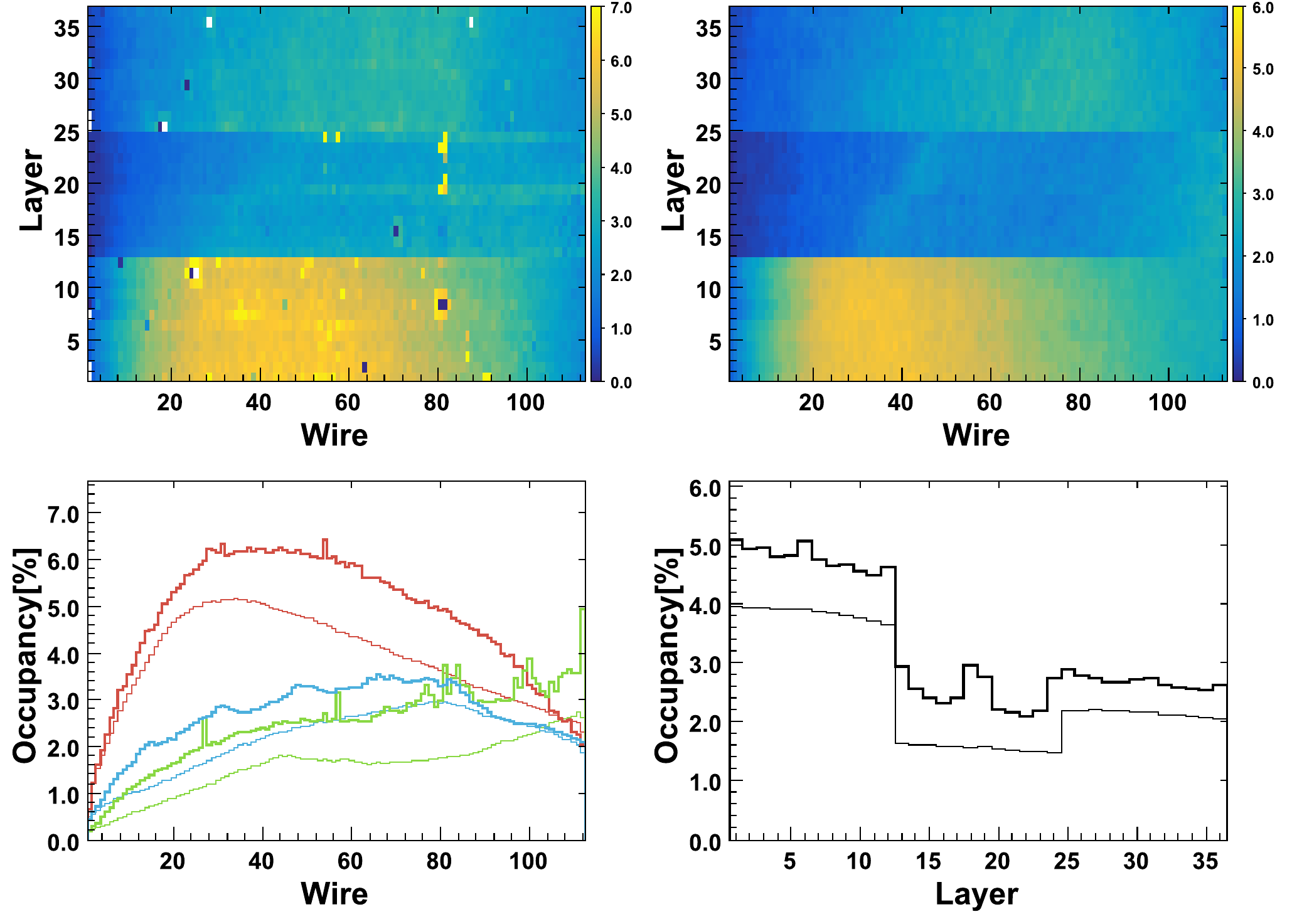}
\caption{Overview of the CLAS12 DC occupancies. Top: 2D wire occupancy as a function of the layer and wire number for data from the RG-A run (left) and from MC simulations (right). Bottom: Integrated occupancies as a function of wire number, for Region~1 (red), Region~2 (green), and Region~3 (blue) (left) and layer number (right). The occupancies from data are represented with thick lines, thin lines represent the occupancies obtained from MC simulations. Wire numbers increase with polar angle, while layer numbers increase moving radially outward from the target.}
\label{fig:dc_data_gemc}
\end{figure*}

\begin{figure*}[htbp]
\begin{center} 
\begin{subfigure}{0.49\linewidth}
\centering
\includegraphics[width=\linewidth]{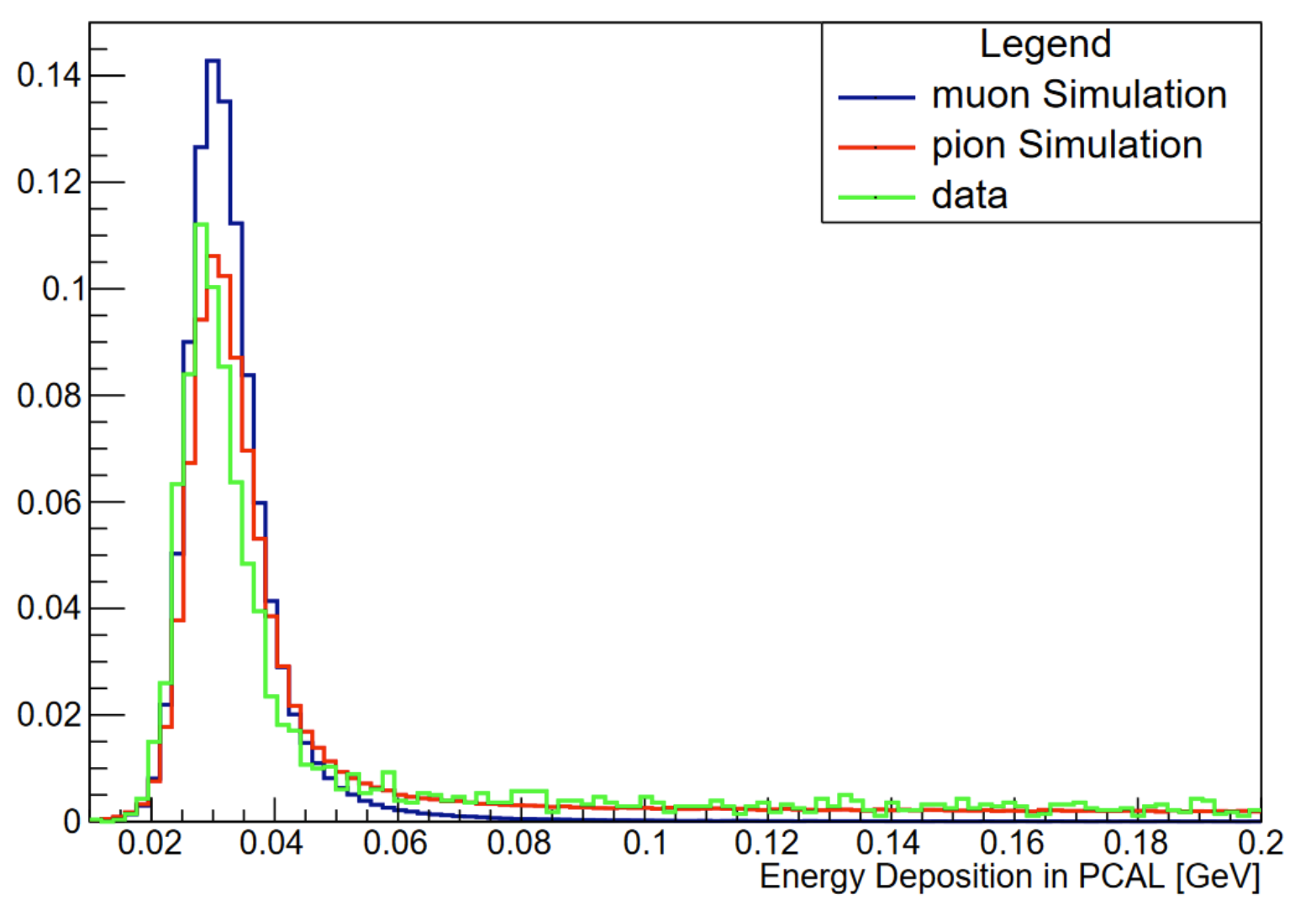}
\caption{}
\label{fig:edep_pcal}
\end{subfigure}
\begin{subfigure}{0.49\linewidth}
\centering
\includegraphics[width=\linewidth]{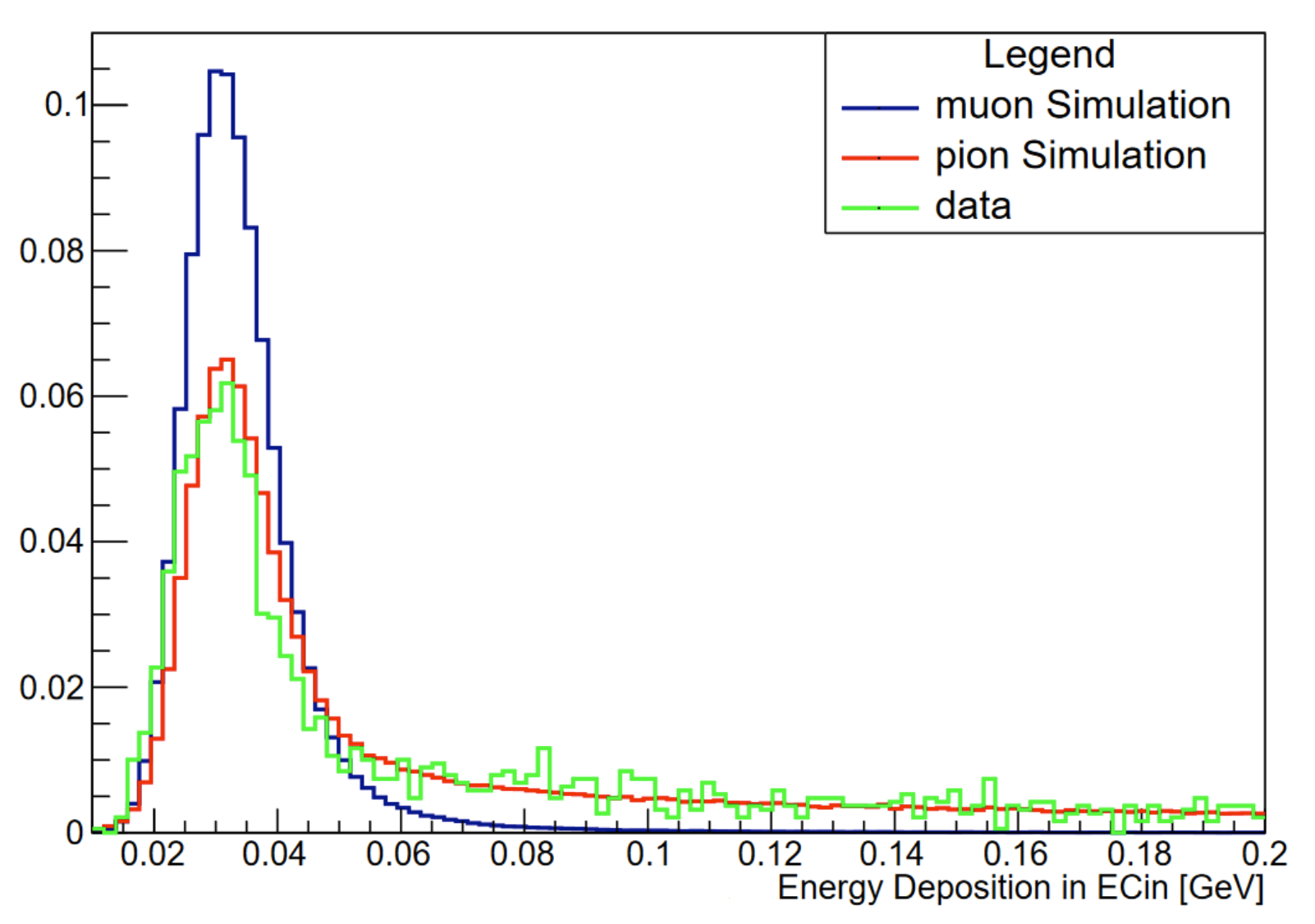}
\caption{}
\label{fig:edep_ecin}
\end{subfigure}
\\
\begin{subfigure}{0.49\linewidth}
\centering
\includegraphics[width=\linewidth]{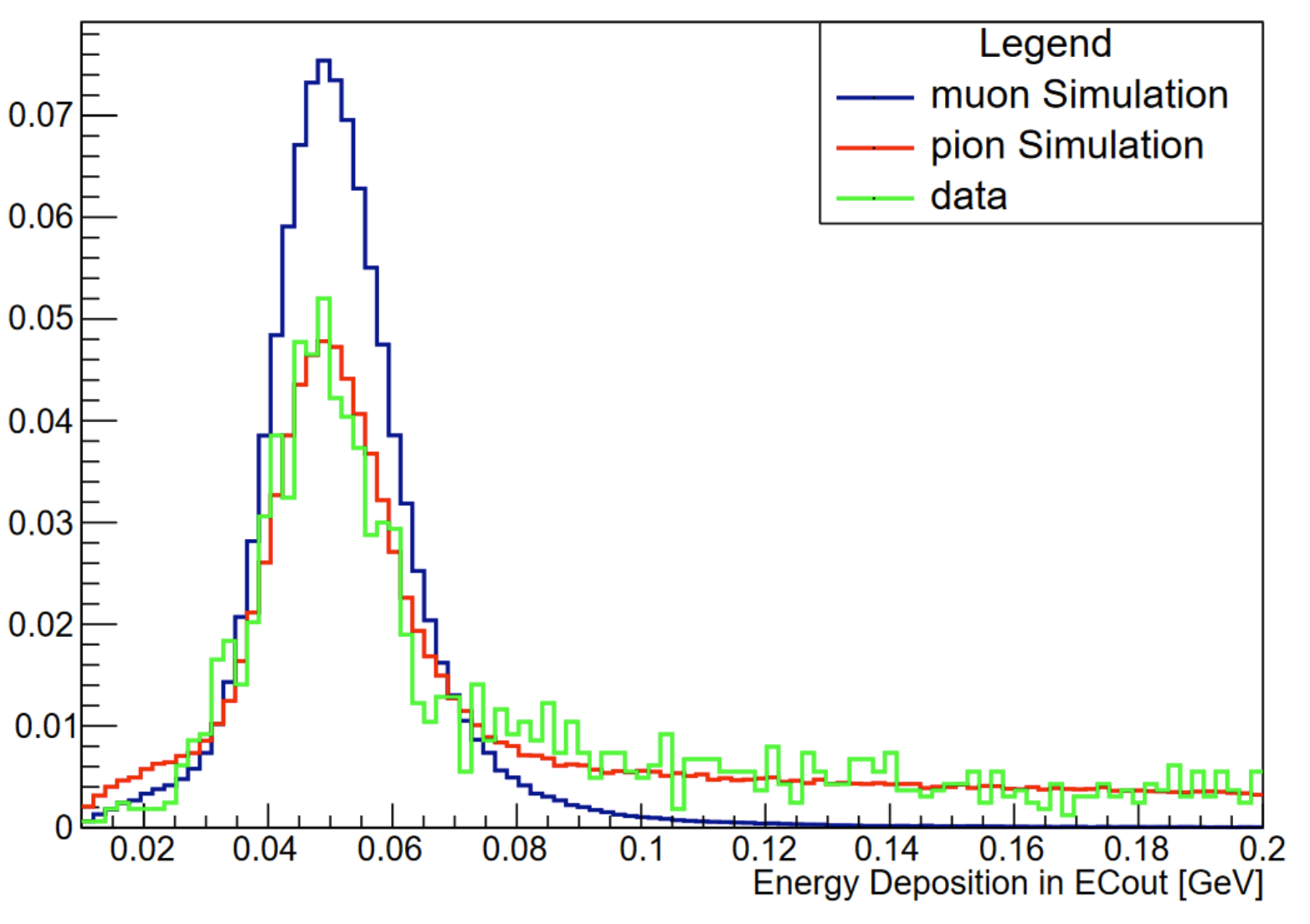}
\caption{}
\label{fig:edep_ecout}
\end{subfigure}

\caption{Energy distributions for minimum ionizing particles ($\pi^\pm$) in each of the ECal modules: PCal (Panel~\ref{fig:edep_pcal}), ECIn (Panel~\ref{fig:edep_ecin}), and ECOut (Panel~\ref{fig:edep_ecout}). Real CLAS12 data, taken on LH2 target, are shown in green. CLAS12 MC simulations for $\pi^\pm$ (red) and for $\mu^\pm$ (blue) are also displayed. All histograms are normalized to the total number of events in each sample. Figure taken from Ref.~\cite{Tyson:2023yer}.}
\label{fig:fecal_mip}
\end{center}
\end{figure*}

The most critical aspects of the MC simulations for this study are background rates, detector occupancies from electron-target interactions, particle energy loss (ensuring an accurate material budget of detectors), and energy and time resolutions. To ensure reliability in all of these areas, GEMC has been optimized over the years and validated against experimental data across various beam energies and target configurations of CLAS12. Figure~\ref{fig:dc_data_gemc} presents a comparison of measured and simulated DC occupancies during the first CLAS12 run (called Run Group A or RG-A), where a 50~nA, 10.6~GeV electron beam scattered on a 5-cm-long LH$_2$ target. The simulated beam background reproduces the observed occupancies within $20\%$. Another comparison, shown in Fig.~\ref{fig:fecal_mip}, illustrates the energy depositions from charged pions for both data and simulation across the three ECal regions (PCal, ECIn, and ECOut) alongside the simulated muon energy loss. The simulation of the calorimeter energy response demonstrates good agreement with the experimental measurements. 

These validations demonstrate that GEMC accurately reproduces the performances of the CLAS12 detector. A GEMC-based simulation package, $\mu$GEMC, has been developed to assess the feasibility and the performances of the $\mu$CLAS12 experiment. Several material (lead and tungsten) and thicknesses of the shield downstream of the wECal have been studied to optimize occupancies in Regions~1 and~2 of the DC, pion-muon separation, muon energy loss, and muon momentum resolution. It was found that a 60-cm-thick lead shield yields acceptable DC occupancies. From an engineering perspective, the lead option is also preferred as it provides a cost‑effective solution that is straightforward to manufacture. 

\subsubsection{Drift Chambers Occupancies}

\begin{figure*}[htbp]
\begin{center}
\includegraphics[width=0.9\linewidth]{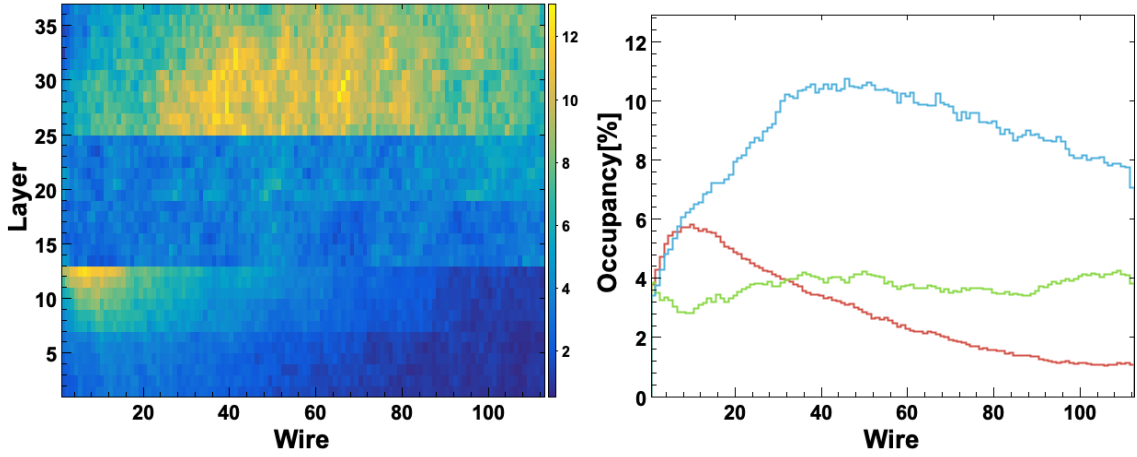}
\caption{DC occupancies at a luminosity of $10^{37}\,\mathrm{cm^{-2}\,s^{-1}}$ with a 60-cm-thick lead shield installed downstream of the PbWO$_4$ calorimeter. Left: 2-D wire occupancy map for layer vs. wire number. Right: average wire occupancies for Region 1 (red), Region 2 (green), and Region 3 (blue). High occupancies in Region 1 are located in the very forward region, while $\mu$CLAS12 aims to detect tracks in the FD above $7^\circ$. The $\sim10\%$ occupancy in Region 3 originates from electrons produced at the downstream end of the torus. }
\label{fig:dc_occ}
\end{center}
\end{figure*}

The DC occupancies have been studied with a shield made of either lead or tungsten. Figure~\ref{fig:dc_occ} shows the occupancies generated from the interaction of an 11~GeV electron beam with an LH$_2$ target at a nominal luminosity of $10^{37}\,\mathrm{cm^{-2}\,s^{-1}}$, and with a 60-cm-thick lead shield. The right sub-figure shows the average occupancies, which are approximately $3\%$, $4\%$, and $9\%$ for Regions 1, 2, and 3, respectively. 

During previous CLAS12 experiments with nuclear targets, DCs have been operated with occupancies close to $10\%$. Although high occupancies will impact tracking efficiency and momentum resolutions, it is important to note that high occupancies in Region~1 are primarily concentrated in the very forward region. Region~1 covers the angular range starting from $5^\circ$, while $\mu$CLAS12 aims to detect tracks in the FD above $7^\circ$. As a result, the high-occupancy area corresponding to the first 15 wires of Region 1 will not significantly affect its performance.   

The primary challenge is Region 3 which covers a substantial portion of the detector and where occupancy is the highest ($<12\%$). MC simulations indicate that a significant portion of the background originates from electrons produced at the downstream end of the torus. Further improvements in shielding and background mitigation are being investigated to ensure optimal detector performance under high-occupancy conditions.

\subsubsection{Rates in the FVT}

A scoring plane was implemented in $\mu$GEMC at a distance of 40~cm from the center of the target to estimate the hit rates in the FVT. Any particle with energy exceeding 10~keV that crossed the plane within the angular range of $7^\circ$ to $30^\circ$ was counted as a hit.

\begin{figure*}[htbp]
\begin{center}
\includegraphics[width=0.9\textwidth]{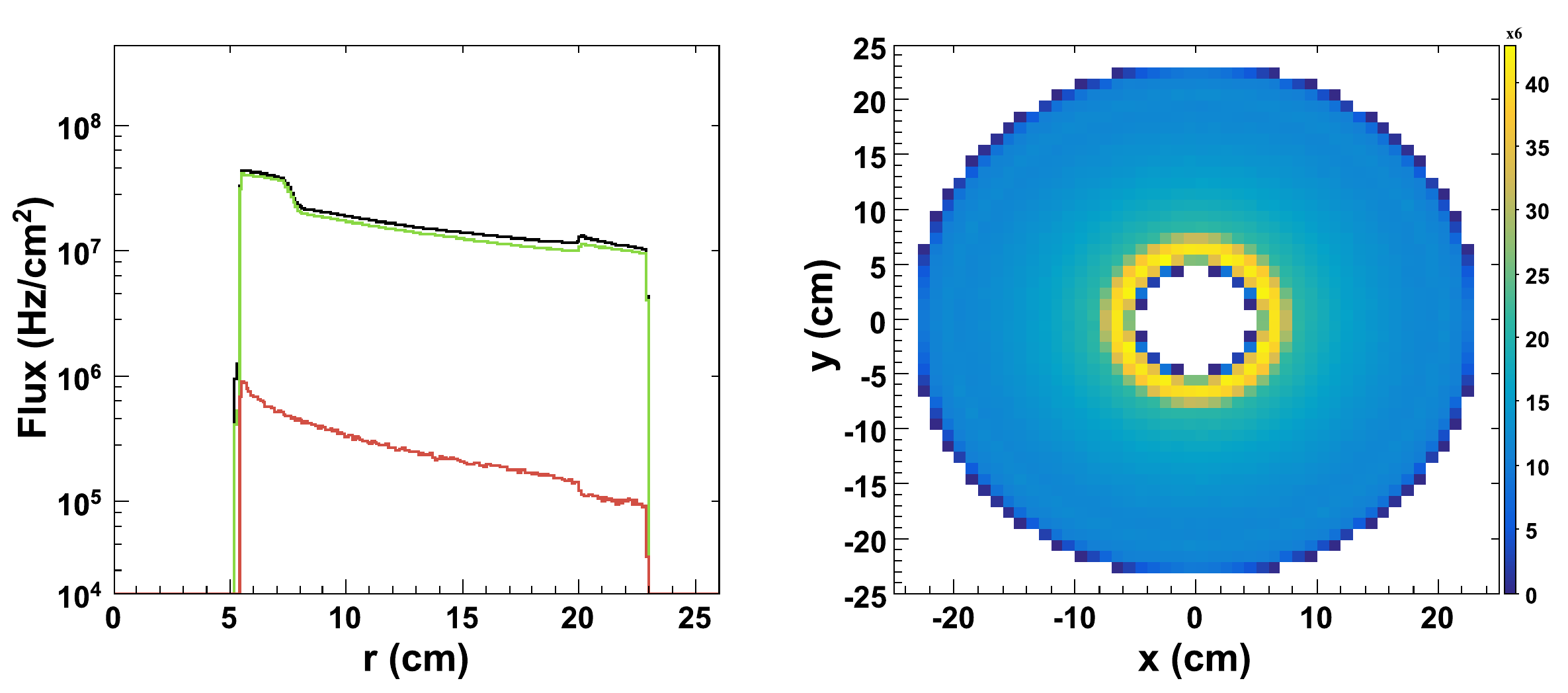}

\caption{Rates of particles at a luminosity of $10^{37}\,\mathrm{cm^{-2}\,s^{-1}}$ crossing the scoring plane located 40~cm downstream from the target center. Left: total (black), photon (green), and charged-particle (red) flux as a function of radius~$r$. Right: total particle rate as a function of the local coordinates $xy$. }
\label{fig:gem_rates}
\end{center}
\end{figure*}

Figure \ref{fig:gem_rates} shows the flux distribution of charged and neutral particles at a luminosity of $10^{37}\,\mathrm{cm^{-2}\,s^{-1}}$. In the region near the beam, the total rate, predominantly composed of photons, is approximately 40~MHz/cm$^2$. Extensive studies with GEM detectors indicate that only $0.5\%$ of photons with energies greater than 10~keV will produce a detectable signal in the tracker. Taking into account this detection factor and the charged-particle rate of about 800~kHz/cm$^2$ in the very forward region, the highest detectable rate is estimated to be less than 1~MHz/cm$^2$.

In the current GEM design, the longest strip covers an area of about 1.2~cm$^2$, stretching from very small angles (with the highest rates) to larger angles (where rates decrease to around 100~kHz/cm$^2$). Averaging the rates over the entire tracker area yields a detectable hit rate of less than 500~kHz/cm$^2$ per GEM module.

\subsubsection{Rates in the wECal}

\begin{figure*}[htbp]
\begin{center}
\includegraphics[width=0.9\textwidth]{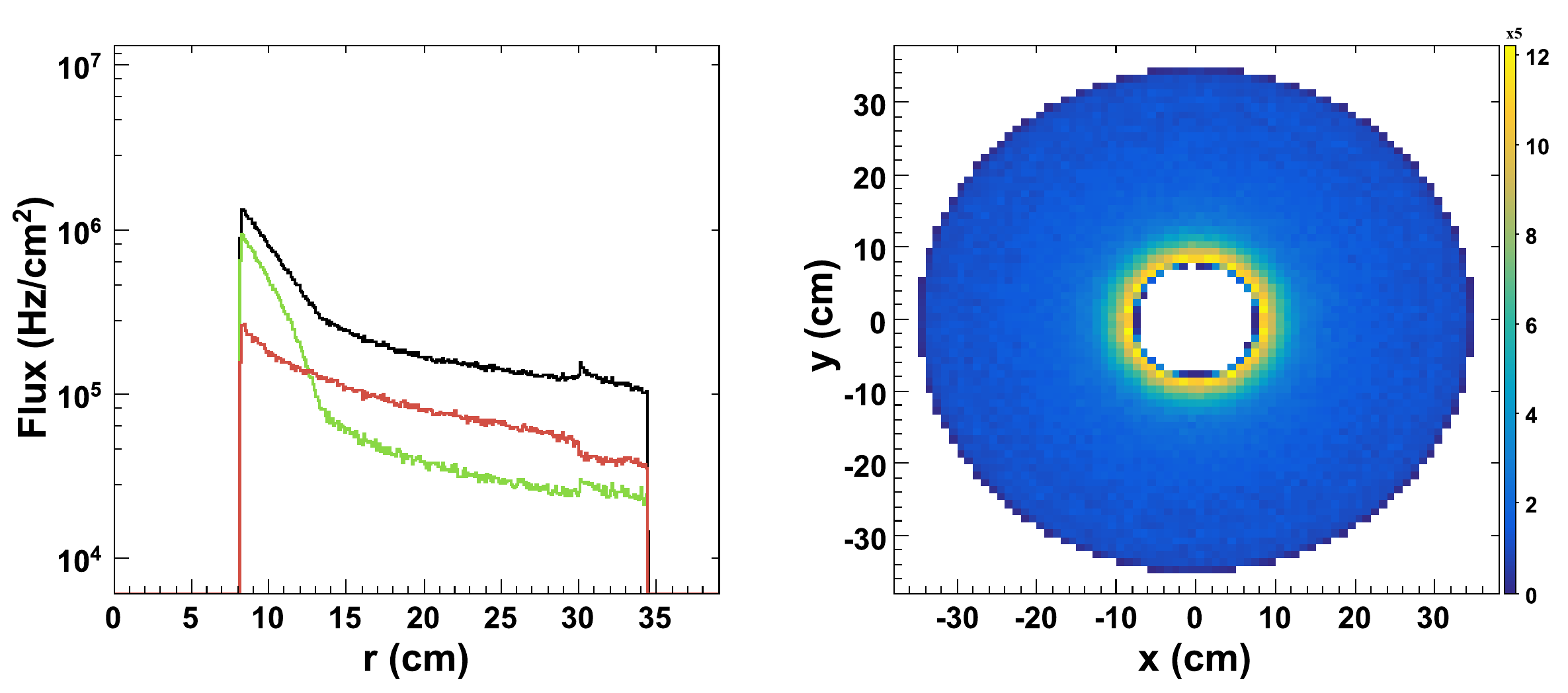}
\caption{Hit rates at the scoring plane located upstream of the wECal, at a luminosity of $10^{37}\,\mathrm{cm^{-2}\,s^{-1}}$. Left: total (black), photon (green), and charged-particle (red) rate as a function of radius $r$, after a 20~MeV energy cut in the wECal. Right: Total rates of charged particles and photons as a function of the local coordinates $xy$, under the same energy cut. }
\label{fig:ecal_rates}
\end{center}
\end{figure*}

The rates in the wECal were estimated using a scoring plane located 60~cm from the center of the target. Figure~\ref{fig:ecal_rates} shows the flux distribution of charged and neutral particles at a luminosity of $10^{37}\,\mathrm{cm^{-2}\,s^{-1}}$ within the acceptance range of the wECal, considering only particles that deposited an energy of at least 20~MeV.

The highest observed rate was approximately 1.2~MHz/cm$^2$ in the region close to the beam, predominantly driven by photons. In contrast, the charged particle rate in this region was found to be about 300~kHz/cm$^2$. Close to the beam, the calorimeter modules will have dimensions of $1.3 \times 1.3 \times 20$~cm$^3$, covering an area of approximately 1.7~cm$^2$. Therefore, the highest hit rate in a single module with a 20~MeV energy threshold was estimated to be less than 2~MHz. This rate is expected to be manageable, as similar calorimeters such as one of HPS, have operated efficiently with comparable rates in modules close to the beam. 

\subsubsection{Rates in the RT and the SH}

\begin{figure*}[htbp]
\begin{center}
\includegraphics[width=0.9\textwidth]{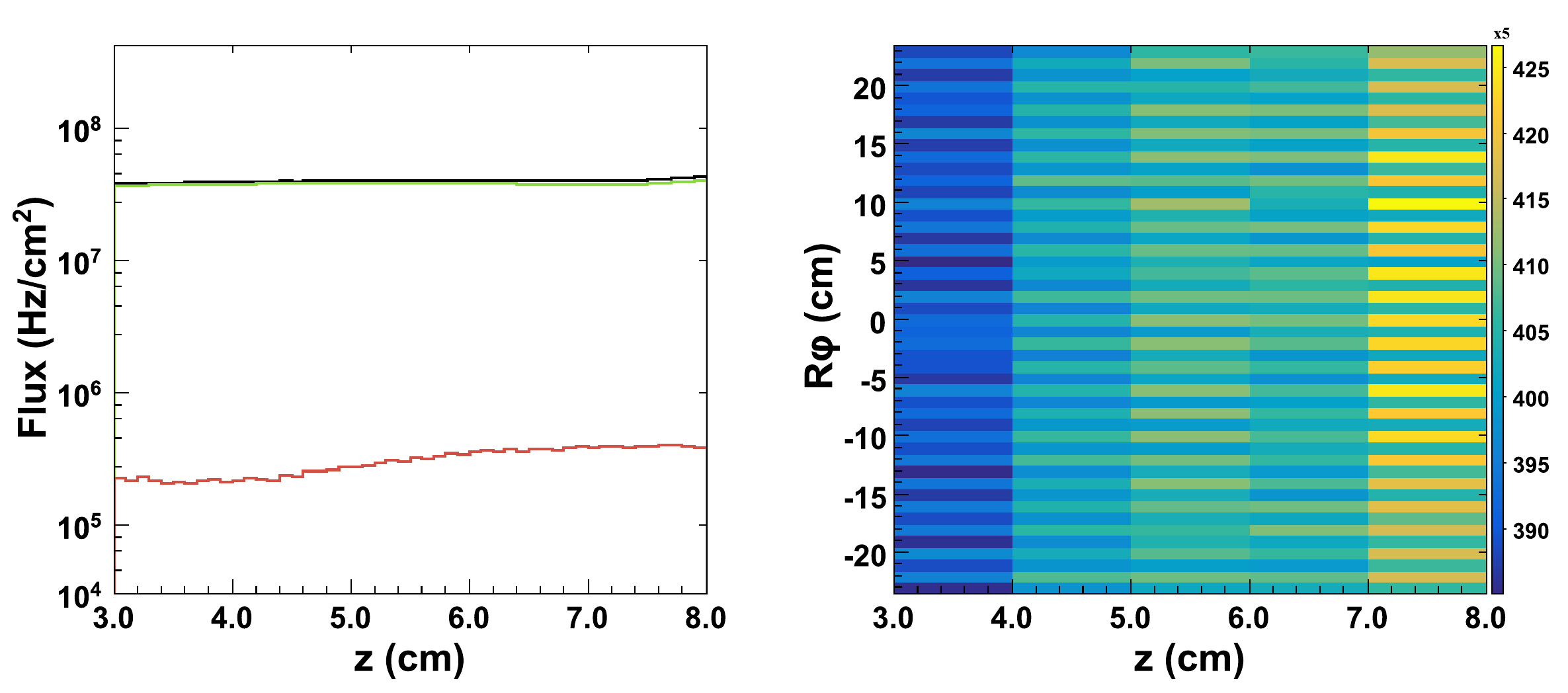}
\caption{Rates of particles at the cylindrical scoring plane located at a 7.5~cm radius from the beam axis, at a luminosity of $10^{37}\,\mathrm{cm^{-2}\,s^{-1}}$. Left: total (black), photon (green), and charged-particle (red) rates as a function of the coordinate $z$ along the beam axis. Right: total particle rate as a function of $z$ and $R\phi$.}
\label{fig:rt_rates}
\end{center}
\end{figure*}

The rates in the RT were estimated using cylindrical scoring planes positioned at 7.5~cm (first tracker layer) and 25~cm (SH) radius from the beam. Figure~\ref{fig:rt_rates} shows the flux distribution of charged and neutral particles at a luminosity of $10^{37}\,\mathrm{cm^{-2}\,s^{-1}}$. The left panel of the figure shows the total rates, the middle panel is the rate of photons, and the right panel is the rate of charged particles. The highest charged particle flux in the first tracking layer of the forward region ($\theta\sim40^\circ$) is about $40$~kHz/cm$^2$, while the photon flux is about $40$~MHz. Considering a $0.5\%$ detection efficiency for photons, the estimated rate per cm$^2$ for the RT is less than $250$~kHz/cm$^2$, which remains within acceptable operational range.

\begin{figure*}[htbp]
\begin{center}
\includegraphics[width=0.9\textwidth]{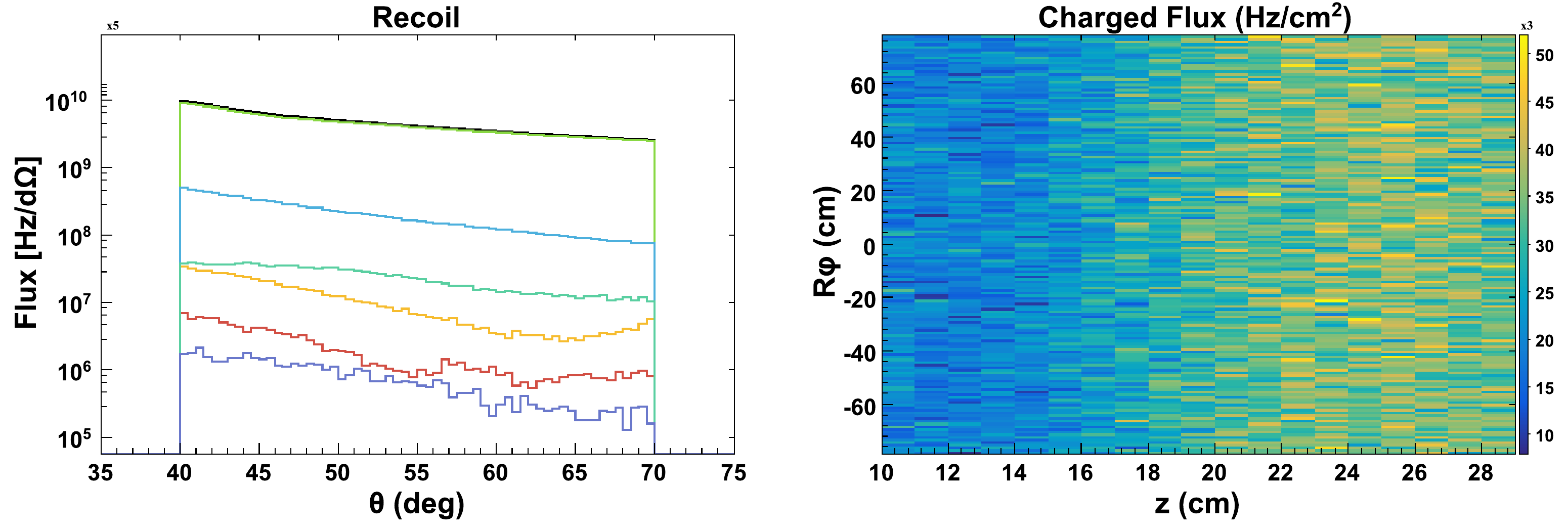}
\caption{Rates of particles at the cylindrical scoring plane located at a 25~cm radius from the beam axis, at a luminosity of $10^{37}\,\mathrm{cm^{-2}\,s^{-1}}$. Left: Rates for different particle types as a function of polar angle. Color lines are: black is the total rate, green - photons, blue - neutrons, yellow - protons, cyan - pions, red - electrons, dark blue - other charged particles. Right:~Azimuthal and longitudinal ($z$) distribution of rates for charged particles.}
\label{fig:sh_rates}
\end{center}
\end{figure*}

The highest rates of charged particles in the scintillation hodoscope are about $80$~kHz/cm$^2$, as shown in Fig.~\ref{fig:sh_rates}. Given the $4$~cm$^2$ area of the forward modules, the anticipated rate of $320$~kHz per module remains well within manageable levels.

\subsubsection{Trigger Rates and Data Acquisition}

The $\mu$CLAS12 trigger will be based on the detection of a single charged track with MIP-like energy deposition in the ECal. To estimate the rate of this single-MIP trigger, CLAS12 data taken with a 5-cm-long LH$_2$ target and a 10.6 GeV electron beam were analyzed. These data were collected using multiple trigger settings, including one that required a single ECal hit with an energy greater than 10~MeV. The raw rate of this trigger at a luminosity of $0.6 \times 10^{35}\,\mathrm{cm^{-2}\,s^{-1}}$ was about 250~kHz, with a pre-scale factor of 2049. The analyzed sample corresponds to a total integrated luminosity of $52.7 \times 10^{37}\,\mathrm{cm^{-2}}$, and yielded approximately $53 \times 10^3$ events with at least one charged track with ECal energy deposition below 300~MeV and momentum greater than $1.5$~GeV/$c$. This last requirement was used to reflect the expected pion and muon momentum thresholds in $\mu$CLAS12. To estimate the corresponding rate in $\mu$CLAS12, the event count was corrected for the pre-scale factor, scaled to the expected luminosity, and further reduced by the GEANT4-based estimate of the survival fraction of charged hadrons through the wECal and lead shield of approximately 1\% (see Section~\ref{sec:pion_survival}). The rate of the single-MIP trigger of $\mu$CLAS12 was therefore estimated to be approximately 21~kHz. 

The current CLAS12 Data Acquisition (DAQ) system supports trigger acceptance rates up to 30~kHz with a live time exceeding 90\% \cite{Boyarinov:2020yry}, comfortably above the expected trigger rate of $\mu$CLAS12. During routine operation, the system typically runs at 20~kHz or higher, with a data throughput of approximately 800~MB/s. Additionally, the transition from CLAS12 to $\mu$CLAS12 is not expected to significantly affect the total number of readout channels, as additional channels from the new detectors will be offset by those removed or replaced.

\subsection{Event Reconstruction and Muon Identification}
\label{sec:eventrec}

\subsubsection{Pion Survival Rates}
\label{sec:pion_survival}

The electromagnetic background generated in the target is largely absorbed by the wECal and downstream shielding, significantly reducing the background in the FD. However, the dominant background, and the main source of contamination in the muon sample, still originates from charged pions that pass through and produce tracks in the DC and deposit energy in the ECal. 

\begin{figure*}[h]
\centering
\begin{subfigure}{0.45\linewidth}
\centering
    \includegraphics[width=\linewidth]{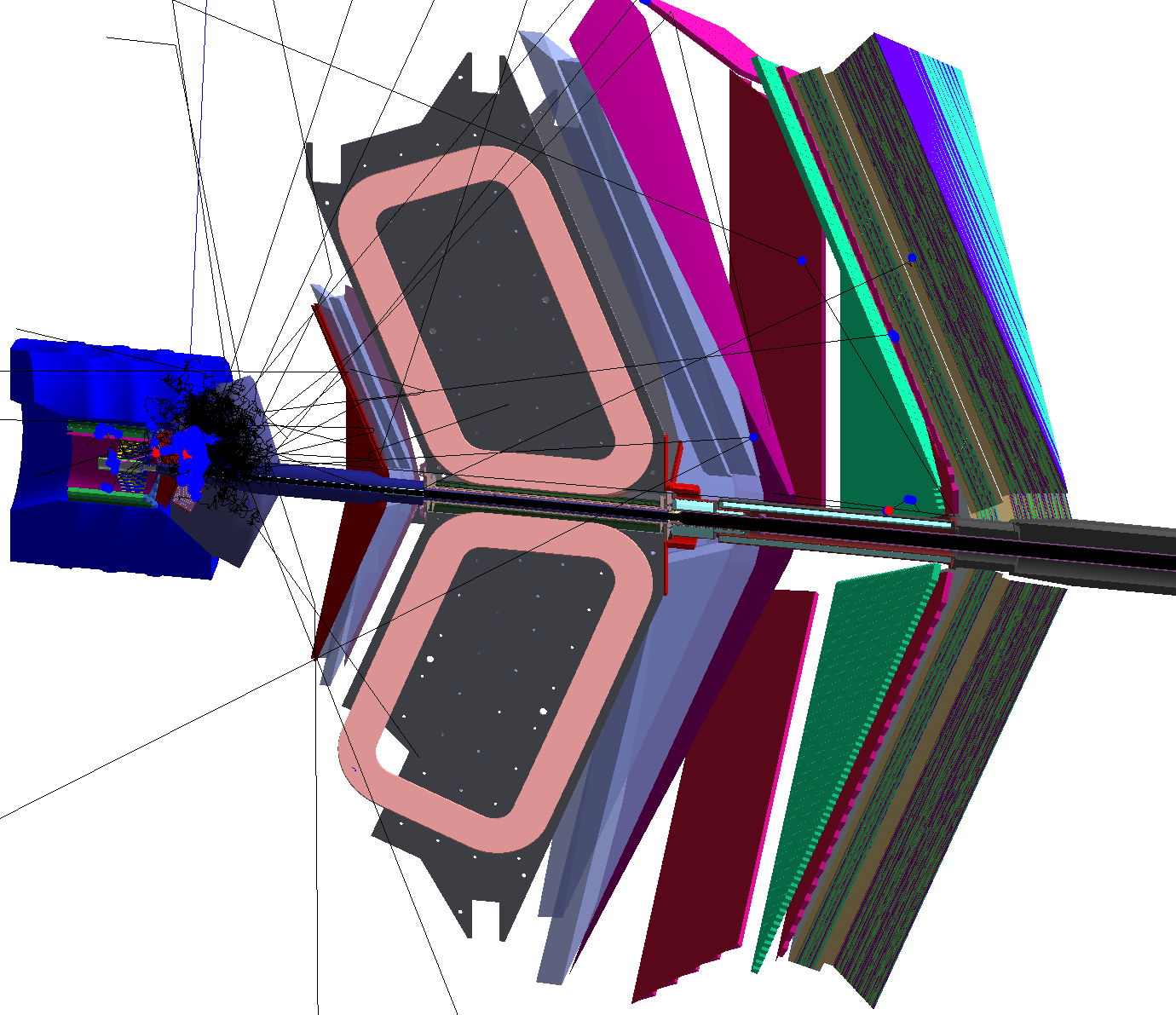}
    \caption{}
    \label{fig:pimu_muclas12:1}
\end{subfigure}
\hspace{0.02\linewidth}
\begin{subfigure}{0.475\linewidth}
\centering
\includegraphics[width=\linewidth]{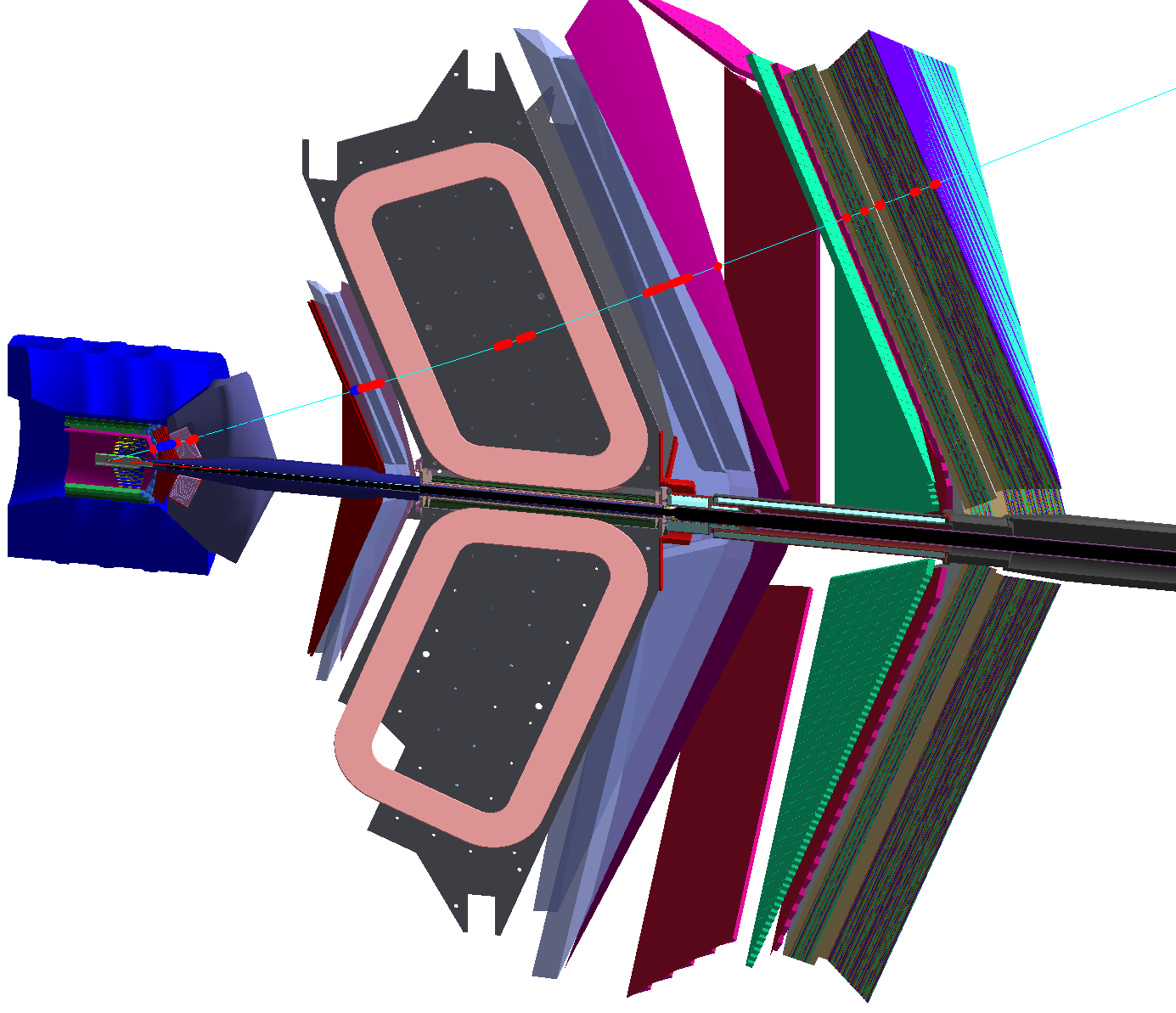}
 \caption{}
    \label{fig:pimu_muclas12:2}
\end{subfigure}
\caption{Simulation of a 6~GeV $\pi^+$ (Panel~\ref{fig:pimu_muclas12:1}) and a 6~GeV $\mu^+$ (Panel~\ref{fig:pimu_muclas12:2}) in the $\mu$CLAS12 GEANT4 model. Nearly all pions undergo hadronic showers in the wECal, while most muons reach the FD and are reconstructed in both the DC and the ECal.}
\label{fig:pimu_muclas12}
\end{figure*}

Figure~\ref{fig:pimu_muclas12} illustrates how a 6~GeV $\pi^+$ and $\mu^+$ interact with the calorimeter and shield before reaching the FD. As shown in Panel~\ref{fig:pimu_muclas12:1}, most pions undergo hadronic showers in the wECal, which prevents them from leaving measurable trajectories in the DC or a MIP signal in the ECal. Only a small fraction (less than $1\%$) of pions reach the FD, either directly or through secondary particles, and leave a MIP signal in the ECal. In contrast, more than 80\% of muons traverse the wECal with only moderate energy loss and remain detectable in both the DC and ECal, as shown in Fig.~\ref{fig:pimu_muclas12:2}.

To estimate pion contamination in the muon sample, $3\times 10^6$ pions of both charges were simulated with uniform momentum and angular distributions. The CLAS12 event reconstruction framework was then used to reconstruct and identify particles reaching the forward calorimeters. The strategy used for muon identification relied on the characteristic energy deposition in the forward calorimeter modules, as shown in Fig.~\ref{fig:fecal_mip}.  An important distinguishing feature between muons and pions is the transverse profile of the energy distribution, and specifically the number of calorimeter strips involved in the energy reconstruction. Pions produce a wider transverse shower profile than muons, involving a larger number of strips. A limit on the number of strips involved in the calorimeter energy reconstruction was used to further suppress pion contamination. In general, the survival rate of pions was found to be less than $0.8\%$. Due to the secondary particles generated in the wECal, a negatively (positively) charged MIP track can occasionally be reconstructed from the original sample of $\pi^+$~($\pi^-$), although with a very low probability, less than $0.2\%$. For physics analysis, two MIP-tracks will be required, ensuring a suppression factor for pion pairs of at least~2$\times 10^{4}$.

\subsubsection{Muon Energy Loss and Momentum Resolution}

Muons with momentum greater than approximately 1.5~GeV/$c$ are capable of penetrating the wECal and the lead shield, undergoing significant ionization energy loss, up to 1~GeV at low energies. Muons that retain enough energy to further traverse the torus field and deposit energy in the ECal will undergo momentum analysis in the DC. However, the reconstructed momentum will correspond to the momentum after energy loss in the wECal and the lead shield. Consequently, momentum corrections are necessary to recover the momenta of the muons at the production vertex, and accurately reconstruct the kinematics of the reactions of interest.  

\begin{figure}[htbp]
    \centering
    \includegraphics[width=\linewidth]{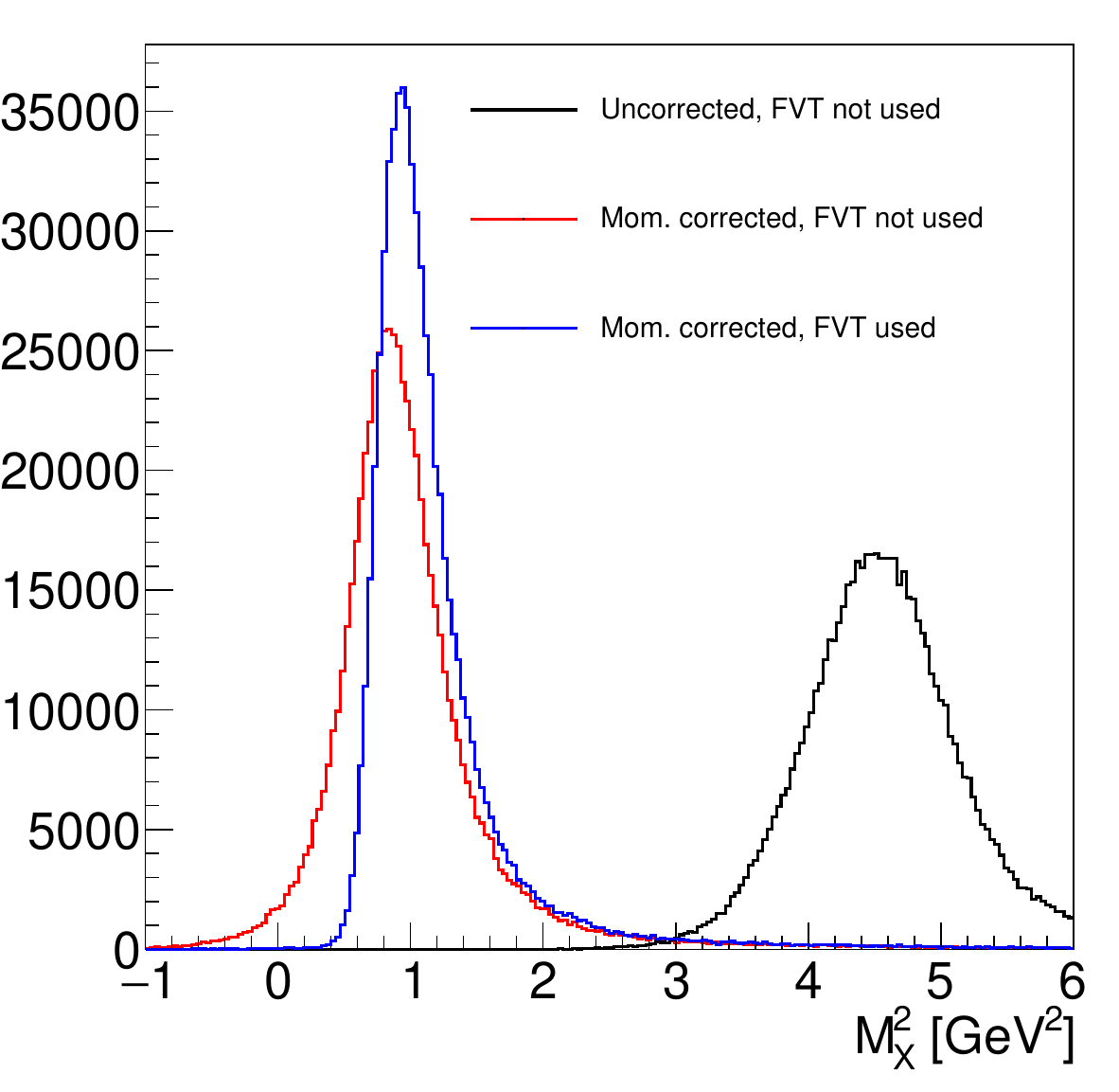}
    \caption{Missing mass squared distribution for the reaction ${ep \rightarrow e^{\prime} \mu^{+} \mu^{-} X}$. 
    The black histogram corresponds to the uncorrected muon momenta, the red histogram is obtained after applying only the momentum corrections to both muons, the blue histogram also includes the effect of the FVT on the angle reconstruction.}

    \label{fig:MX2Corrections}
\end{figure}

The importance of these momentum corrections is illustrated using one of the key kinematic variables for identifying the reaction ${ep \rightarrow e^{\prime} \mu^{+} \mu^{-} p'}$, the missing mass of the final state ${e^{\prime} \mu^{+} \mu^{-} X}$, where the proton is identified using missing momentum analysis. Figure~\ref{fig:MX2Corrections} shows the missing mass squared distributions of the reconstructed ${e^{\prime} \mu^{+} \mu^{-}}$ final state from simulated BH events. The black distribution corresponds to the missing mass of the proton calculated using the reconstructed muon momenta, while the red distribution includes muon momentum corrections. Significant improvements in both the position and the width of the peak are observed after applying the corrections. In addition to energy loss, multiple scattering will also significantly degrades the angular resolution of the muons. The primary purpose of the FVT is to mitigate this effect and precisely measure the angles of muons and electrons at the interaction vertex. The impact of the FVT on the angle reconstruction was assessed by incorporating it into the $\mu$CLAS12 tracking procedure. As shown by the blue histogram of Fig.~\ref{fig:MX2Corrections}, the FVT does significantly improve the missing mass resolution.

\subsection{Physics Backgrounds}
\label{physbg}

Three significant background processes produce the same final-state particles as the primary reaction of interest, ${ep \rightarrow e^\prime \mu^{+} \mu^{-} (p')}$ where the proton is not detected, and may contribute to the measured signal. 

The first major background consists of inelastic muon pair production processes of the type ${ep\rarr e^\prime\mu^{+}\mu^{-}(X\ne p)}$. These produce di‑muon final states in which the missing hadronic mass is sufficiently close to the proton mass to satisfy the exclusivity requirements. An example of such a background is ${ep\rarr e^\prime\mu^{+}\mu^{-}(\pi N)}$.
   
The second major background arises from pion‑pair production, ${ep\rarr e^\prime\pi^{+}\pi^{-}(X)}$. When both pions are identified as MIPs, the resulting final state can satisfy the missing‑mass requirements.
    
Finally, the third major background consists of accidental coincidences between two MIPs in the FD with an electron-like hit in the wECal. This occurs when a muon pair or a pion pair is produced by an electron scattered at an angle close to $0^\circ$ and which is not detected, while another inclusive electron from the same beam bunch is detected by the wECal.

\subsubsection{Inelastic Muon Pair Production}

The GRAPE event generator~\cite{GRAPE_gen}, extensively used in HERA data analysis, is capable of generating both elastic ${ep \rightarrow e^\prime \mu^{+} \mu^{-} p'}$ and inelastic ${ep \rightarrow e^\prime \mu^{+} \mu^{-} ({X \neq p'})}$ reactions. To estimate the contribution from inelastic muon pair production, events were generated with an invariant mass cut of ${M_{\mu^{+} \mu^{-}} > 1.2}$ GeV, corresponding to the region of interest of $\mu$CLAS12. These events were passed through $\mu$GEMC and subsequently reconstructed using COATJAVA.

\begin{figure}[htbp]
    \centering
    \includegraphics[width=\linewidth]{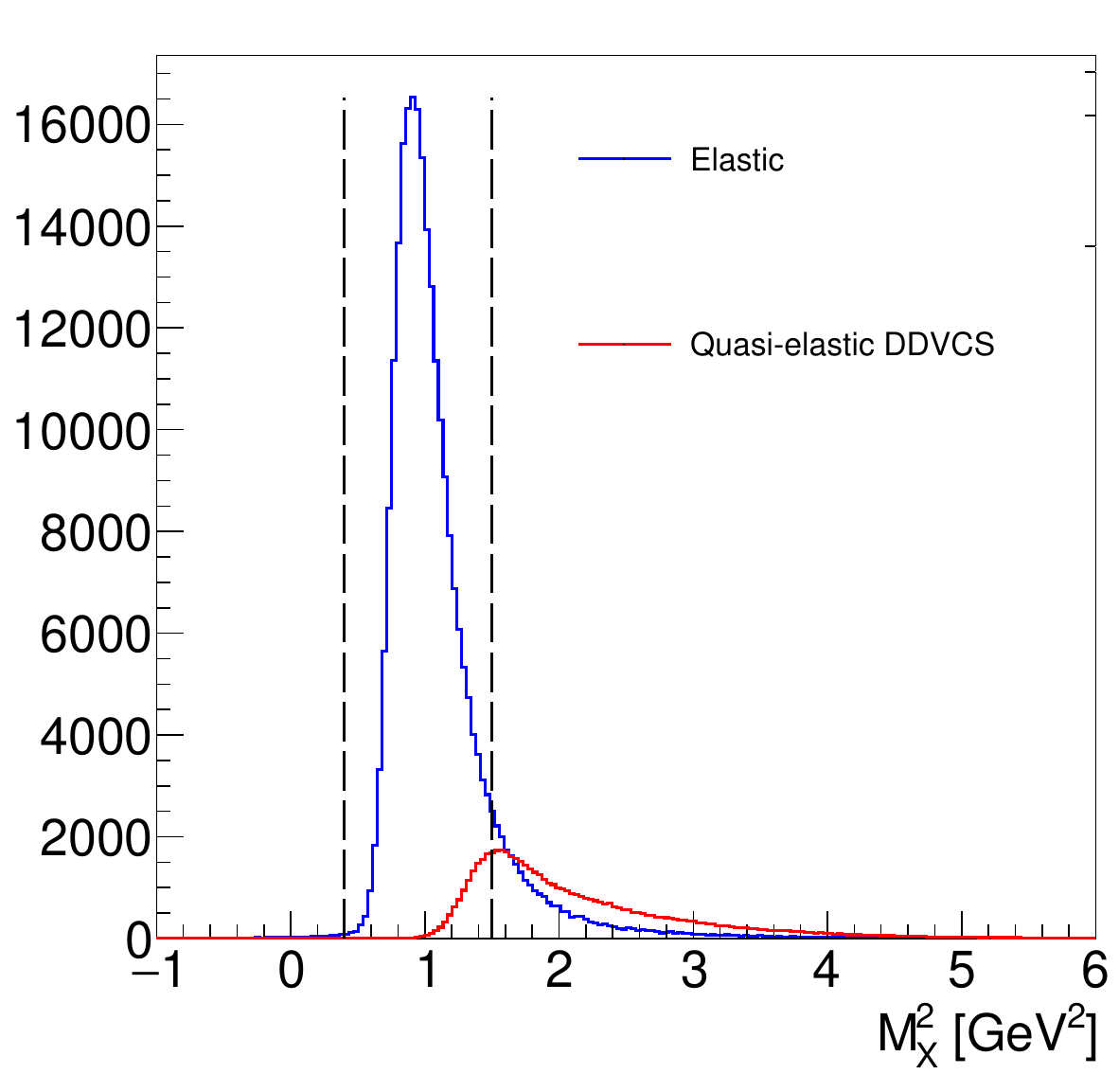}
    \caption{Missing mass squared distribution for elastic (blue) and inelastic (red) muon pair production.
    Vertical dashed lines represent the missing mass cut.} 
    \label{fig:Mx2_Elast_quasiElast}
\end{figure}

The contribution of inelastic events to the elastic final state was evaluated by analyzing the missing mass distribution of $e^\prime \mu^{+} \mu^{-}$ events after applying momentum corrections for muons and smearing the electron momentum to the expected resolution of the wECal $\sigma/\sqrt{E}$=$4\%$. Figure~\ref{fig:Mx2_Elast_quasiElast} shows normalized missing mass squared distributions for both elastic (blue) and inelastic (red) events detected in $\mu$CLAS12. The dashed vertical lines indicate the selection criteria applied on the missing mass squared to select elastic events, $0.4~\mathrm{GeV}^{2} < M_{X}^{2} < 1.5~\mathrm{GeV^{2}}$. The contamination of inelastic events within this range was found to be approximately 5.5\%.


\subsubsection{Pion Pair Production}

To estimate the pion pair background, CLAS12 electroproduction data taken with a 10.6~GeV electron beam scattering on a 5-cm-long LH$_2$ target were analyzed. A dataset corresponding to an integrated luminosity of approximately $\mathrm{30.3\;fb^{-1}}$ was examined. Events containing at least one pion pair were selected, totaling 1.2 billion events, for a cross section of approximately 41~nb. The selected events were processed through $\mu$GEMC and reconstructed using the CLAS12 event reconstruction framework. This approach provides the most accurate estimate of the pion pair electroproduction background, as any pions detectable as MIPs in \uCLAS\ would also be detected in CLAS12 under similar conditions.

\begin{figure}[htbp]
    \centering
    \includegraphics[width=\linewidth]{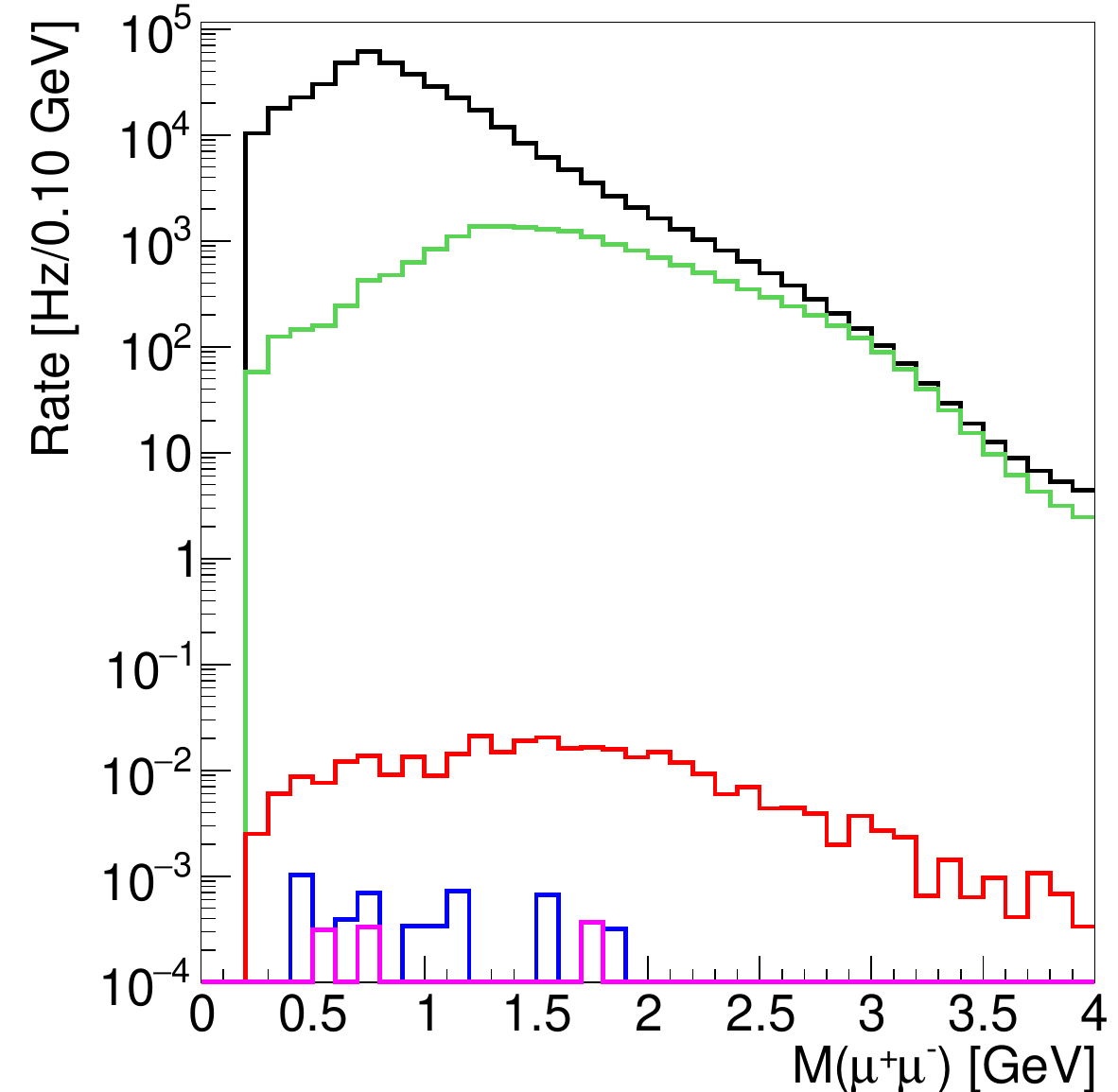} 
    \caption{Rates of pion pairs as a function of the two MIP invariant mass at different stages of the analysis. Description of histograms from highest  to lowest total rate are as follows: the black distribution (391.4~kHz) represents all initial $\pi^+\pi^-$ pairs selected from CLAS12 data; the green distribution shows events where both of pions have more than 2~GeV/$c$ momentum (17.45~kHz), the red shows events in which at least one pair of opposite-charge MIP particles was detected in $\mu$CLAS12 (0.3123~Hz); the blue corresponds to events where an electron was also detected in the wECal (0.0052~Hz). The pink histogram, with only three events, represents the subset of three-particle final states that satisfy the missing mass cut, ${0.4~ {\rm GeV}^{2} < M_{X}^{2} < 1.5~{\rm GeV}^{2}}$ (0.001~Hz).}
    \label{fig:pionBackground_RGA}
\end{figure}

Figure~\ref{fig:pionBackground_RGA} shows the expected rates as a function of the invariant mass of two MIP particles at various analysis stages, normalized to a luminosity of $10^{37}\,\mathrm{cm^{-2}\,s^{-1}}$. The black histogram represents all initial opposite-charge pion pairs, the green distribution corresponds to events where both pions have momenta larger than 2~GeV/$c$, while the red histogram shows events where both pions are reconstructed as MIPs in \uCLAS. The blue histogram corresponds to events where an electron is also identified in the wECal in addition to the MIPs. Finally, the pink histogram shows events that satisfy the missing mass squared criteria for the $ep \rightarrow e^{\prime} \mathrm{MIP^{+} MIP^{-}}X$ final state described in the previous section.

The analyzed statistics yielded three pion pair events with a detected electron that could be reconstructed as $e^{\prime} \mu^{+} \mu^{-}$ and also passed the missing mass squared cut, corresponding to a rate of 0.001Hz. Furthermore, a single event was found in the region of interest, above 1.2 GeV, yielding a rate of pion contamination of about 0.33 mHz. This corresponds to about 1\% contamination to the  $e^{\prime} \mu^{+} \mu^{-}$ final state in the  \mbox{$M_{\mu^{+} \mu^{-}} > 1.2\;{\rm GeV}$} region.

\subsubsection{Accidental Coincidences}

Accidental coincidences occur when a reconstructed MIP particle pair and an electron from an unrelated event are detected within the time window of a single beam bunch. 
The main sources of MIP pairs in \uCLAS\ are final states with $\mu^{+}\mu^{-}$ or $\pi^{+}\pi^{-}$.

\paragraph{Accidental coincidences with muon pairs:}
\label{sec:acc_muons}

\begin{figure}[htbp]
    \centering
    \includegraphics[width=1\linewidth]{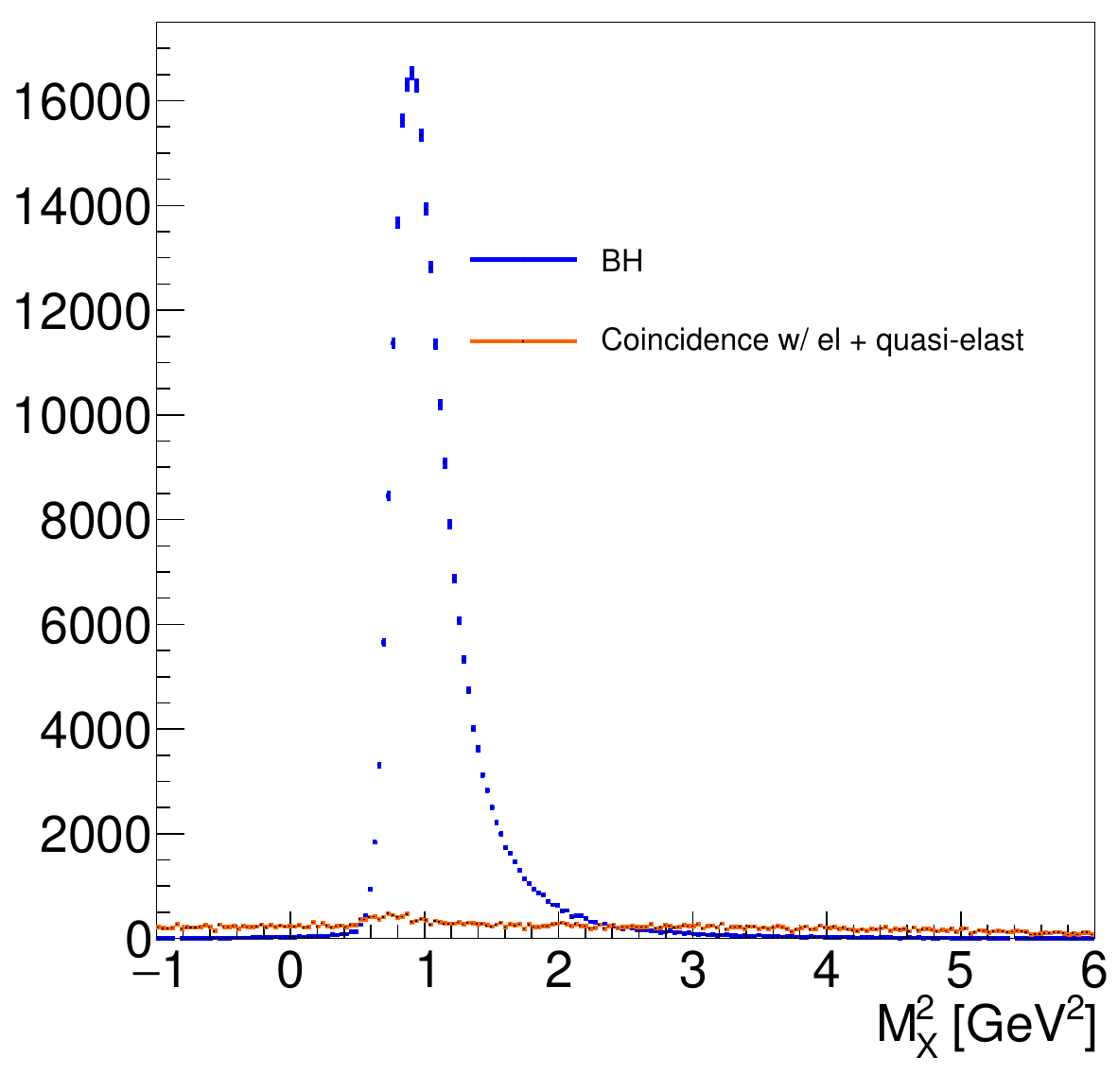}
    \caption{Missing mass squared distribution for electron-di-muon events corresponding to 200 days of running at a luminosity of $10^{37}\,\mathrm{cm^{-2}\,s^{-1}}$. The blue histogram represents true electroproduced BH events, the orange histogram represents the contribution from the coincidence of inclusive electrons with elastic and inelastic  $\mathrm{\mu^{+}\mu^{-}}$ pairs. 
    In the range $0.4~\mathrm{GeV}^2 < M_{X}^{2} < 1.5~\mathrm{GeV}^2$, the total contribution from accidentals is $\mathrm{4.6\%}.$}
    \label{fig:accidentalCoincidences}
\end{figure}

To estimate the fraction of muon pair coincidences, inclusive electron events were generated using the IncEG generator~\cite{Klimenko:incEG}, which accurately reproduces the CLAS12 inclusive cross section results~\cite{CLAS:2025zup}. Additionally, both elastic and inelastic muon pairs were generated with GRAPE without imposing any constraints on the scattered electron momentum. The inclusive electron events from IncEG and all particles from the GRAPE generator were combined to create mixed events. These events were then processed through the $\mu$GEMC simulation and reconstructed using COATJAVA. The cross sections for di-muon production and inclusive electron scattering were used to calculate the coincidence rate ${R_{C}}$ as
\begin{equation}
    {R_{C}} = \sigma_{\rm Incl} \sigma_{\rm dilepton} \mathcal{L}^2 \times \mathrm{\tau}
\end{equation}
where $\mathcal{L}$ is the instantaneous luminosity, $\mathrm{\sigma_{Incl}}$ is the inclusive cross section, $\mathrm{\sigma_{dilepton}}$ is the dilepton production cross section from GRAPE and the $\mathrm{\tau}$ is the coincidence time window equal to 4 ns, the time between two beam bunches delivered to CLAS12. Figure~\ref{fig:accidentalCoincidences} shows the missing mass squared distribution of electron-di-muon final states for 200~days of running at a luminosity of $10^{37}\,\mathrm{cm^{-2}\,s^{-1}}$.
The distribution includes both true BH di-muons (in blue) and the contribution from the coincidence of inclusive electrons with elastic and inelastic di-muon pairs (in orange). In the missing mass squared range $0.4~\mathrm{GeV}^2 < M_{X}^{2} < 1.5~\mathrm{GeV}^2$, the total contribution from accidentals is $\mathrm{4.6\%}$.


\paragraph{Accidental coincidences with pion pairs:}
\label{sec:acc_pions}
The accidental coincidence rate for pion pairs with an inclusive electron was estimated from the coincidence rate of muon pairs with an inclusive electron.
The total detection rate of $ep\rarr \mu^{+}\mu^{-}(X)$ in \uCLAS\ with ${M_{\mu^{+}\mu^{-}} > 1.2\; \mathrm{GeV}}$ is 1.24~Hz.
The rate of ${\pi^{+}\pi^{-}}$ pairs
in the same mass range was obtained from the integral of the red histogram above $M_{\mu^{+}\mu^{-}} > 1.2\; \mathrm{GeV}$ in Fig.~\ref{fig:pionBackground_RGA}, and found to be 0.31~Hz.
This is four times smaller than the muon pair detection rate. Consequently, there will be four times less ${\pi^{+}\pi^{-}}$ pair accidental coincidences with electrons compared to coincidences of
${\mu^{+}\mu^{-}}$ pairs with electrons. This leads to a contribution to the true ${ep \rightarrow e'\mu^{+}\mu^{-}(p')}$ final state that is below the percent level.

\section{Planned Measurements}
\label{measurements}

$\mu$CLAS12 will measure the production of muon pairs in electron-proton scattering with an $11$~GeV longitudinally polarized electron beam. Multiple processes, DDVCS, TCS, and \jpsi~production, will be studied using the exclusive production of muon pairs. Observables including cross sections, beam-spin asymmetries, and angular asymmetries will be measured across a wide range of initial photon energies, invariant squared four-momentum transfer $t$, and spacelike and timelike virtualities of the incoming and outgoing photons. For the DDVCS and \jpsi~measurements, the scattered electron and the muon pairs will be detected in $\mu$CLAS12, while the recoil proton will remain undetected 
\begin{equation}
ep\to e^\prime \mu^+ \mu^- (p^\prime),
\end{equation}
but identified in the missing mass analysis as
 \begin{equation}
   M^2_{ X}=(k+p-k^\prime-p_{\mu^+}-p_{\mu^-})^2\approx M^2_{p},
 \end{equation}
where all four-momenta are defined in Fig.~\ref{fig:ddvcs}. For the TCS measurements, the scattered proton and the pair of muons will be detected
\begin{equation}
ep\to (e^\prime) \mu^+ \mu^- p^\prime. 
\end{equation} 
The undetected scattered electron kinematics will be deduced by missing momentum analysis, using an approach similar to that used in the published CLAS12 TCS analysis~\cite{clas12tcs}.
Events will be selected by requiring the missing mass of the scattered electron $M^2_{ X}$ to be consistent with the electron mass,
\begin{equation}
   M^2_{ X}=(k+p-p^\prime-p_{\mu^+}-p_{\mu^-})^2\approx M^2_{e},
\end{equation} 
and the transverse missing momentum fraction to be close to zero
\begin{equation}
   \frac{P^\bot_{\rm X}}{P_{ X}}\approx 0.
\end{equation}
Additionally, semi-exclusive ($e^\prime \mu^{\pm} p^\prime X$) final states will also be analyzed to develop momentum corrections and for systematic studies.

\subsection{Double Deeply Virtual Compton Scattering}

\begin{figure*}[htbp]
\begin{center} 

\begin{subfigure}{0.49\linewidth}
\centering  
\includegraphics[width=\linewidth]{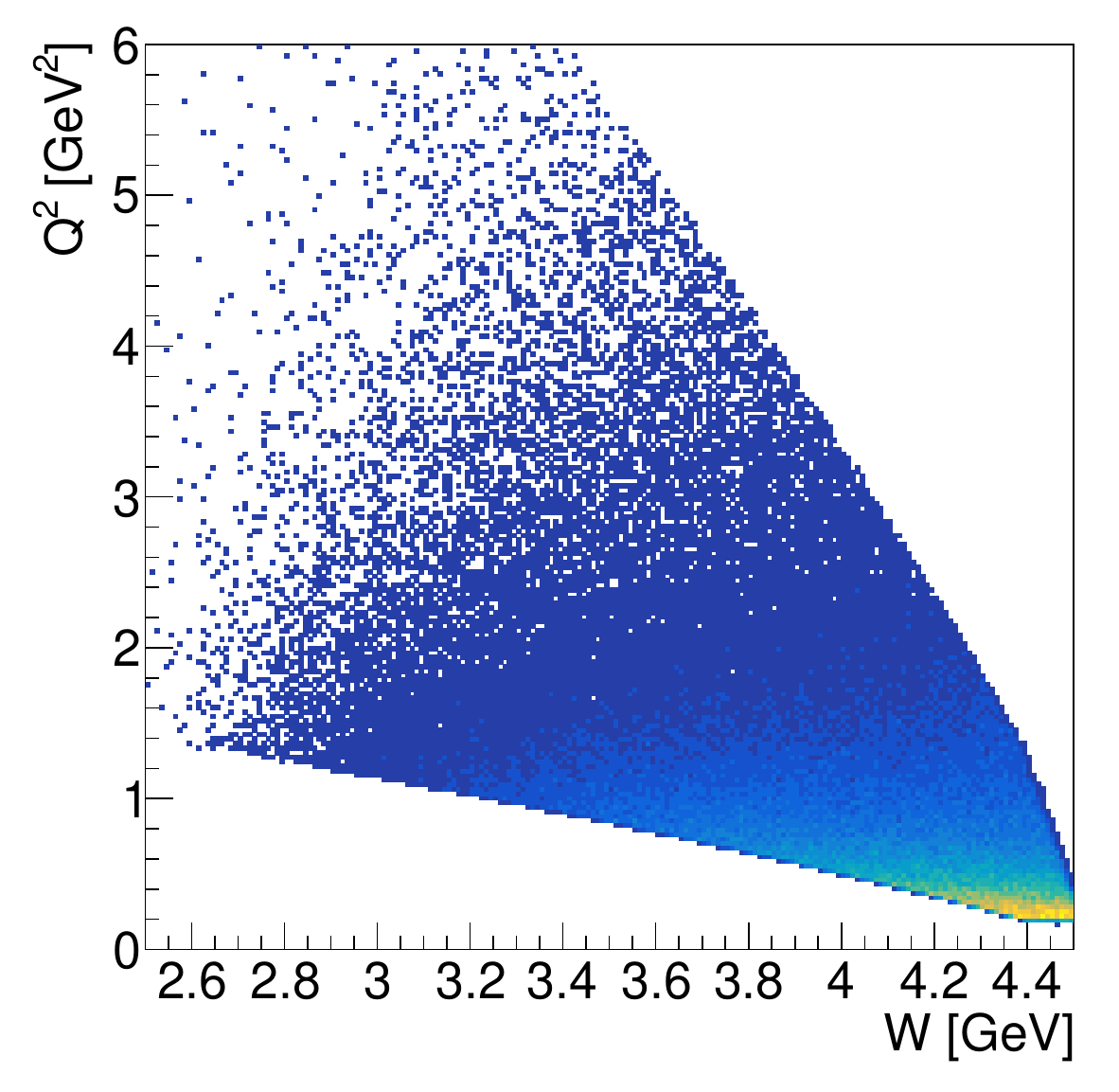}
 \caption{}
 \label{fig:ekine:1}
\end{subfigure}
\begin{subfigure}{0.49\linewidth}
\centering 
\includegraphics[width=\linewidth]{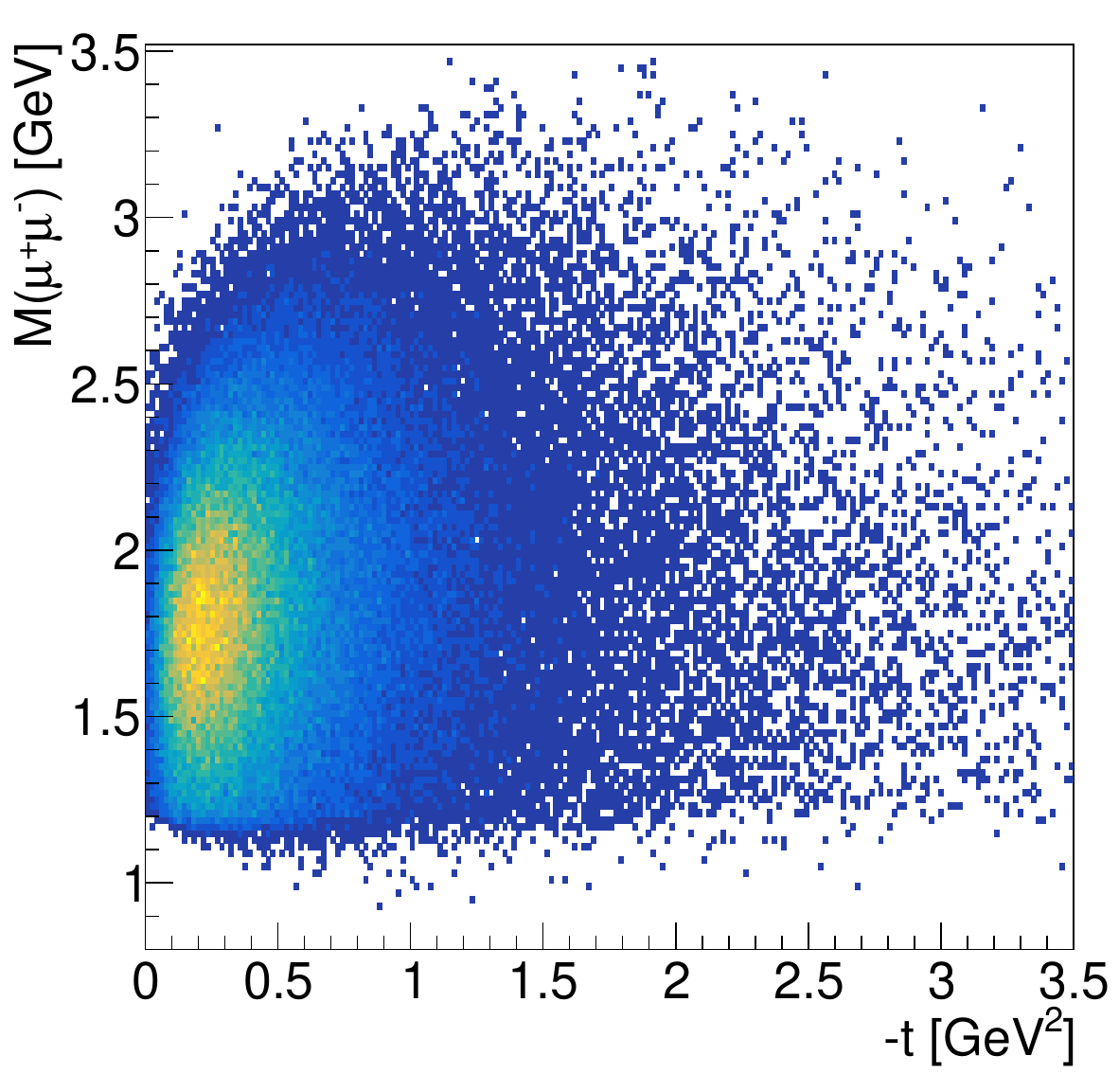}
\caption{}
 \label{fig:ekine:2}
\end{subfigure}

\end{center}
\caption{Kinematic coverage of $\mu$CLAS12 for di-muon electroproduction. Panel~\ref{fig:ekine:1}: $Q^2$ vs. $W$ distribution, with limits defined by the detection of the scattered electron in the wECal. Panel~\ref{fig:ekine:2}: Invariant mass distribution of lepton pairs detected in the $\mu$CLAS12 FD as a function of $t$.}
\label{fig:ekine}
\end{figure*}

\begin{figure}[!htb]
 \centering
\includegraphics[width=1\linewidth]{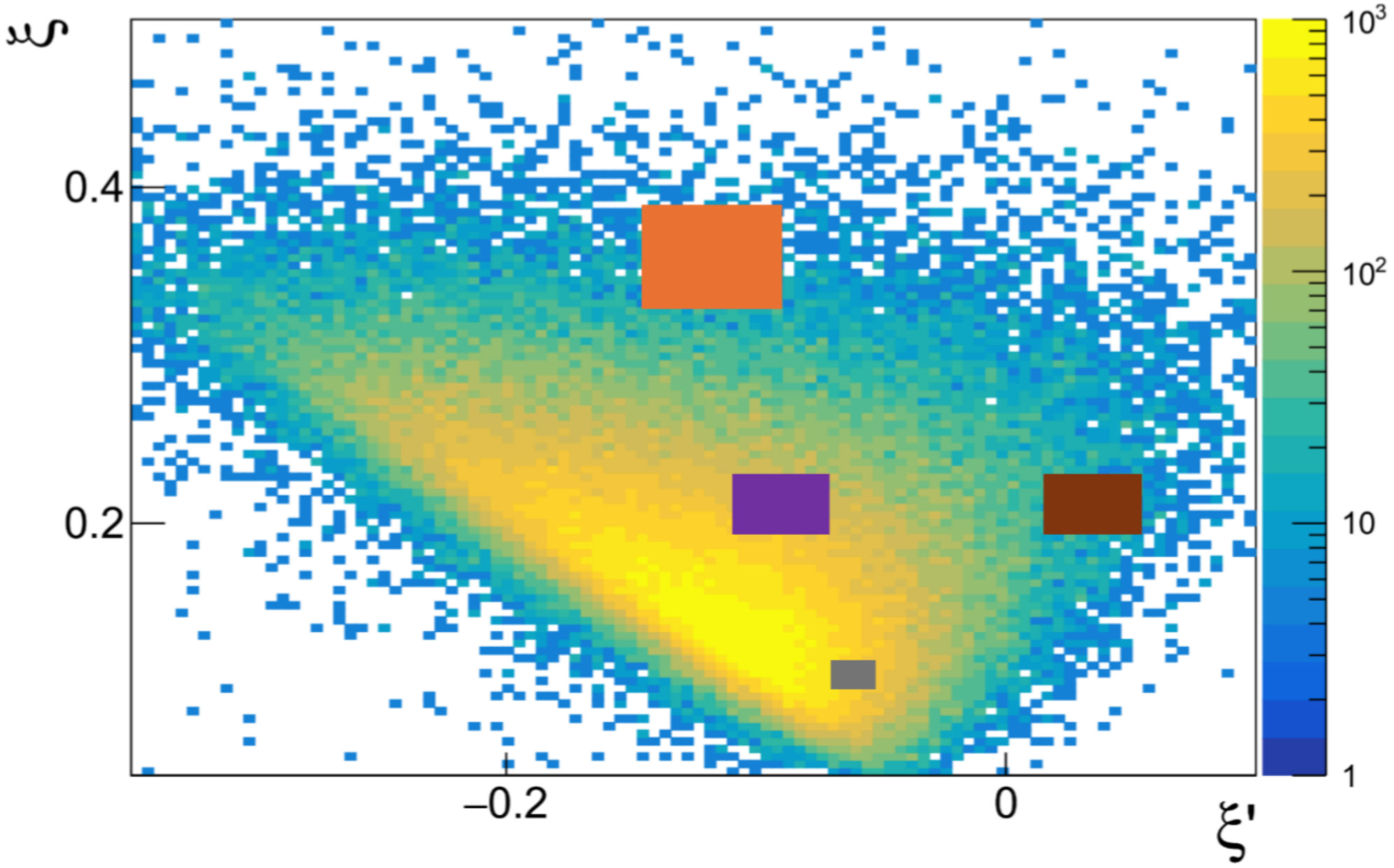}
 \caption{$\xi$ vs. $\xi^\prime$ distribution for reconstructed DDVCS events. The boxes represent the kinematic bins used to illustrate the expected beam-spin asymmetries in the timelike ($Q^{\prime 2} > Q^2$) and spacelike ($Q^{\prime 2} < Q^2$) regions.}
 \label{fig:xi_xip_bins}
\end{figure}

The kinematic coverage of $\mu$CLAS12, utilizing a LH$_2$ target and an 11 GeV electron beam, for the electroproduction of di-muons in terms of $W$, $Q^2$, $t$, and $M_{\mu^+\mu^-}$ is illustrated in Fig.~\ref{fig:ekine}. The distributions were obtained using BH events produced by the GRAPE event generator and processed through $\mu$GEMC. The scattered electrons were reconstructed in the wECal, with a momentum detection threshold of $0.5$ GeV/$c$ in the angular range $8^\circ <\theta_{e^\prime}<30^\circ$. Muon kinematics were obtained from the FD of $\mu$CLAS12, accounting for energy loss in the wECal and the lead shield. The accessible phase space in $\xi$ and $\xi^\prime$ is shown in Fig.~\ref{fig:xi_xip_bins}. The $\mu$CLAS12 acceptance predominantly favors the timelike region largely due to the limit at large \mbox{$Q^2\gg Q^{\prime 2}_{\rm min}\ge 1.4$ GeV$^2$}. However, expected statistics will also enable studies of DDVCS in the spacelike region of $\xi^\prime <0.1$. More than $0.5\times 10^6$ events are expected to be collected for the DDVCS analysis.

\subsubsection{Observables}

\begin{figure*}[htbp]
\centering
\includegraphics[width=\linewidth]{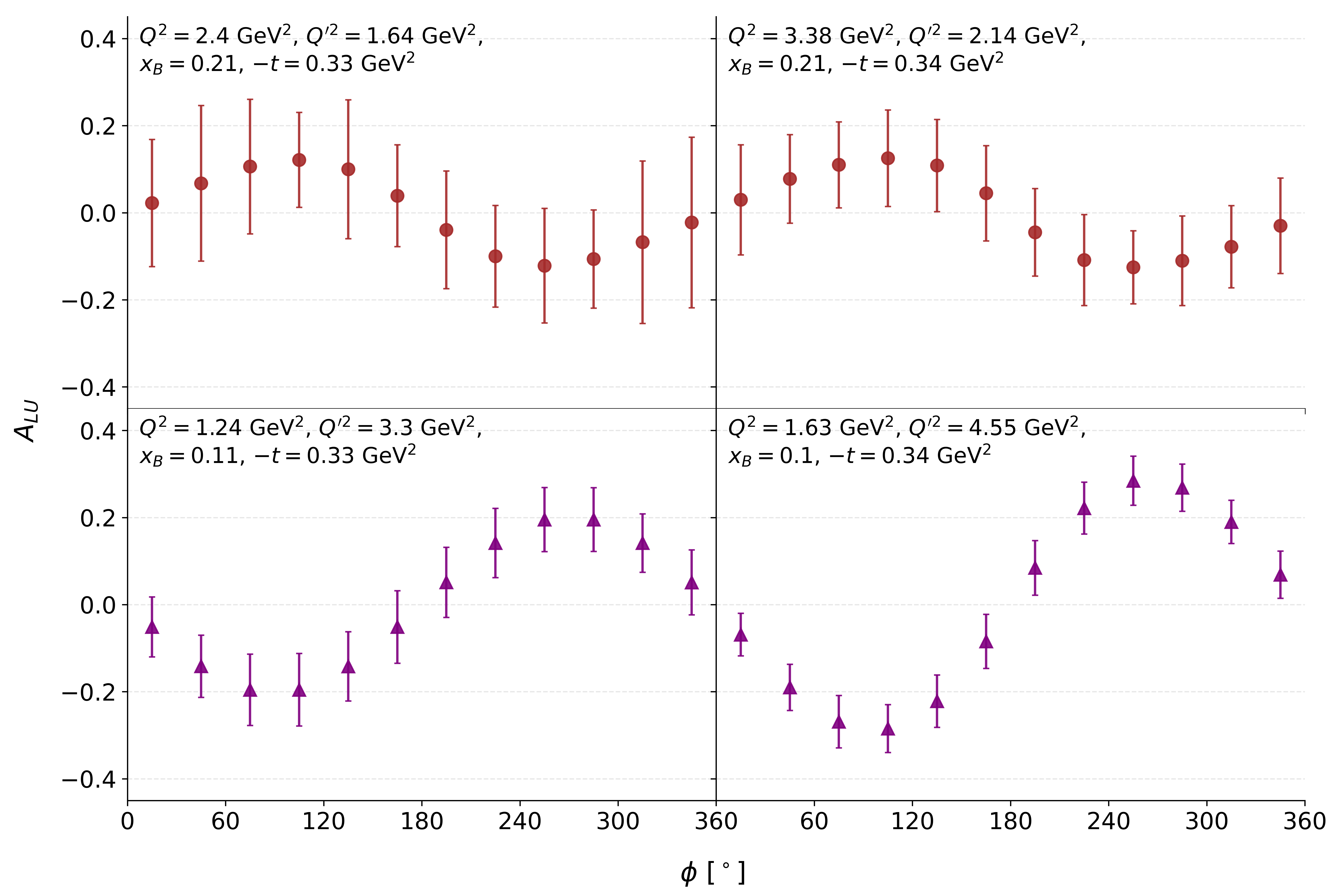}
 \caption{Expected DDVCS $A_{LU}$ as a function of the angle $\phi$. The top plots show the expected $A_{LU}$ in the spacelike region (brown box of Fig.~\ref{fig:xi_xip_bins}) for two different kinematic points, the bottom plots for two bins in the timelike region (purple box of Fig.~\ref{fig:xi_xip_bins}). The expected change of sign between the two kinematic region is clearly visible.}
\label{fig:bsa_sl_tl}
\end{figure*}

The primary goal of DDVCS studies is the measurement of beam-spin asymmetries $A_{LU}$ as a function of the angle between the leptonic and hadronic planes, $\phi$, in a wide range of skewness $\xi$ and the generalized Bjorken variable $\xi^\prime$. The asymmetry is defined as
\begin{equation}
 A_{LU} = \frac{1}{P_b}\frac{N^{+} - N^{-}}{N^{+} + N^{-}},
 \label{eq:alu}
\end{equation}
where $N^{-}$ and $N^{+}$ are the acceptance-corrected number of events with positive and negative beam helicities, respectively, and $P_b$ is the average beam polarization. 
The asymmetry will be measured in multiple bins, covering both the spacelike ($Q^{\prime 2} < Q^{2}$) and timelike ($Q^{\prime 2} > Q^{2}$) regions. A key objective is to observe the sign change of the asymmetry between these regions. For the violet (timelike region) and brown (spacelike region) boxes in Fig.~\ref{fig:xi_xip_bins}, the asymmetry $A_{LU}$ was extracted from simulated data for two different mean values of ${Q}^{\prime 2}$ and ${Q}^2$. These asymmetries were generated using the VGG model \cite{marcprl2}. The obtained $A_{LU}$ values, along with the expected statistical uncertainties, are shown in Fig.~\ref{fig:bsa_sl_tl}. As expected, the statistical uncertainties of $A_{LU}$ are larger in the spacelike region than in the timelike region. However, in both cases, the expected statistical uncertainties are small enough to extract the $\sin{\phi}$ modulation, $A^{\sin{\phi}}_{LU}$, with sufficient precision.   

\begin{figure}[hbtp]
 \centering
\includegraphics[width=1\linewidth]{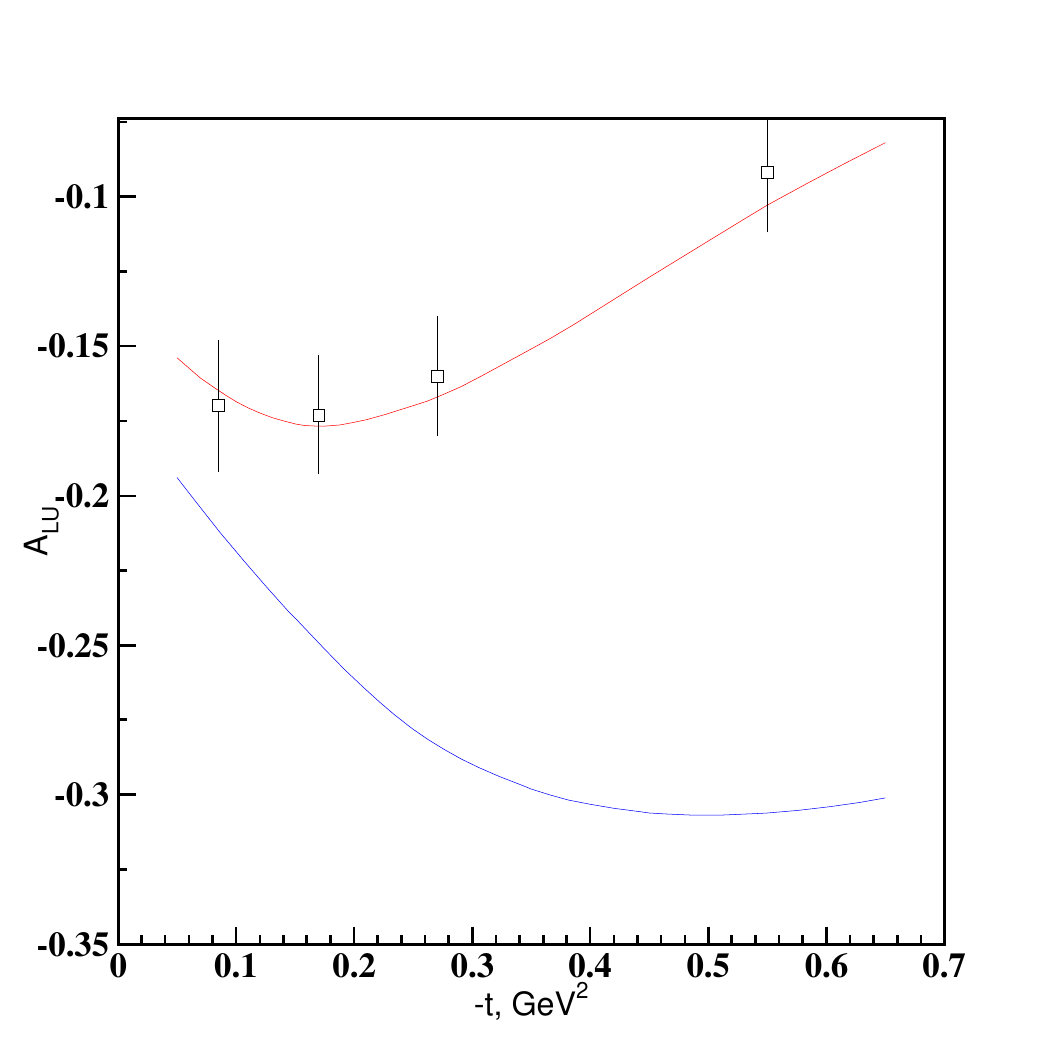}
\caption{Projected DDVCS $A_{LU}$ as a function of $-t$. The extracted asymmetries from 1000 pseudo-experiments are shown by the black points. The RMS of the extracted asymmetries is taken as the uncertainty on the measurement. The red and blue lines correspond to predictions from the VGG and the GK19 models, respectively.}
 \label{fig:bsa_tdep}
\end{figure}

Another key measurement is the extraction of the $t$-dependence of $A_{LU}$. To demonstrate its feasibility, a representative kinematic region was selected, corresponding to average values of $\xi^\prime $=$ -0.062$ and $\xi $=$ 0.1$ as indicated by the gray box in Fig.~\ref{fig:xi_xip_bins}. The $t$-range was divided into four bins centered at $-0.058$, $-0.17$, $-0.27$, and $-0.55$~GeV$^2$. The event yields for each bin were obtained from a full simulation of BH events generated using the GRAPE event generator, passed through $\mu$GEMC and reconstructed using COATJAVA. Asymmetries were modeled using the VGG model for GPDs. For each ($Q^2$, $Q^{\prime 2}$, $x_B$, $t$, $\phi$)-bin, pseudo-data samples were generated by smearing the event counts $N^+$ and $N^-$ within one standard deviation. For each $t$-bin, 1000 pseudo-experiments of $A_{LU}(\phi)$ were generated and fitted with the function $A_{LU}(\phi)$=$A_{LU}^{90^\circ}\sin{\phi}$. The $t$-dependence of the extracted $A_{LU}^{90^\circ}$ values and the estimated statistical uncertainties are shown in Fig.~\ref{fig:bsa_tdep}. The extracted asymmetries align closely with the input from the VGG model, confirming the robustness of the extraction method. The statistical precision achieved in each $t$-bin is sufficient to clearly distinguish between different GPD model predictions, demonstrating the capability of $\mu$CLAS12 to provide meaningful constraints on theoretical frameworks through precise measurements of beam-spin asymmetries.

\begin{figure}[hbtp]
 \centering
\includegraphics[width=\linewidth]{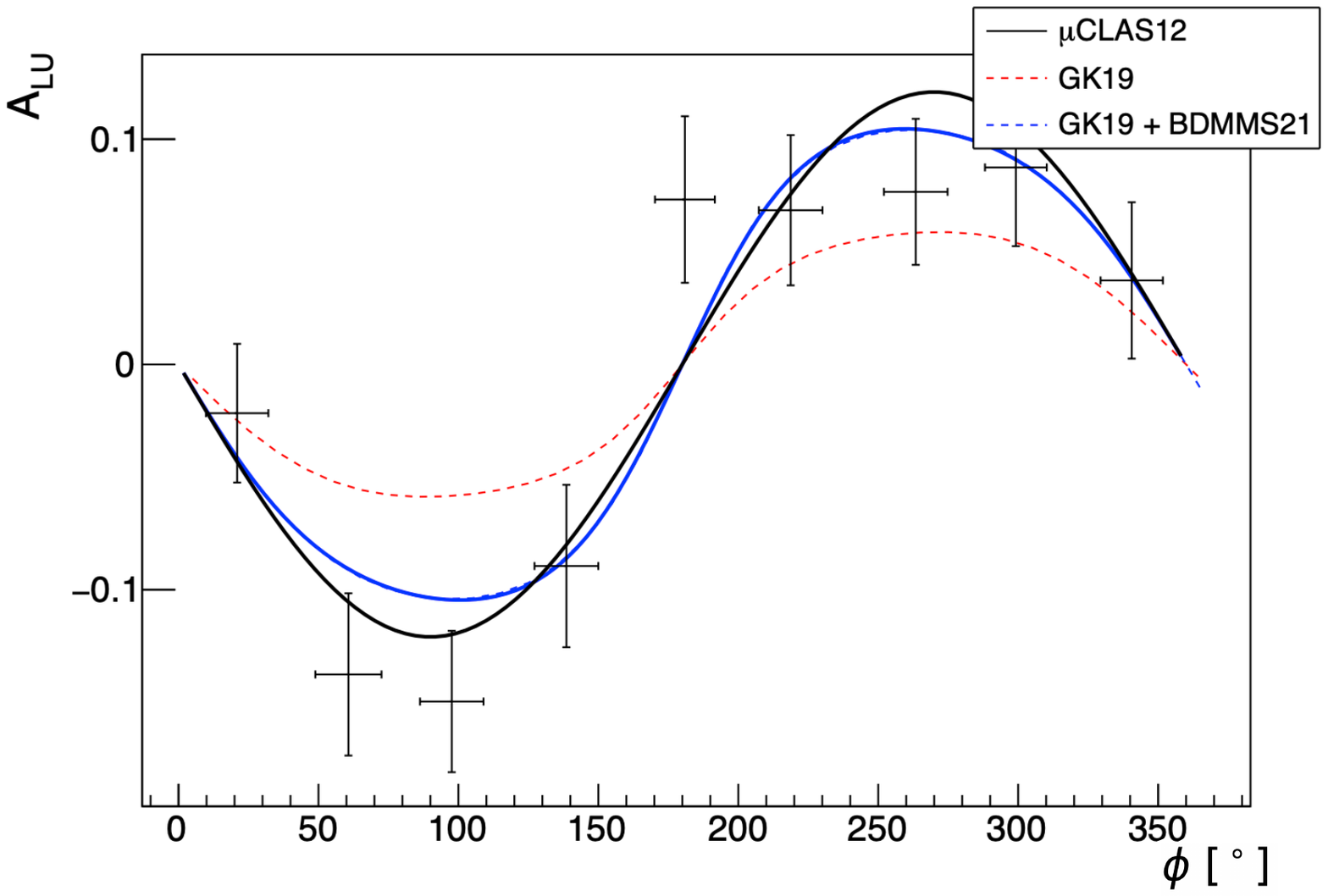}
 \caption{Comparison of the expected $A_{LU}$ asymmetry generated using the GK19 model from PARTONS (red curve), with the prediction including an additional SGPD, implemented via the GPDBDMMS21 module of the PARTONS software~\cite{partons} (blue curve). Black points represent projected pseudo-data, with the black curve showing the fit using the function $A_{LU}(\phi)$=$A_{LU}^{90^\circ}\sin{\phi}$. The kinematic bin corresponds to ${\xi} $=$ 0.36$, ${\xi}^\prime $=$ -0.0821$, and ${t} $=$ -0.82$~GeV$^2$. }
 \label{fig:bsa_sgpd}
\end{figure}

As discussed in Section~\ref{sec:GDP_exp}, DVCS and TCS observables access only two of the three variables that define GPDs. The variable $x$ is integrated out in the CFFs for these processes, leading to non‑unique solutions when reconstructing GPDs from experimental data. Indeed, SGPDs can be added to regular GPDs while still reproducing DVCS and TCS measurements, thereby complicating their interpretation in terms of GPDs. On the other hand, DDVCS observables retain sensitivity to the variable $x$, allowing all three GPD variables to vary independently in the $|x|\le \xi$ region. This allows GPDs to be mapped in three dimensions, leading to a more detailed and precise picture of nucleon structure. Figure~\ref{fig:bsa_sgpd} shows an example of the planned $A_{LU}$ measurement in the ($\xi,\xi^\prime$)-bin identified by the orange box in Fig.~\ref{fig:xi_xip_bins}. Asymmetries generated using the GK19 model~\cite{Kroll:2019wug} from PARTONS~\cite{partons} and asymmetries generated with the same model that incorporates an additional SGPD can be well distinguished. As in the $t$-dependence studies, $10^4$ pseudo-asymmetries were generated according to the expected statistics, and fitted with the $A_{LU}(\phi)$=$A_{LU}^{90^\circ}\sin{\phi}$ function. In this particular bin, the $A_{LU}^{90^\circ}$ was extracted with a $13.6$\% accuracy, and a $4.5$\% relative shift. This highlights the capability of DDVCS measurements to mitigate ambiguities associated with SGPD contributions to existing DVCS observables, thereby enhancing the reliability of GPD extractions. Yet, the DDVCS measurements of $\mu$CLAS12 will not completely resolve the SGPD problem. A full separation of SGPD effects will remain unattainable in certain regions of the accessible phase space and for some classes of SGPDs, as shown in Ref.~\cite{bertone}. Even so, these measurements will substantially constrain SGPDs and support more robust GPD modeling.

\subsubsection{Systematic Uncertainties}
\label{sec:syst_DDVCS}

The main sources of systematic uncertainties in the DDVCS $A_{LU}$ measurements are the knowledge of beam polarization, $P_b$, the background in the event sample, and radiative effects. 

The beam polarization will be measured using the Hall B M{\o}ller polarimeter. This polarimeter has been in service since the start of the Hall B physics program, and its accuracy has been studied and checked relative to other such measurements at CEBAF. It is estimated to be below $3\%$ of the measured polarization value, as shown in Ref.~\cite{clas12beam}. 

The backgrounds that will contribute to the final state of interest are extensively discussed in Section~\ref{physbg}. The total estimated background is about $12$\%, where the largest contribution, $5.5$\%, comes form inelastic muon pair production. The accidental background will not have a significant beam spin asymmetry, considering that the difference in the accumulated charge between both beam-helicity orientation is expected to be negligible. Similarly, the inelastic muon pair contribution is not expected to have an asymmetry and thus will only dilute the measured DDVCS asymmetry. Finally, studies with pion pair production from CLAS12 hydrogen data did not yield a statistically significant asymmetry at $Q^{\prime 2}>1.2$ GeV, and it is expected that this background will only contribute to the dilution of the measured asymmetry. 

\begin{figure}[h!]
\centering
\includegraphics[width=\linewidth]{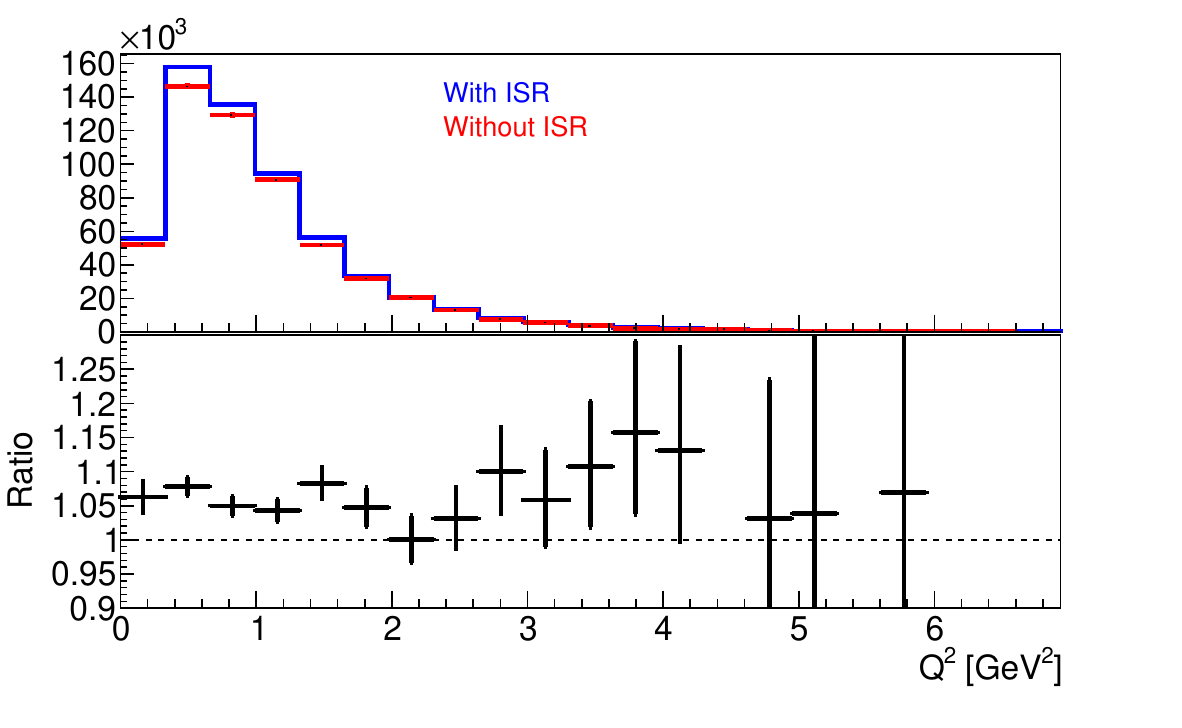}
\caption{The effect of initial state radiation (ISR) for the $ep\rightarrow e^\prime\mu^+\mu^-(p')$ using the GRAPE event generator. Top: number of reconstructed BH events generated with (blue) and without ISR (red). Bottom: ratio of reconstructed events with and without ISR.}
\label{fig:isr}
\end{figure}

The radiative effects for DDVCS at JLab kinematics have been studied in Ref.~\cite{Heller:2021}. At JLab kinematics, the radiative corrections to the cross section of the reaction $ep\rightarrow e^\prime \mu^+\mu^-(p^\prime) $ can reach up to 35\%. However, for polarization asymmetries, radiative effects are expected to be below $5\%$. In addition, the impact of initial state radiation (ISR) was investigated using the GRAPE event generator. Samples with and without ISR were processed through the $\mu$CLAS12 simulation and reconstruction chain. Figure~\ref{fig:isr} shows the number of reconstructed events in both cases and their ratio as a function of $Q^2$. The ISR effect is approximately $10\%$ at low $Q^2$, gradually increasing to $15\%$ above $3$~GeV$^2$.

\subsection{Near-Threshold \jpsi\ Production}

The exclusive production of vector mesons has long been identified as one of the primary ways to access the gluon content of the nucleon. This section demonstrates that $\mu$CLAS12 will be capable of measuring the production of \jpsi~with a pair of muons in the final state, with expected statistics significantly larger than those accumulated by current \jpsi~experiments at JLab~\cite{gluexjp:2019,gluexjp:2023,hallc:007,007:2026dow, CLAS:2026lls}. Additionally, this measurement will provide data for initial photon virtualities up to 1~GeV$^2$, offering further leverage to understand the gluon content of the proton.

\begin{figure*}[htbp]
\begin{center} 
\begin{subfigure}{0.49\linewidth}
\centering  
\includegraphics[width=\linewidth]{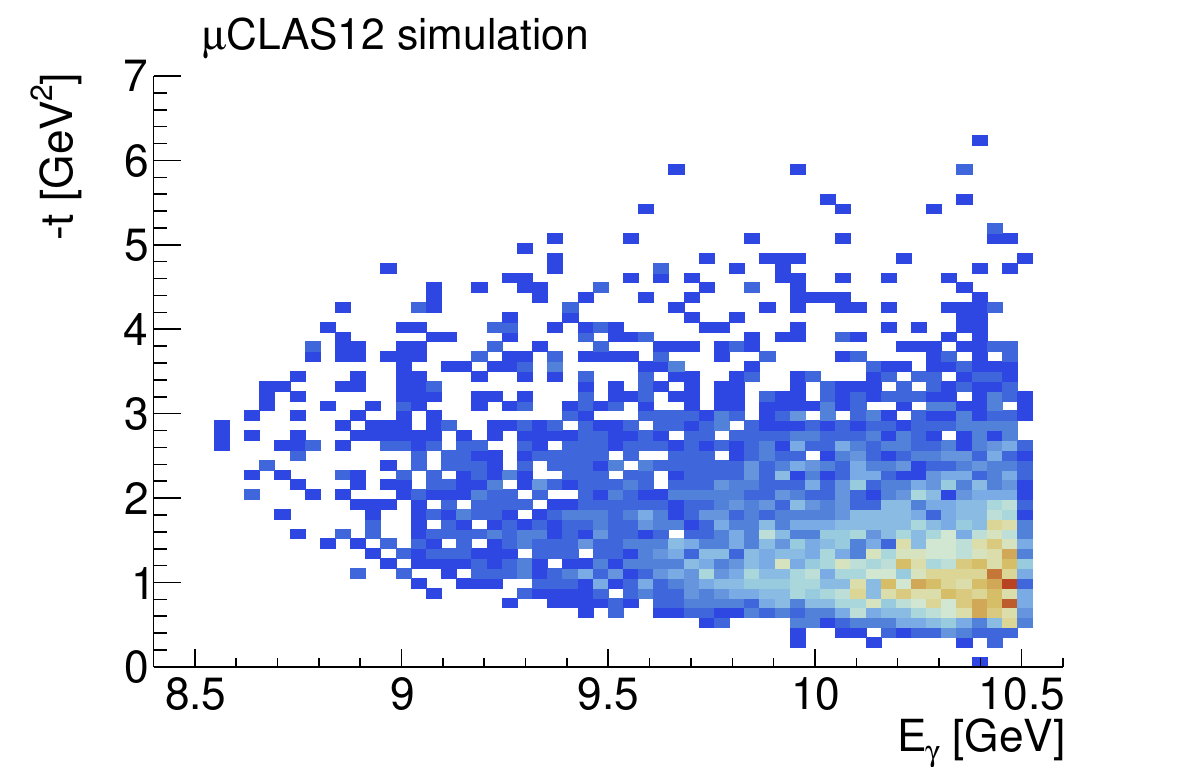}
\caption{}
 \label{fig:jp_ekine:1}
\end{subfigure}
\begin{subfigure}{0.49\linewidth}
\centering  
\includegraphics[width=\linewidth]{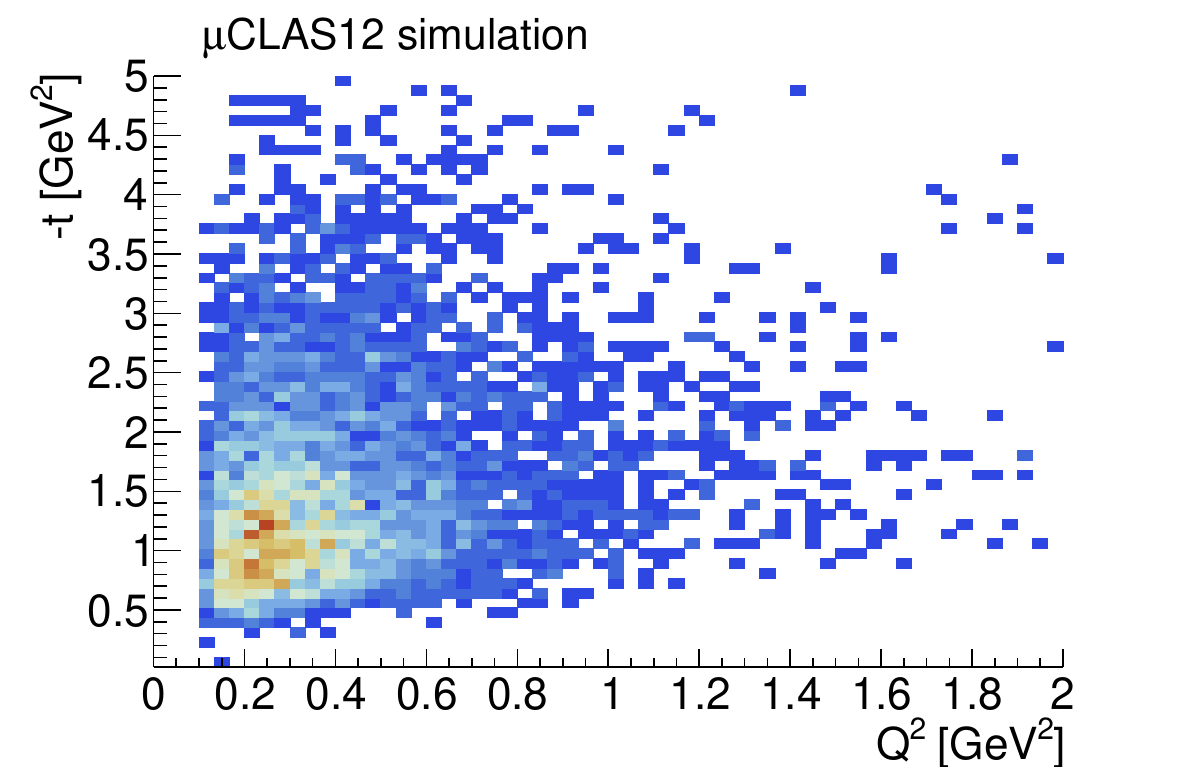}
\caption{}
 \label{fig:jp_ekine:2}
\end{subfigure}

\begin{subfigure}{0.49\linewidth}
\centering  
\includegraphics[width=\linewidth]{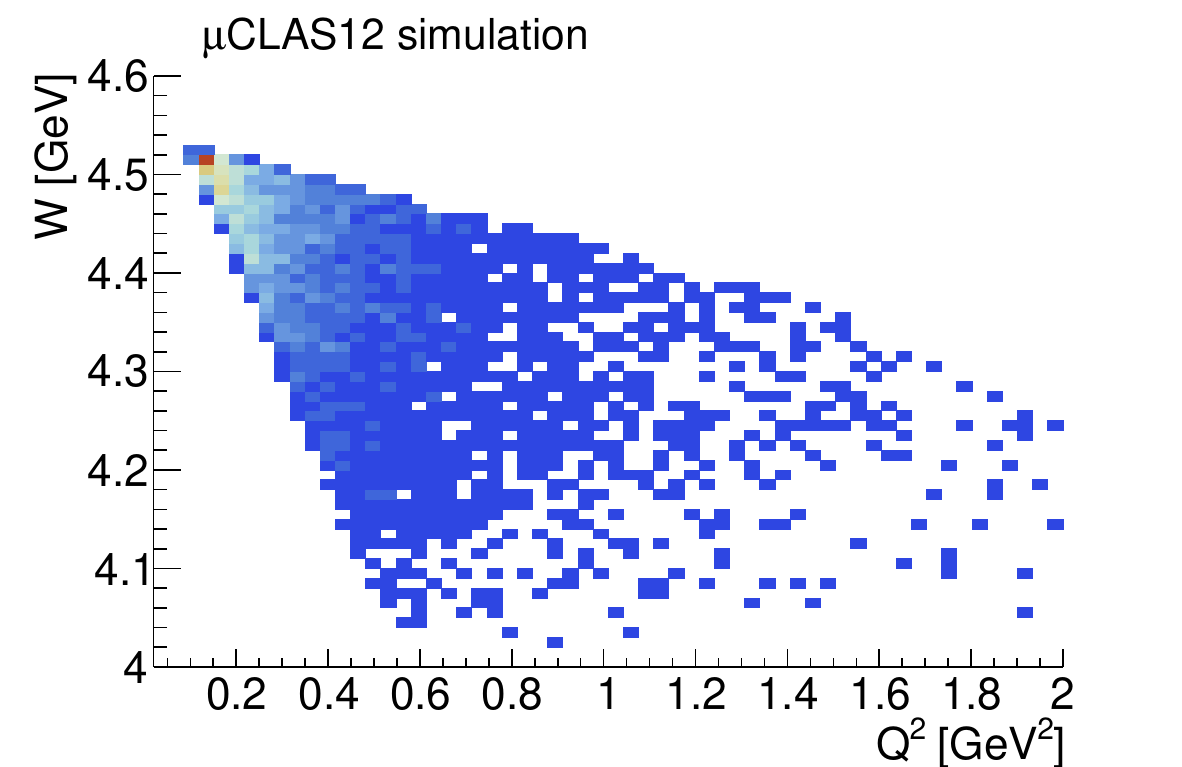}
\caption{}
 \label{fig:jp_ekine:3}
\end{subfigure}
\caption{Kinematics of \jpsi~electroproduction with a 11~GeV beam. Panel~\ref{fig:jp_ekine:1}: $-t$ vs $E_{\gamma}$, Panel~\ref{fig:jp_ekine:2}: $-t$ vs $Q^2$, Panel~\ref{fig:jp_ekine:3}: $W$ vs $Q^2$. All distributions are produced using \jpsi~events with final state particles in the acceptance of $\mu$CLAS12.}
\label{fig:jp_ekine}
\end{center}
\end{figure*}

The model developed in Ref.~\cite{PhysRevD.100.034019} and implemented in the \textit{elSpectro} event generator \cite{elSpectro_generator}, was used to simulate \jpsi~events produced with an 11~GeV beam. MC samples describing the various backgrounds of this measurement were also produced, particularly to describe the mass continuum at lower invariant mass. The generated events were passed through $\mu$GEMC, and events with two detected muons and an electron with momentum above 0.5 GeV/$c$ and within the geometric acceptance of the wECal were kept for the rest of the analysis. The energy of the electron was smeared by 4\%/$\sqrt{E}$ to best match the expected performance of the wECal. Figure~\ref{fig:jp_ekine} shows the kinematic reach of $\mu$CLAS12 for the selected \jpsi~events. The accessible $t$-range extends from 0.5 to 4~GeV$^2$, since the scattered proton will not be detected. This gives access to a region where current JLab experiments have accumulated only limited data to measure the $t$-dependent cross section, which is a key observable for understanding the gluon distribution in the proton. The initial photon virtuality spans approximately 0.1 to 1~GeV$^2$, allowing studies of the $x_B$-dependence of the gluon content of the proton. The initial photon energies $E_\gamma$ will cover the threshold region up to 10.5~GeV. The kinematic range where open-charm contributions are expected to be largest will be fully included, allowing for a precise investigation of the impact of such contributions. The $W$-coverage extends up to 4.5~GeV allowing to investigate the production of pentaquarks, as demonstrated in Section~\ref{sec:pentaquarks}. 

\begin{figure*}[htbp]
\centering
  \begin{subfigure}{0.49\linewidth}
\centering  
   \includegraphics[width=\linewidth]{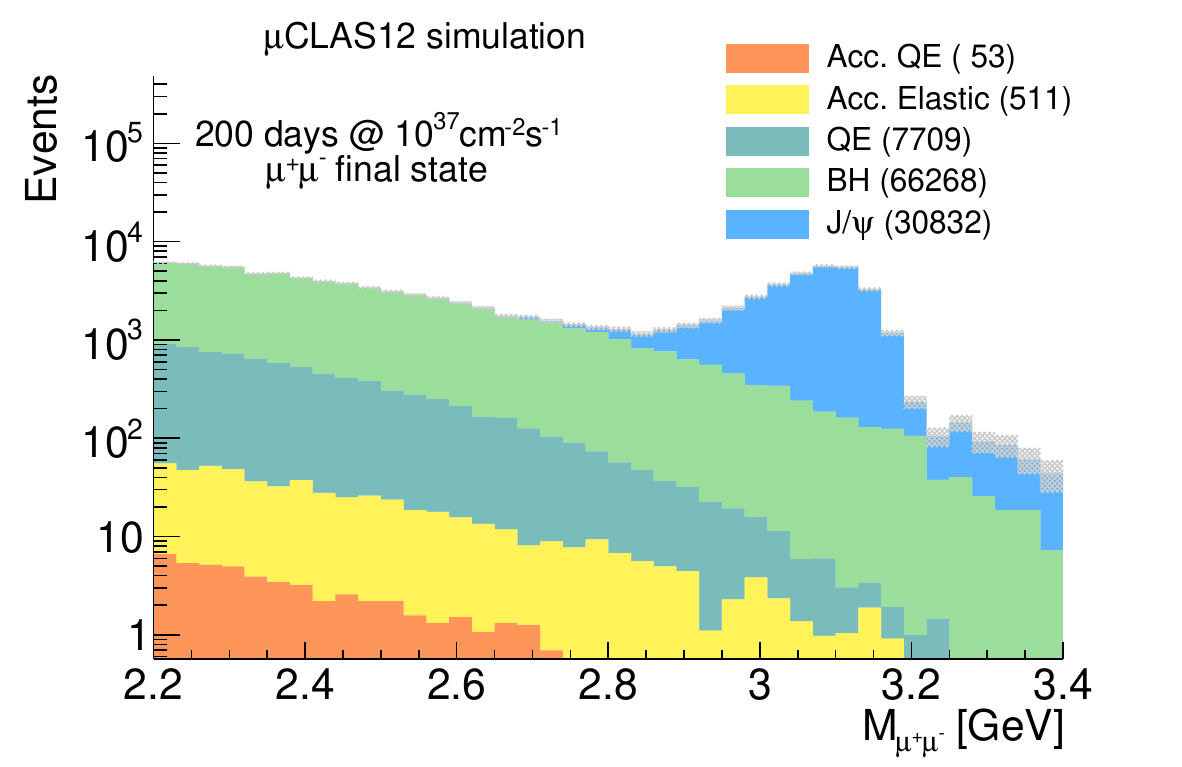}
\caption{}
 \label{fig:Jpsi_M:1}
\end{subfigure}
   \begin{subfigure}{0.49\linewidth}
\centering  
   \includegraphics[width=\linewidth]{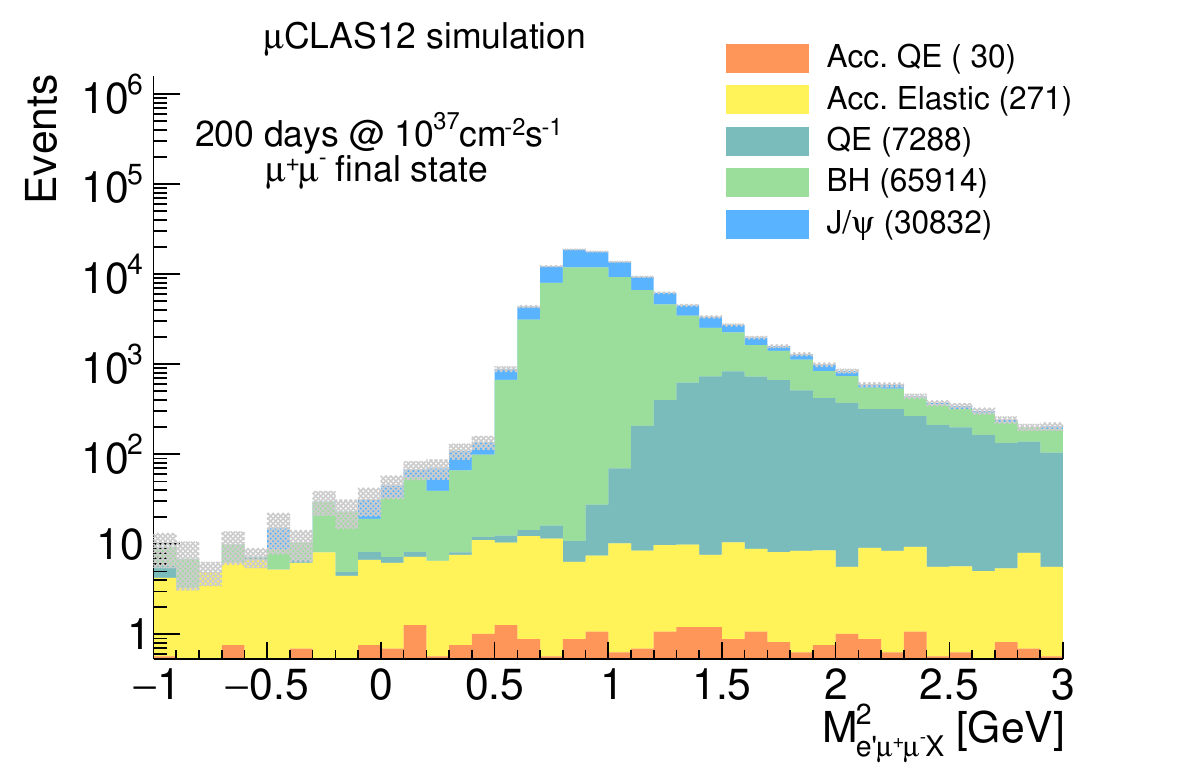}
   \caption{}
 \label{fig:Jpsi_M:2}
\end{subfigure}
\caption{Panel~\ref{fig:Jpsi_M:1}: Invariant mass distribution of reconstructed muon pairs in the $J/\psi$ mass region, with an expected yield of approximately $3\times 10^4$ events. Panel~\ref{fig:Jpsi_M:2}: Missing mass distribution of the undetected proton in the 2.2 to 3.4~GeV invariant mass range. All contributions ($J/\psi$ signal, BH, quasi-elastic BH, and accidental coincidences with elastic and quasi-elastic events) are displayed as a stacked histogram, and their respective number of events are given in parentheses in the legend.}
\label{fig:Jpsi_M}
\end{figure*}

With the planned operating conditions of $\mu$CLAS12, a total of $3\times 10^4$ \jpsi~is expected. This projection assumes an electron detection efficiency in the calorimeter fiducial volume close to 100\%. With a realistic identification efficiency of 90\%, the expected yield is greater than $2.7\times 10^4$, which is 40 times larger than the statistics accumulated by CLAS12. Additionally, the backgrounds under the \jpsi~peak have been estimated. Contributions from elastic and inelastic BH processes, and the accidental coincidence backgrounds were considered. As seen in Fig.~\ref{fig:Jpsi_M:1}, the \jpsi~peak is clearly visible above these backgrounds. Although not used in this projection, an additional leverage to reduce the background is to use the missing mass of the $e'\mu^+\mu^-X$ system that peaks at the proton mass. As shown in Fig.~\ref{fig:Jpsi_M:2}, selecting events with a missing mass consistent with that of the proton would reduce both inelastic and accidental backgrounds. 

\subsubsection{Observables}

\begin{figure}[htbp]
\centering
   \includegraphics[width=\linewidth]{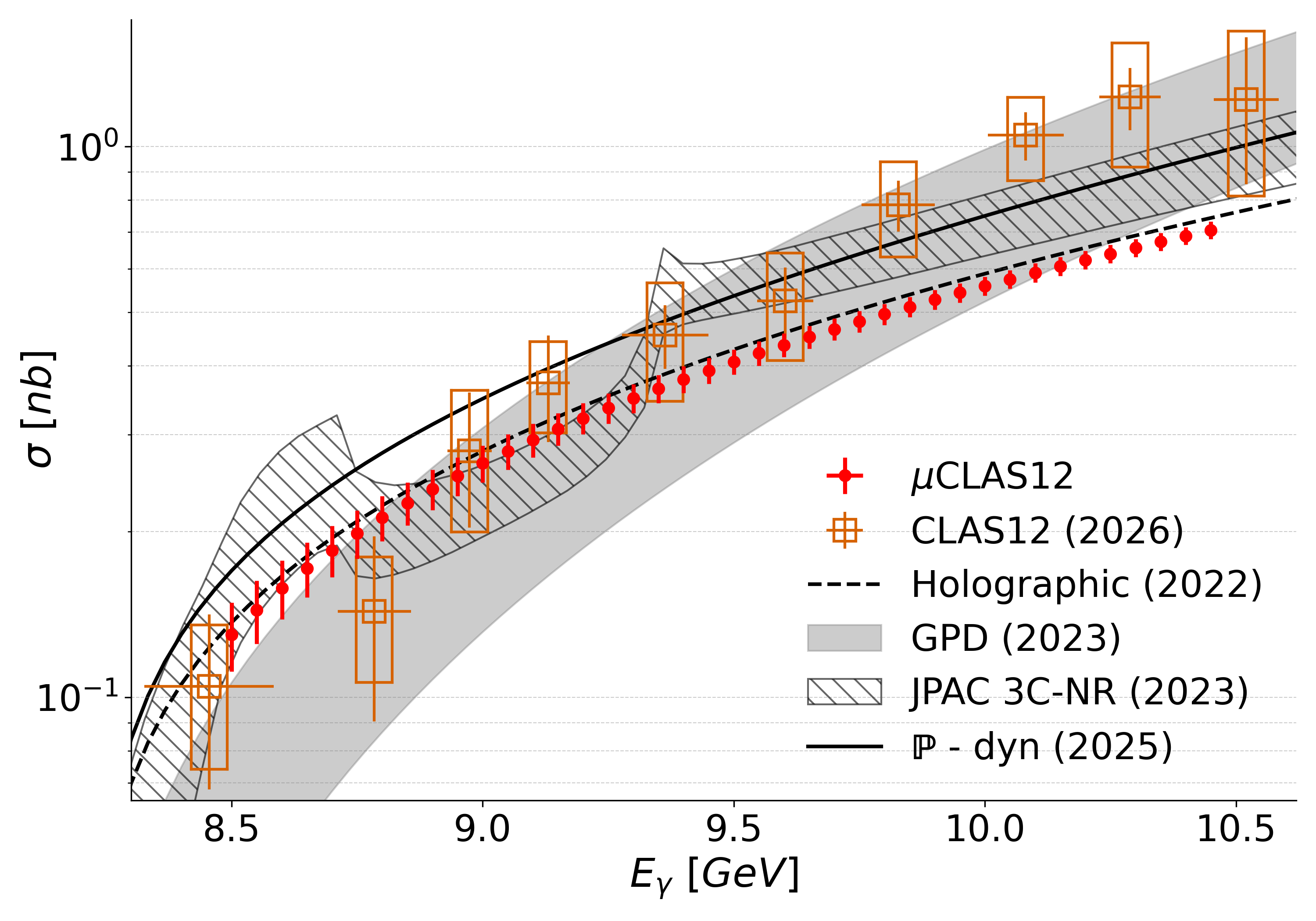}
\caption{Expected statistical uncertainties for the total cross section as a function of initial photon energy $E_\gamma$. The full red points represent the expected statistical uncertainties achievable with $\mu$CLAS12. The holographic model from Ref.~\cite{Mamo:2022eui} was used for the projection, after being adjusted for the virtuality of the incoming photon. The CLAS12 data points are shown in orange. Several models are also displayed: dashed line: Holographic from Ref.~\cite{Mamo:2022eui}, gray band: GPD-based from Ref.~\cite{Guo:2023pqw}, hashed band: JPAC from Ref.~\cite{PhysRevD.108.054018}, solid line: Pomeron-based from Ref.~\cite{Tang:2025qqe}.}
\label{fig:Projection_tot}
\end{figure}

CLAS12 measured the $E_\gamma$-dependence of the $J/\psi$ photoproduction cross section using data taken in 2018 and 2019~\cite{CLAS:2026lls}. During this time, a total of 700~\jpsi~were collected. $\mu$CLAS12 will perform a similar extraction with significantly improved statistical precision, as shown in Fig.~\ref{fig:Projection_tot}. $\mu$CLAS12 will reach the precision necessary to distinguish between the main models developed to describe the near-threshold \jpsi~production. In particular, the measurement will precisely cover the energy range where open charm contributions are expected to dominate, from 8.7 to 9.4 GeV. 

\begin{figure*}[htbp]
\centering
  \centering
   \includegraphics[width=\linewidth]{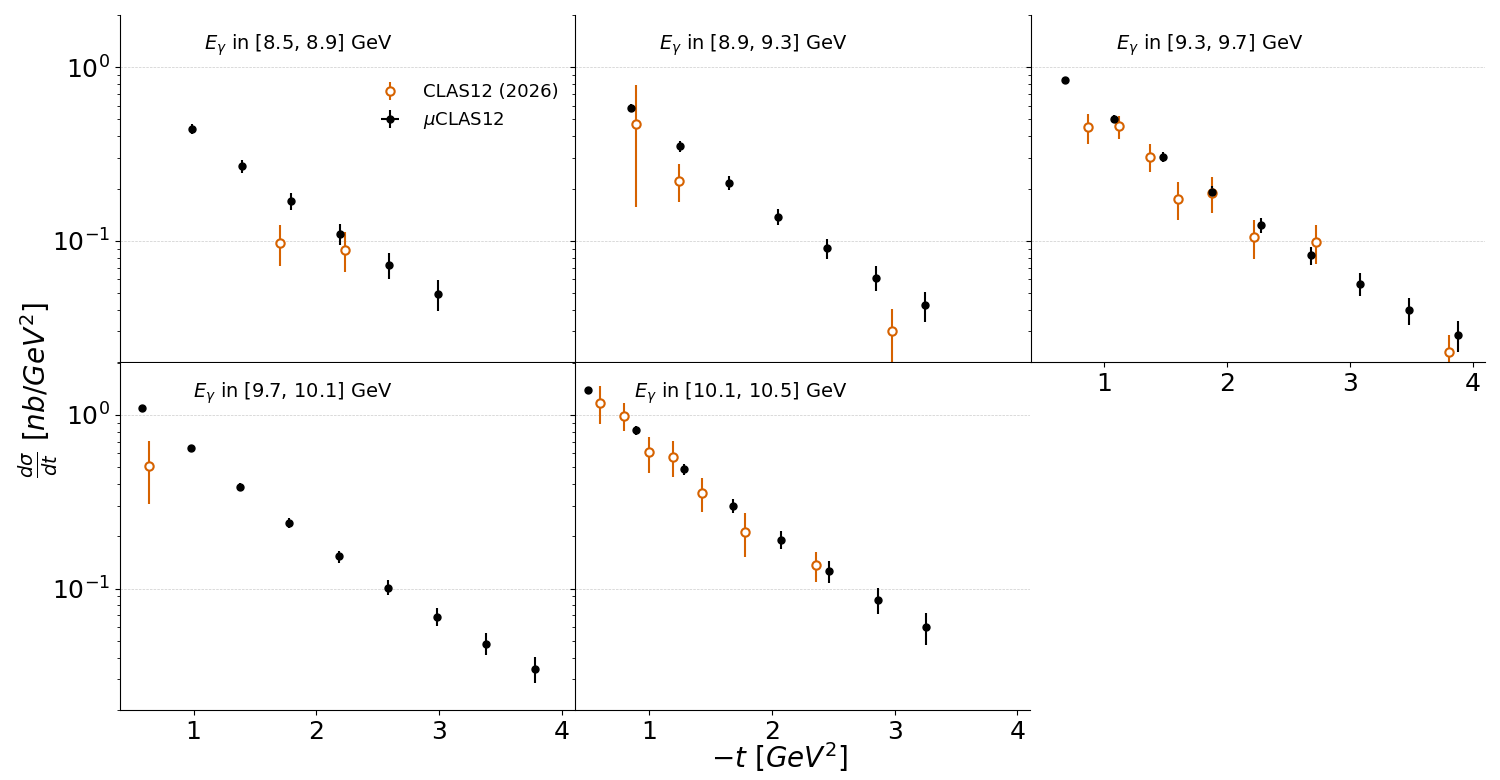}
\caption{Expected statistical uncertainties for the differential cross section as a function of $-t$. The achievable statistical uncertainties are shown by the black points. These projections are compared to the error bars obtained from the analysis of the CLAS12 data in orange.}
\label{fig:Projection_t}
\end{figure*}

\begin{figure}[htbp]
\centering
  \centering
   \includegraphics[width=\linewidth]{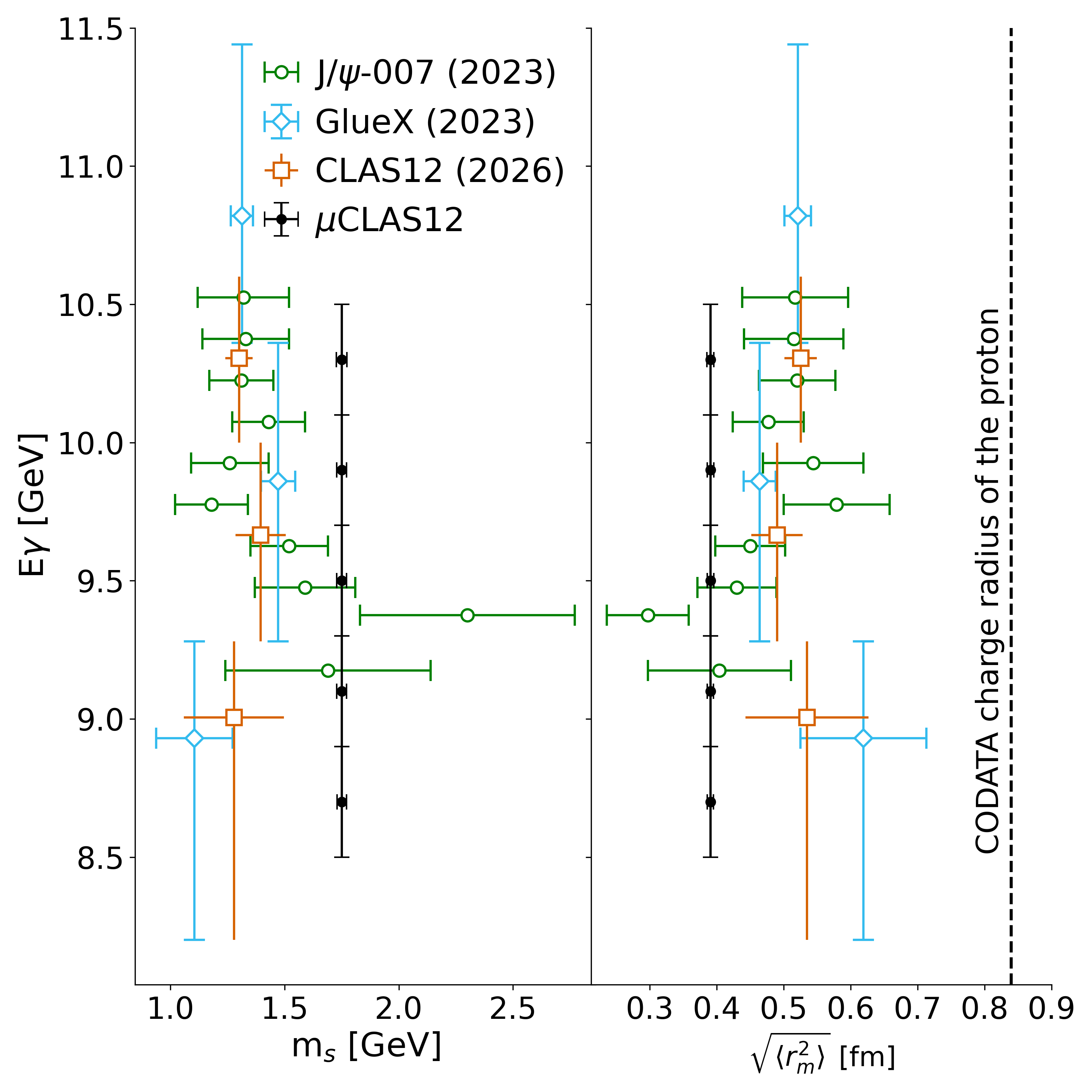}
\caption{Slope parameter of the dipole model $m_S$ (left) and the corresponding gluon mass radius of the proton $\sqrt{\langle r^2_m \rangle}$ (right), as a function of $E_{\gamma}$. The achievable statistical uncertainties are indicated with black points for an arbitrary value of the mass radius. These projections are compared with existing extractions from the $J/\psi-007$, GlueX, and CLAS12 Collaborations.}
\label{fig:Projection_rm}
\end{figure}

The measurement of the $t$-dependence of the cross section is critical to understand the gluon content of the proton. This observable has also been extracted from CLAS12 data in Ref.~\cite{CLAS:2026lls}. $\mu$CLAS12 will perform the same measurement with substantially improved statistical precision and extended $t$-coverage, as highlighted in Fig.~\ref{fig:Projection_t}. The expected uncertainties are compared with those obtained by the CLAS12 experiment, demonstrating the relevance of the measurement. From the $t$- and $E_\gamma$-dependencies of the cross section, it is possible to extract the gluons GFFs, the gluon mass and scalar radius, and the gluon pressure profiles in the proton. Given the error bars shown in Fig.~\ref{fig:Projection_t}, it is expected that the measurements presented in this article will have an important impact on these extractions. To illustrate this statement, the gluon mass radius of the proton was extracted using the dipole model of Ref.~\cite{PhysRevD.104.054015}. In this model, the $t$-dependence of the cross section is fitted with
\begin{equation}
\frac{d\sigma}{dt} = \left. \frac{d\sigma}{dt} \right| _0  \frac{1}{(1-t/m_S^2)^4},
\end{equation}
and the parameter $m_S$ can be related to the gluon mass radius of the proton as \mbox{$\sqrt{\langle r^2_m \rangle}$=${\sqrt{12}}/{m_S}$}. The dipole parameter $m_S$, along with its projected uncertainty, was extracted as a function of the photon energy in Fig.~\ref{fig:Projection_rm}, highlighting that the \jpsi~dataset of $\mu$CLAS12 will probe the gluon content of the proton with the best accuracy to date at JLab.

Finally, the angular distributions of the muons in the rest frame of the $J/\psi$ provide information on the longitudinal to transverse cross section ratio \mbox{$R$=$\sigma_L/\sigma_T$}~\cite{wolfschil}. Taking into account the large $J/\psi$ sample that will be recorded by $\mu$CLAS12, a measurement of this ratio is also expected.  

\subsubsection{Systematic Uncertainties}

The main sources of systematic uncertainty for the measurement of $J/\psi$ cross section will be the uncertainty in determining the accumulated charge, the radiative effects, and the understanding of the detection efficiency of $\mu$CLAS12.

In the inclusive cross section measurement using CLAS12 data \cite{CLAS:2025zup}, the accumulated charge uncertainty is estimated to be less than 1.5\%. For $\mu$CLAS12, the accumulated charge is expected to be measured with a similar precision.

The GlueX Collaboration estimated that radiative corrections introduce an 8.3\% systematic uncertainty to the measured cross section~\cite{gluexjp:2023} . The CLAS12 analysis also estimated this uncertainty to be about 10\%. Since the phase space probed by $\mu$CLAS12 is similar to that of these previous measurements, the systematic uncertainty associated with radiative effects is expected to also be at the level of 10\%.

A dominant systematic uncertainty in this cross section measurement is expected to arise from matching the simulated to the real detector efficiency. The CLAS12 $J/\psi$ analysis estimated this uncertainty to be about 10\%. Considering the ongoing efforts of the CLAS Collaboration to publish cross sections for multi-particle final states with CLAS12, this systematic is expected to decrease as the understanding of the detector performance improves.

Since the final states for DDVCS and \jpsi~are identical, the detector efficiency and resolution for exclusive $J/\psi$ production are very similar to those of DDVCS events in the range of invariant mass of interest. The narrow $J/\psi$ peak will enable an easier identification of the reaction and a more reliable yield extraction than the DDVCS-BH continuum. Hence, the $J/\psi$ production reaction will serve as an important benchmark for the DDVCS measurements. The $\phi(1020)$ could, in principle, also be used in a similar way at the lower end of the invariant mass range. Therefore, the measurement of the $J/\psi$ cross section will help in understanding the DDVCS data and in better constraining systematic uncertainties, such as acceptance and muon identification.

\subsubsection{Search for Pentaquarks}
\label{sec:pentaquarks}

In 2015, the LHCb Collaboration published the discovery of three exotic structures in the decay channel $J/\psi~p$, which have been referred to as charmonium-pentaquark states \cite{Aaij:2015tga}. The minimum quark content of these states is $c\bar c u u d$, and these states were labeled as $P_c(4312)$, $P_c(4440)$, and $P_c(4457)$. Since these states were observed in the $J/\psi~p$ final state, the photoproduction process $\gamma^* p\to P_c\to J/\psi~p'$, where these states would appear as resonances in the $s$-channel for photon energy around $10$~GeV, is expected~\cite{Kubarovsky:2015aaa, PhysRevD.100.034019, Wang:2015jsa, Karliner:2015voa}, but has not yet been observed.

\begin{figure}[htbp]
\centering
  \centering
\includegraphics[width=\linewidth]{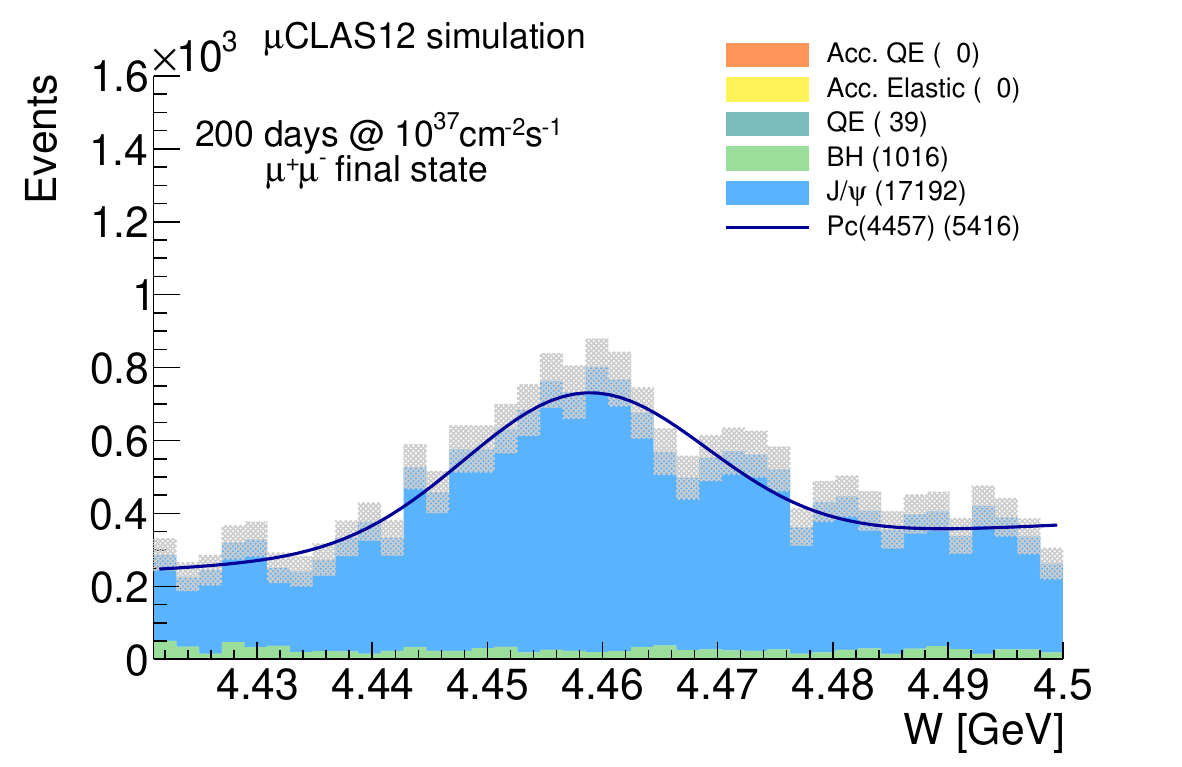}
\caption{Hadronic mass $W$ in the $P_c(4457)$ region. The invariant mass of the muon pair is restricted in the 2.7 to 3.2~GeV range. With the  planned integrated luminosity, an 11 GeV beam, and a 2\% branching ratio in the model of Ref.~\cite{PhysRevD.100.034019}, approximately 4500 $P_c(4457)$ are expected to be produced.}
\label{fig:Pc}
\end{figure}

A direct search of these pentaquarks will be possible with $\mu$CLAS12, using the $W$ spectrum of \jpsi~events, as the pentaquarks would manifest themselves as peaks at their respective mass. To assess whether $\mu$CLAS12 will produce enough statistics and that the electron momentum resolution will allow for pentaquarks peaks to be distinguished, an extensive simulation with the \textit{elSpectro} event generator was performed. The model presented in Ref.~\cite{PhysRevD.100.034019} was used, with a 2\% branching ratio. In Fig.~\ref{fig:Pc}, the obtained $W$ spectrum is shown for events where a \jpsi~has been identified in $\mu$CLAS12. The $P_c(4457)$ is visible, as the electron momentum resolution in the wECal is sufficient not to smear the peak. With the 2\% branching ratio used for this simulation, it is estimated that about 4500 $P_c(4457)$ will be produced over the 200 days of the experiment. Hence, either a direct measurement of the production cross section or an upper limit to it is planned.

\subsection{Timelike Compton Scattering}

The TCS reaction will be measured in the quasi-real photoproduction regime, where the beam electron radiates a quasi-real photon. The reaction will be identified by requiring a pair of muons in the FD and a proton in the recoil detector. The kinematics of the undetected scattered electron can be calculated from the momenta of the detected particle. To select the quasi-real events, the missing mass of the undetected scattered electron and the virtuality of the initial photon can be constrained to be small. This analysis strategy was used in the first ever TCS measurement using CLAS12 data in Ref.~\cite{clas12tcs}.

\begin{figure*}[htbp]
\centering
  \begin{subfigure}{0.49\linewidth}
\centering  
   \includegraphics[width=\textwidth]{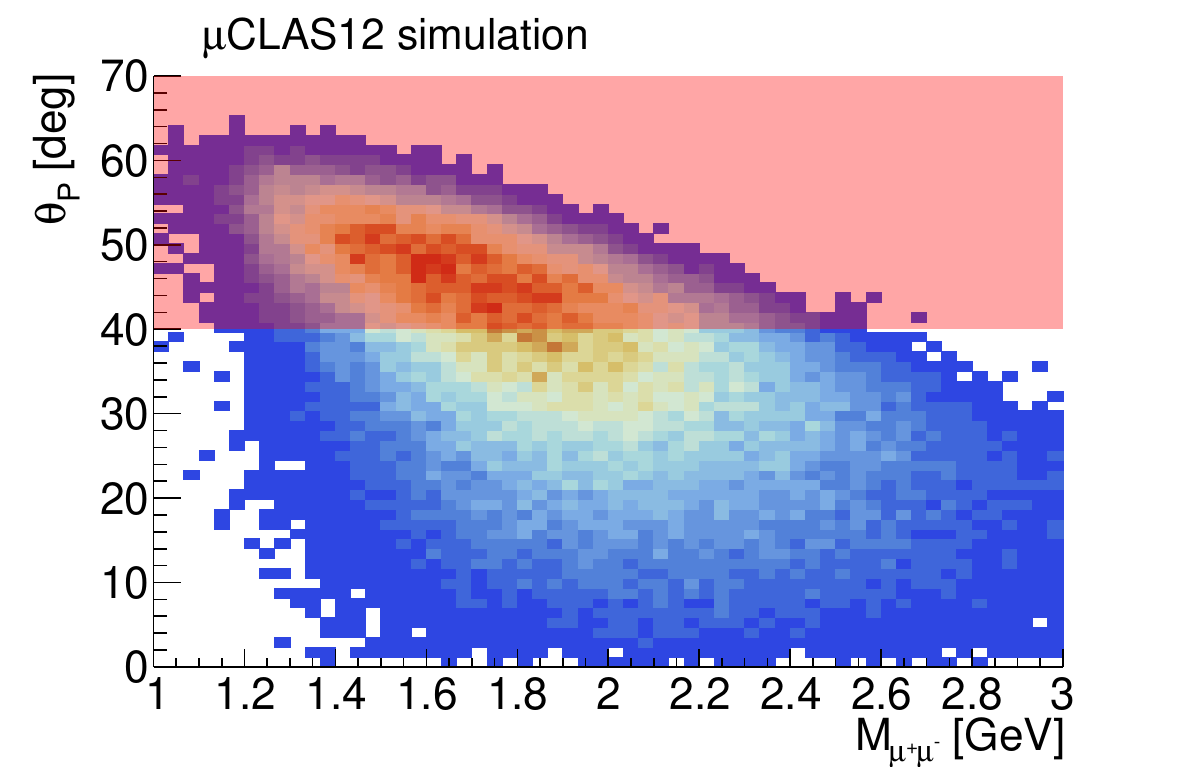}
\caption{}
 \label{fig:TCS_PS:1}
\end{subfigure}
   \begin{subfigure}{0.49\linewidth}
\centering  
   \includegraphics[width=\textwidth]{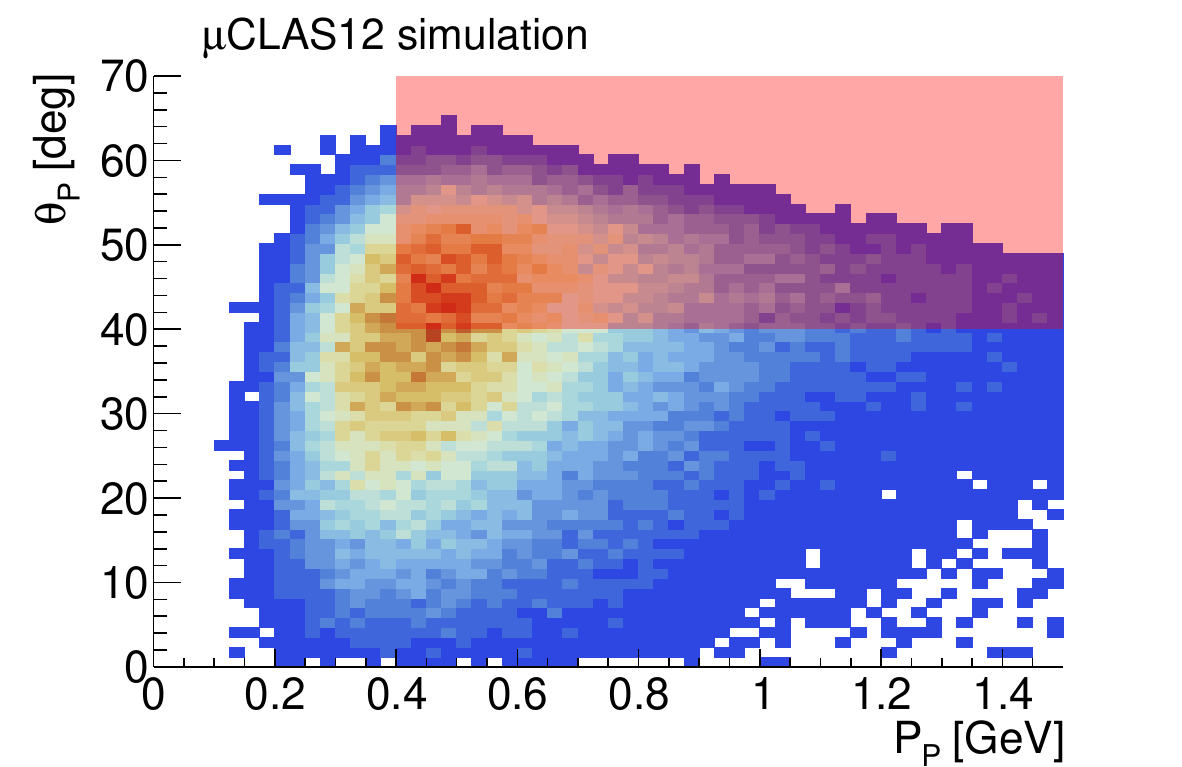}
   \caption{}
 \label{fig:TCS_PS:2}
\end{subfigure}
\caption{Panel~\ref{fig:TCS_PS:1}: Proton polar angle as a function of the invariant mass of the muon pair. Panel~\ref{fig:TCS_PS:2}: Polar angle as a function of momentum for the proton. Events displayed were required to have a muon pair detected in $\mu$CLAS12. The red rectangles show the acceptance limit of the recoil detector. Events within this region were used to estimate the measurement yield.}
\label{fig:TCS_PS}
\end{figure*}

To estimate the kinematic coverage and the rate of TCS events in $\mu$CLAS12, a sample of $10\times 10^6$ BH events was processed through $\mu$GEMC. Events with two identified muons in the FD were selected, and the kinematics of the generated recoil proton was restricted to the active area of the RD. Figure~\ref{fig:TCS_PS:1} shows the polar angle of the proton as a function of the invariant mass of the muon pair. In the case of the TCS measurement, an invariant mass above 1.5~GeV was selected to ensure that the GPD formalism applies. In the current CLAS12 configuration, the CVT can detect protons with momenta larger than 0.35~GeV/$c$. Considering that the RT will have a similar geometry, only protons with a minimum momentum of 0.35~GeV/$c$ were selected. Figure~\ref{fig:TCS_PS:2} shows the generated proton angles and momenta, and the phase space covered by the recoil detector. The total accumulated statistics for 200~days with a luminosity of $10^{37}\,\mathrm{cm^{-2}\,s^{-1}}$ was estimated to be $7.7\times 10^6$ events. Thus, this measurement will have a three-order-of-magnitude increase in statistics with respect to the first CLAS12 TCS publication.

\begin{figure}[htbp]
\centering
  \centering
   \includegraphics[width=\linewidth]{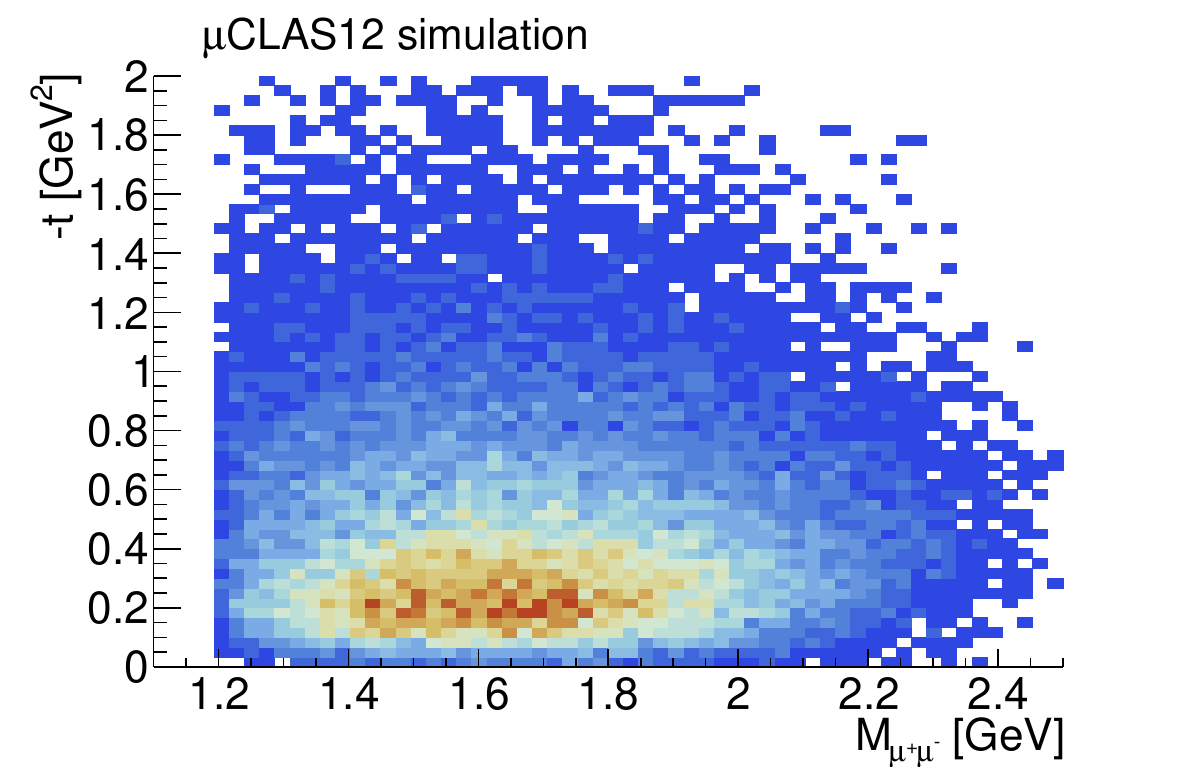}
\caption{Mandelstam $t$ as a function of the invariant mass of the muon pair, for proton within the acceptance of the recoil detector.}
\label{fig:TCS_tM}
\end{figure}

Figure~\ref{fig:TCS_tM} shows $t$ as a function of the invariant mass of the muon pair for events with a proton in the acceptance limit of the recoil tracker. With $\mu$CLAS12, it will be possible to access a wide range of invariant masses, up to 2.3 GeV, with a large coverage of $t$, especially in the region below 0.4 GeV$^2$, where measurements will be most relevant for the extraction of GPDs.

\subsubsection{Observables}

The photon beam-polarization asymmetry ($A_{\odot U}$) and the forward-backward asymmetry ($A_{FB}$) of TCS will be measured by $\mu$CLAS12, in a wide range of $E_\gamma$, $t$, and $Q^{\prime 2}$, and with a very large collected statistics sample compared to the published CLAS12 results.

\begin{figure*}[htbp]
\begin{center} 
\includegraphics[width=\linewidth]{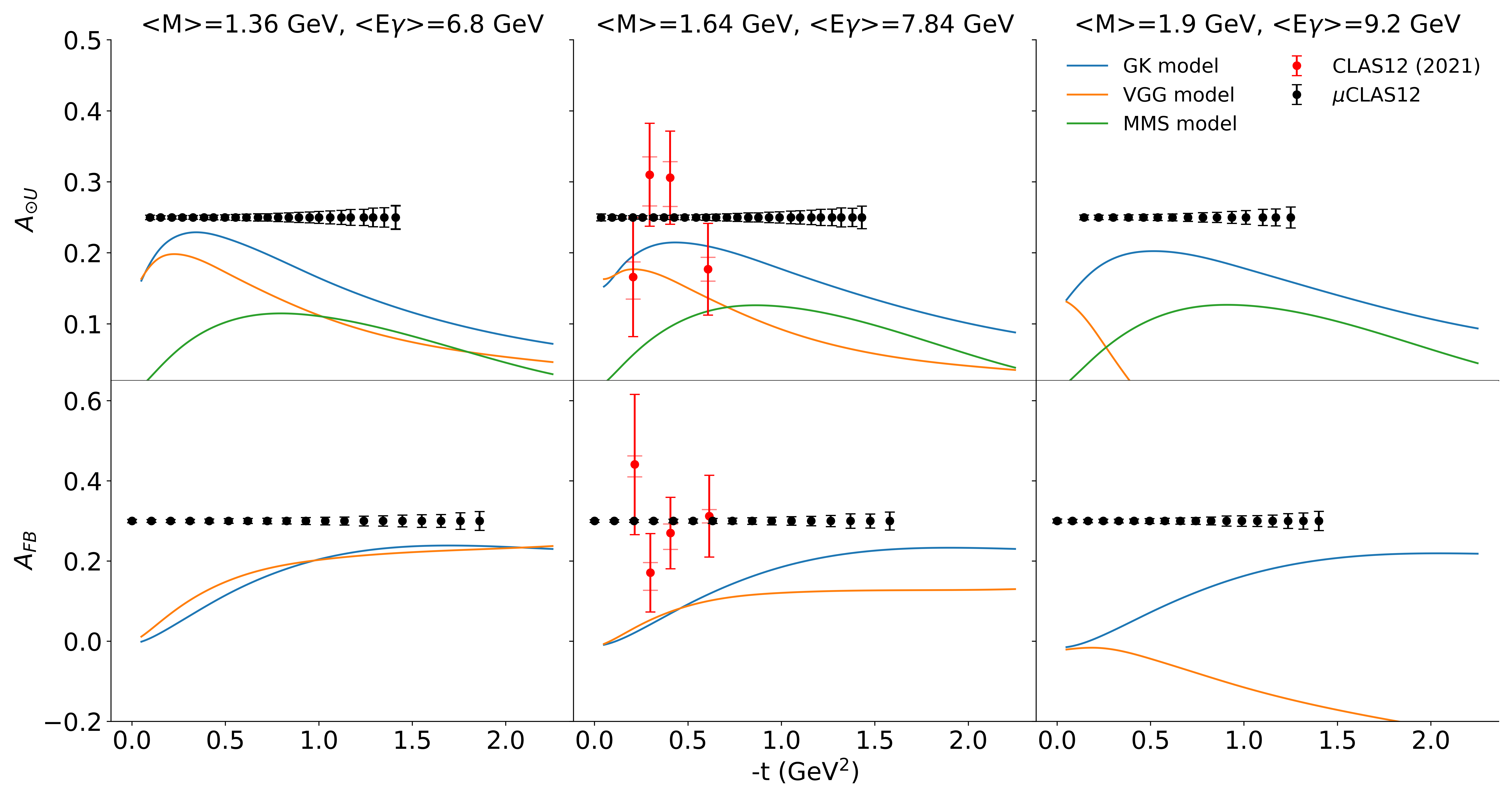}
\caption{Expected statistical error bars of the TCS $A_{\odot U}$ (top) and $A_{FB}$ (bottom), for three invariant mass bins (left: $[1.2, 1.5]$ GeV; middle: $[1.5, 1.8]$ GeV; right: $[1.8, 2.4]$ GeV) as a function of $t$. Model predictions based on the GK19, VGG, and MMS~\cite{PhysRevD.88.014001} models implemented in PARTONS are displayed. For comparison, the published CLAS12 measurements of the two observables \cite{clas12tcs} are shown in red in the bin nearest to their average kinematic value.}
\label{fig:TCS_Proj}
\end{center}
\end{figure*}

In the case of $A_{\odot U}$, the experimental asymmetry is defined as
\begin{equation}
    A_{\odot U} = \frac{1}{P_b}\frac{N^+-N^-}{N^++N^-},
\end{equation}
where ${P_b}$ is the average polarization of the beam, and $N^+$ and $N^-$ are the number of events with a right- and left-handed circular polarization of the initial real photon, respectively. The polarization of the real photon will be estimated using the polarization of the initial beam electron and the well-known polarization transfer calculable in QED from Ref.~\cite{PhotonPolarization}. The projected statistical error bars for $A_{\odot U}$ are shown in Fig. \ref{fig:TCS_Proj}, in three invariant mass bins and as a function of $t$. These projections were obtained by fitting 1000 independent pseudo-data distributions in $\phi$, with 50  bins, and where each $\phi$-asymmetry was randomly distributed within its expected uncertainty. As the phase space covered by $\mu$CLAS12 will overlap that covered by CLAS12, it will not only be possible to cross-check the planned results with those published by CLAS12, but also greatly extend the $t$ coverage and the precision of the published CLAS12 results. 

\begin{figure}[htbp]
\centering
  \centering
   \includegraphics[width=1\linewidth]{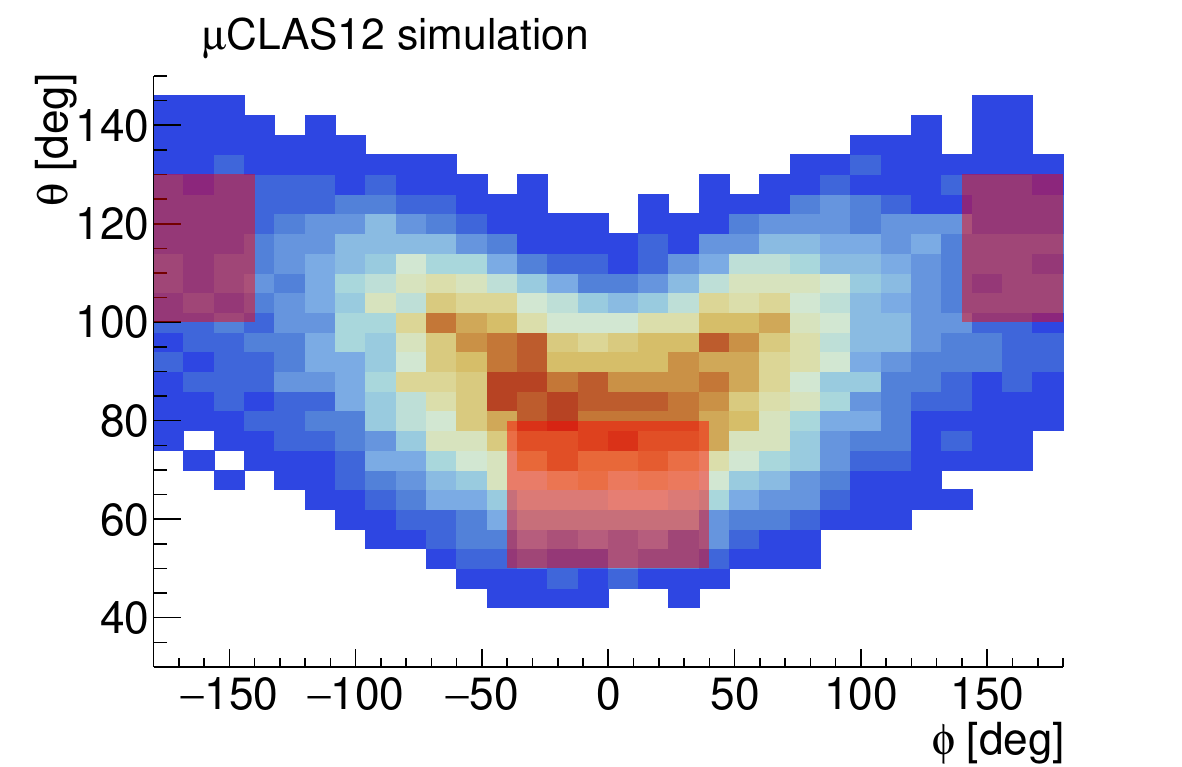}
\caption{Angular coverage of $\mu$CLAS12 for TCS events. The red rectangles highlight the angular bin where the forward-backward asymmetry can be compared to the published CLAS12 results.}
\label{fig:TCS_CM}
\end{figure}

The TCS $A_{FB}$, which allows direct access to the real part of the CFF~$\mathcal{H}$, is given by
\begin{equation}
A_{FB}=\frac{N_F-N_B}{N_F+N_B},
\end{equation}
where $N_F$ and $N_B$ are the acceptance-corrected number of events in the forward and backward bins, respectively. The center-of-mass angular range of these two bins is related as $\phi_B$=$180^\circ + \phi_F$ and $\theta_B$=$180^\circ-\theta_F$. The angular acceptance of $\mu$CLAS12 is very similar to that of CLAS12. Hence, it will be possible to measure this asymmetry in the same bin as CLAS12 (\mbox{$-40^\circ<\phi_F<40^\circ$}, \mbox{$50^\circ<\theta_F<80^\circ$}), with a much improved precision. The TCS center-of-mass angular coverage of $\mu$CLAS12 is shown in Fig. \ref{fig:TCS_CM}, where the angular bin limits described above are shown to be well within acceptance. The projected statistical error bars for $A_{FB}$ are shown in Fig. \ref{fig:TCS_Proj} as a function of $t$ for three invariant mass bins. The projections are compared with the statistical uncertainties published by CLAS12, highlighting the precision of the $\mu$CLAS12 measurement.

\subsubsection{Systematic Uncertainties}

The main sources of systematic uncertainties for the measurement of the TCS photon polarization asymmetry will arise from the determination of the initial beam polarization, the radiative effects, and the understanding of the background in the event sample; as for the DDVCS measurement (see Section~\ref{sec:syst_DDVCS} for details).

In the case of the forward-backward asymmetry, the understanding of the acceptance and efficiency of the detector will dominate the systematic uncertainties. As mentioned for the $J/\psi$ measurement, it is expected to be less than 10\% for cross section measurements.

%

\section{Summary}

In this article, the studies of DDVCS, TCS, and \jpsi~electroproduction on the proton using an  $11$~GeV electron beam and the $\mu$CLAS12 detector in Experimental Hall B at JLab are presented. The planned modifications to the CLAS12 detector serve two primary purposes. First, they will enable the CLAS12 FD to operate at luminosities two orders of magnitude higher than its design luminosity. Second, they will effectively convert the CLAS12 FD into a high-efficiency muon detector. In this configuration, scattered electrons will be detected and identified using a new, compact, PbWO$_4$ electromagnetic calorimeter. A new vertex tracking system and a compact central detector will be incorporated, to better reconstruct forward-going tracks and to measure recoil protons.

To achieve the objectives of $\mu$CLAS12, 200 days of production running at a luminosity of $10^{37}\,\mathrm{cm^{-2}\,s^{-1}}$, using an 11 GeV, 7.5 $\mu$A longitudinal polarized ($>$85\%) electron beam incident on a 5-cm-long LH$_2$ target, were awarded by the JLab Program Advisory Committee in July 2025. In addition, 15 days will be dedicated to the commissioning of $\mu$CLAS12, and 30 days will be devoted to empty-target and low-luminosity calibration runs. The awarded beam time is summarized in Table~\ref{tab:BeamTimeRequest}, which also provides a breakdown of the planned running periods and the main parameters of the beam and target.

\begin{table*}[hbtp]
    \centering
\begin{tabularx}{\textwidth}{>{\centering\arraybackslash}X >{\centering\arraybackslash}X >{\centering\arraybackslash}X >{\centering\arraybackslash}X >{\centering\arraybackslash}X >{\centering\arraybackslash}X }
\hline\hline
Beam   & Beam    & Beam         & Target   & Target   & Beam time   \\
Energy & Current & Polarization & Material & Length  & (days) \\
(GeV)  & ($\mu$A)    &              &          & (cm) &        \\
\hline

\multicolumn{6}{c}{Commissioning}      \\
11 &  & & &   5  &    15    \\\hline
\multicolumn{6}{c}{Calibration}       \\
11 & 7.5 & & Empty target &   5  &    10    \\
11 & $<$1 & & LH2 &   5  &    20    \\\hline
\multicolumn{6}{c}{Production}       \\
11    &   7.5   & $>85$\%  & LH2   & 5        & 200    \\
   
\hline
\multicolumn{5}{c}{Total time}            & 245 \\
\hline\hline
\end{tabularx}
\caption{\centering  Summary of the beam time awarded to the $\mu$CLAS12 experiment.}
    \label{tab:BeamTimeRequest}
\end{table*}

With the $\mu$CLAS12 experiment, the beam-spin asymmetry of DDVCS will be measured for both spacelike and timelike virtualities of the incoming and outgoing virtual photons. These measurements will uniquely map the full kinematic dependencies of GPDs and provide unparalleled insight into GPDs, the 3D-structure of the proton, and its mechanical properties.

The di-muon final state will also be used to measure the cross section for near-threshold $J/\psi$ production with unmatched statistical precision. This will make it possible to establish the production mechanism in this kinematic regime with high accuracy, providing particularly strong sensitivity to potential open-charm and pentaquark contributions. Ultimately, the $J/\psi$ dataset collected by $\mu$CLAS12 will be instrumental in understanding the gluon content of the proton.

Finally, $\mu$CLAS12 will produce substantial new TCS data, greatly extending the ongoing TCS program at JLab. This will allow for the extraction of observables such as photon polarization and forward–backward asymmetries in significantly finer kinematic bins than currently feasible, thereby contributing to a deeper understanding of the GPDs, and of the internal structure of the proton.

\section*{Acknowledgments}
This work was supported in part by the U.S. Department of Energy, the Italian Istituto Nazionale di Fisica Nucleare (INFN), the French Centre National de la
Recherche Scientifique (CNRS), the French Commissariat {\`a} l’Energie Atomique (CEA), the National Science Centre, Poland (grant
OPUS-27 No.~2024/53/B/ST2/00968).
This material is based on work supported by the U.S. Department of Energy, Office of Science, Office of Nuclear Physics under Contract No. DE-AC05-06OR23177.

\bibliographystyle{epj}
\bibliography{References}

\begin{thebibliography}{168}

\bibitem{mueller1994wave}
D.~M{\"u}ller et~al., Fortsch. Phys. \textbf{42}, 101 (1994)

\bibitem{Ji:1996nm}
X.D. Ji, Phys. Rev. D \textbf{55}, 7114 (1997)

\bibitem{Radyushkin:1997}
A.V. Radyushkin, Phys. Rev. D \textbf{56}, 5524 (1997)

\bibitem{collins:1997}
J.C. Collins et~al., Phys. Rev. D \textbf{56}, 2982 (1997)

\bibitem{ARRINGTON2022103985}
J.~Arrington et~al., Prog. Part. Nucl. Phys. \textbf{127}, 103985 (2022)

\bibitem{Radyushkin:1996ru}
A.V. Radyushkin, Phys. Lett. B \textbf{385}, 333 (1996)

\bibitem{clasdvcs1}
S.~Stepanyan et~al. (CLAS Collaboration), Phys. Rev. Lett. \textbf{87}, 182002 (2001)

\bibitem{carlos}
C.~Mu{\~n}oz~Camacho et~al. (Hall A Collaboration), Phys. Rev. Lett. \textbf{97}, 262002 (2006)

\bibitem{fx}
F.X. Girod et~al. (CLAS Collaboration), Phys. Rev. Lett. \textbf{100}, 162002 (2008)

\bibitem{Maxime}
M.~Defurne et~al. (Hall A Collaboration), Phys. Rev. C \textbf{92}, 055202 (2015)

\bibitem{Jo:2015ema}
H.S. Jo et~al. (CLAS Collaboration), Phys. Rev. Lett. \textbf{115}, 212003 (2015)

\bibitem{erin}
E.~Seder et~al. (CLAS Collaboration), Phys. Rev. Lett. \textbf{114}, 032001 (2015)

\bibitem{Pisano:2015iqa}
S.~Pisano et~al. (CLAS Collaboration), Phys. Rev. D \textbf{91}, 052014 (2015)

\bibitem{halladvcs:1}
F.~Georges et~al. (Hall A Collaboration), Phys. Rev. Lett. \textbf{128}, 252002 (2022)

\bibitem{clas12dvcs:1}
G.~Christiaens et~al. (CLAS Collaboration), Phys. Rev. Lett. \textbf{130}, 211902 (2023)

\bibitem{CLAS:2024qhy}
A.~Hobart et~al. (CLAS Collaboration), Phys. Rev. Lett. \textbf{133}, 211903 (2024)

\bibitem{H1:2001nez}
C.~Adloff et~al. (H1 Collaboration), Phys. Lett. B \textbf{517}, 47 (2001)

\bibitem{H1:2005gdw}
A.~Aktas et~al. (H1 Collaboration), Eur. Phys. J. C \textbf{44}, 1 (2005)

\bibitem{ZEUS:2003pwh}
S.~Chekanov et~al. (ZEUS Collaboration), Phys. Lett. B \textbf{573}, 46 (2003)

\bibitem{HERMES:2001bob}
A.~Airapetian et~al. (HERMES Collaboration), Phys. Rev. Lett. \textbf{87}, 182001 (2001)

\bibitem{HERMES:2006pre}
A.~Airapetian et~al. (HERMES Collaboration), Phys. Rev. D \textbf{75}, 011103 (2007)

\bibitem{HERMES:2008abz}
A.~Airapetian et~al. (HERMES Collaboration), JHEP \textbf{06}, 066 (2008)

\bibitem{Joerg:2016hhs}
P.~Joerg (COMPASS Collaboration), PoS \textbf{DIS2016}, 235 (2016)

\bibitem{COMPASS:2018pup}
R.~Akhunzyanov et~al. (COMPASS Collaboration), Phys. Lett. B \textbf{793}, 188 (2019), [Erratum: Phys.Lett.B 800, 135129 (2020)]

\bibitem{Berger:2001xd}
E.R. Berger et~al., Eur. Phys. J. C \textbf{23}, 675 (2002)

\bibitem{Goritschnig}
A.T. Goritschnig et~al., Phys. Rev. D \textbf{89}, 094031 (2014)

\bibitem{Boer:2015hma}
M.~Bo{\"e}r et~al., Eur. Phys. J. A \textbf{51}, 103 (2015)

\bibitem{Boer:2015gv}
M.~Bo{\"e}r et~al., Eur. Phys. J. A \textbf{52}, 33 (2016)

\bibitem{clas12tcs}
P.~Chatagnon et~al. (CLAS Collaboration), Phys. Rev. Lett. \textbf{127}, 262501 (2021)

\bibitem{herve}
H.~Moutarde, Phys. Rev. D \textbf{79}, 094021 (2009)

\bibitem{partons}
B.~Berthou et~al., Eur. Phys. J. C \textbf{78}, 478 (2018)

\bibitem{bertone}
V.~Bertone et~al., Phys. Rev. D \textbf{103}, 114019 (2021)

\bibitem{moffat}
E.~Moffat et~al., Phys. Rev. D \textbf{108}, 036027 (2023)

\bibitem{Guidal:2002kt}
M.~Guidal et~al., Phys. Rev. Lett. \textbf{90}, 012001 (2003)

\bibitem{ddvcs_bm1}
A.V. Belitsky et~al., Phys. Rev. Lett. \textbf{90}, 022001 (2003)

\bibitem{ddvcs_bm2}
A.V. Belitsky et~al., Phys. Rev. D \textbf{68}, 116005 (2003)

\bibitem{deja:2023}
K.~Deja et~al., Phys. Rev. D \textbf{107}, 094035 (2023)

\bibitem{Martinez-Fernandez:2025gub}
V.~Martinez-Fernandez et~al., Phys. Rev. D \textbf{111}, 074034 (2025)

\bibitem{clas12}
V.D. Burkert et~al. (CLAS Collaboration), Nucl. Instrum. Meth. A \textbf{959}, 163419 (2020)

\bibitem{Sharabian:2020whm}
Y.G. Sharabian et~al., Nucl. Instrum. Meth. A \textbf{968}, 163824 (2020)

\bibitem{Kobzarev:1962wt}
I.Y. Kobzarev et~al., Zh. Eksp. Teor. Fiz. \textbf{43}, 1904 (1962)

\bibitem{Pagels:1966zza}
H.~Pagels, Phys. Rev. \textbf{144}, 1250 (1966)

\bibitem{Polyakov:2018zvc}
M.V. Polyakov et~al., Int. J. Mod. Phys. A \textbf{33}, 1830025 (2018)

\bibitem{Burkert2018}
V.D. Burkert et~al., Nature \textbf{557}, 396 (2018)

\bibitem{Lorce:2021xku}
C.~Lorc\'e et~al., JHEP \textbf{11}, 121 (2021)

\bibitem{Burkert2023}
V.D. Burkert et~al., Rev. Mod. Phys. \textbf{95}, 041002 (2023)

\bibitem{Burkardt:2000za}
M.~Burkardt, Phys. Rev. D \textbf{62}, 071503 (2000)

\bibitem{Belitsky:2003nz}
A.V. Belitsky et~al., Phys. Rev. D \textbf{69}, 074014 (2004)

\bibitem{revdiehl}
M.~Diehl, Phys. Rept. \textbf{388}, 41 (2003)

\bibitem{Ji:1996ek}
X.D. Ji, Phys. Rev. Lett. \textbf{78}, 610 (1997)

\bibitem{Pire:2011st}
B.~Pire et~al., Phys. Rev. D \textbf{83}, 034009 (2011)

\bibitem{Braun:2025xlp}
V.M. Braun et~al., Phys. Rev. D \textbf{111}, 076011 (2025)

\bibitem{Martinez-Fernandez:2025jvk}
V.~Mart{\'\i}nez-Fern{\'a}ndez et~al. (2025), \texttt{2509.06669}

\bibitem{Martinez-Fernandez:2025rcg}
V.~Mart{\'\i}nez-Fern{\'a}ndez et~al. (2025), \texttt{2509.05059}

\bibitem{PhysRevD.70.117504}
A.~Bacchetta et~al., Phys. Rev. D \textbf{70}, 117504 (2004)

\bibitem{Polyakov:2002yz}
M.V. Polyakov, Phys. Lett. B \textbf{555}, 57 (2003)

\bibitem{Pasquini:2014vua}
B.~Pasquini et~al., Phys. Lett. B \textbf{739}, 133 (2014)

\bibitem{Grocholski:2019pqj}
O.~Grocholski et~al., Eur. Phys. J. C \textbf{80}, 171 (2020)

\bibitem{Mueller:2012sma}
D.~Mueller et~al., Phys. Rev. D \textbf{86}, 031502 (2012)

\bibitem{fitmick}
M.~Guidal, Eur. Phys. J. A \textbf{37}, 319 (2008)

\bibitem{fithermes}
M.~Guidal et~al., Eur. Phys. J. A \textbf{42}, 71 (2009)

\bibitem{fittsa}
M.~Guidal, Phys. Lett. B \textbf{689}, 156 (2010)

\bibitem{fitall}
M.~Guidal, Phys. Lett. B \textbf{693}, 17 (2010)

\bibitem{kum2014}
K.~Kumeri{\v{c}}ki et~al., Phys. Part. Nucl. \textbf{45}, 723 (2014)

\bibitem{jifit}
K.~Shiells et~al., JHEP \textbf{08}, 048 (2022)

\bibitem{mswbsfit}
H.~Moutarde et~al., Eur. Phys. J. C \textbf{78}, 890 (2018)

\bibitem{kum2008}
K.~Kumericki et~al., Nucl. Phys. B \textbf{794}, 244 (2008)

\bibitem{fitmuller}
K.~Kumeri{\v{c}}ki et~al., Nucl. Phys. B \textbf{841}, 1 (2010)

\bibitem{kum2011}
K.~Kumericki et~al., JHEP \textbf{07}, 073 (2011)

\bibitem{mswfit}
H.~Moutarde et~al., Eur. Phys. J. C \textbf{79}, 614 (2019)

\bibitem{mksfit}
M.~{\v{C}}ui{\'c} et~al., Phys. Rev. Lett. \textbf{125}, 232005 (2020)

\bibitem{PhysRevD.74.054027}
V.~Guzey et~al., Phys. Rev. D \textbf{74}, 054027 (2006)

\bibitem{lattice}
M.~Constantinou, Eur. Phys. J. A \textbf{57}, 77 (2021)

\bibitem{BELITSKY2002323}
A.V. Belitsky et~al., Nucl. Phys. B \textbf{629}, 323 (2002)

\bibitem{zhao:2021}
S.~Zhao et~al., Eur. Phys. J. A \textbf{57}, 240 (2021)

\bibitem{alvarado2025}
J.S. Alvarado et~al., Phys. Rev. C \textbf{111}, 065205 (2025)

\bibitem{Accardi:2020swt}
A.~Accardi et~al., Eur. Phys. J. A \textbf{57}, 261 (2021)

\bibitem{Accardi:2023chb}
A.~Accardi et~al., Eur. Phys. J. A \textbf{60}, 173 (2024)

\bibitem{AbdulKhalek:2021gbh}
R.~Abdul~Khalek et~al., Nucl. Phys. A \textbf{1026}, 122447 (2022)

\bibitem{JeffersonLabSoLID:2022iod}
J.~Arrington et~al. (SoLID Collaboration), J. Phys. G \textbf{50}, 110501 (2023)

\bibitem{KHARZEEV1999568}
D.~Kharzeev et~al., Nucl. Phys. A \textbf{661}, 568 (1999)

\bibitem{PhysRevD.100.014032}
Y.~Hatta et~al., Phys. Rev. D \textbf{100}, 014032 (2019)

\bibitem{PhysRevD.103.096010}
Y.~Guo et~al., Phys. Rev. D \textbf{103}, 096010 (2021)

\bibitem{PhysRevD.104.054015}
D.E. Kharzeev, Phys. Rev. D \textbf{104}, 054015 (2021)

\bibitem{Mamo:2022eui}
K.A. Mamo et~al., Phys. Rev. D \textbf{106}, 086004 (2022)

\bibitem{Guo:2023pqw}
Y.~Guo et~al., Phys. Rev. D \textbf{108}, 034003 (2023)

\bibitem{Shanahan:2018pib}
P.E. Shanahan et~al., Phys. Rev. D \textbf{99}, 014511 (2019)

\bibitem{Pefkou:2021fni}
D.A. Pefkou et~al., Phys. Rev. D \textbf{105}, 054509 (2022)

\bibitem{Hackett:2023rif}
D.C. Hackett et~al., Phys. Rev. Lett. \textbf{132}, 251904 (2024)

\bibitem{PhysRevLett.35.1616}
B.~Gittelman et~al., Phys. Rev. Lett. \textbf{35}, 1616 (1975)

\bibitem{PhysRevLett.35.483}
U.~Camerini et~al., Phys. Rev. Lett. \textbf{35}, 483 (1975)

\bibitem{Zeus:2002fa}
S.~Chekanov et~al. (ZEUS Collaboration), Eur. Phys. J. C \textbf{24}, 345 (2002)

\bibitem{H1:2005dtp}
A.~Aktas et~al. (H1 Collaboration), Eur. Phys. J. C \textbf{46}, 585 (2006)

\bibitem{H1:2013okq}
C.~Alexa et~al. (H1 Collaboration), Eur. Phys. J. C \textbf{73}, 2466 (2013)

\bibitem{ALICE:2014eof}
B.B. Abelev et~al. (ALICE Collaboration), Phys. Rev. Lett. \textbf{113}, 232504 (2014)

\bibitem{ALICE:2018oyo}
S.~Acharya et~al. (ALICE Collaboration), Eur. Phys. J. C \textbf{79}, 402 (2019)

\bibitem{LHCb:2018rcm}
R.~Aaij et~al. (LHCb Collaboration), JHEP \textbf{10}, 167 (2018)

\bibitem{Adderley:2024czm}
P.A. Adderley et~al., Phys. Rev. Accel. Beams \textbf{27}, 084802 (2024)

\bibitem{gluexjp:2019}
A.~Ali et~al. (GlueX Collaboration), Phys. Rev. Lett. \textbf{123}, 072001 (2019)

\bibitem{gluexjp:2023}
S.~Adhikari et~al. (GlueX Collaboration), Phys. Rev. C \textbf{108}, 025201 (2023)

\bibitem{hallc:007}
B.~Duran et~al., Nature \textbf{615}, 813 (2023)

\bibitem{007:2026dow}
S.~Joosten et~al. (007 Collaboration) (2026), \texttt{2602.14416}

\bibitem{E12_12_001}
I.~Albayrak et~al., \emph{{Jefferson Lab PAC 39 Proposal: Timelike Compton Scattering and J/$\psi$ photoproduction on the proton in $e^{-}e^{+}$ pair production with CLAS12 at 11 GeV}}, \url{https://www.jlab.org/exp_prog/proposals/12/PR12-12-001.pdf} (2012)

\bibitem{E12_12_001A}
M.~Battaglieri et~al., \emph{{Near threshold J/$\psi$ photoproduction and study of LHCb pentaquarks with CLAS12}}, \url{https://www.jlab.org/exp_prog/proposals/17/E12-12-001A.pdf} (2017)

\bibitem{CLAS:2026lls}
P.~Chatagnon et~al. (CLAS Collaboration) (2026), \texttt{2602.22128}

\bibitem{PhysRevD.104.066023}
K.A. Mamo et~al., Phys. Rev. D \textbf{104}, 066023 (2021)

\bibitem{PhysRevD.101.086003}
K.A. Mamo et~al., Phys. Rev. D \textbf{101}, 086003 (2020)

\bibitem{PhysRevD.103.094010}
K.A. Mamo et~al., Phys. Rev. D \textbf{103}, 094010 (2021)

\bibitem{Lorce:2018egm}
C.~Lorc{\'e} et~al., Eur. Phys. J. C \textbf{79}, 89 (2019)

\bibitem{Ji:2021mtz}
X.~Ji, Front. Phys. (Beijing) \textbf{16}, 64601 (2021)

\bibitem{Tang:2025qqe}
L.~Tang et~al. (2025), \texttt{2510.08845}

\bibitem{Du:2020bqj}
M.L. Du et~al., Eur. Phys. J. C \textbf{80}, 1053 (2020)

\bibitem{PhysRevD.108.054018}
D.~Winney et~al. (JPAC Collaboration), Phys. Rev. D \textbf{108}, 054018 (2023)

\bibitem{Eides:2015dtr}
M.I. Eides et~al., Phys. Rev. D \textbf{93}, 054039 (2016)

\bibitem{Kubarovsky:2015aaa}
V.~Kubarovsky et~al., Phys. Rev. D \textbf{92}, 031502 (2015)

\bibitem{Guo:2015umn}
F.K. Guo et~al., Phys. Rev. D \textbf{92}, 071502 (2015)

\bibitem{Blin:2016dlf}
A.N. Hiller~Blin et~al., Phys. Rev. D \textbf{94}, 034002 (2016)

\bibitem{Strakovsky:2023kqu}
I.~Strakovsky et~al., Phys. Rev. C \textbf{108}, 015202 (2023)

\bibitem{Fair:2020yfx}
R.~Fair et~al., Nucl. Instrum. Meth. A \textbf{962}, 163578 (2020)

\bibitem{gemc}
M.~Ungaro et~al., Nucl. Instrum. Meth. A \textbf{959}, 163422 (2020)

\bibitem{Thomadakis:2022zcd}
P.~Thomadakis et~al., Comput. Phys. Commun. \textbf{271}, 108201 (2022)

\bibitem{Gnanvo:2024jag}
K.~Gnanvo et~al., PoS \textbf{QNP2024}, 014 (2025)

\bibitem{eai:tenorio}
M.~Tenorio-Pita, \emph{{Enhancing Lepton Identification in CLAS12 using Machine Learning Techniques}}, \url{https://misportal.jlab.org/mis/physics/clas12/viewFile.cfm/2024-005.pdf?documentId=172}

\bibitem{Tyson:2023yer}
R.~Tyson, Ph.D. thesis, University of Glasgow, Glasgow U. (2023)

\bibitem{Asryan:2020iqj}
G.~Asryan et~al., Nucl. Instrum. Meth. A \textbf{959}, 163425 (2020)

\bibitem{Antonioli:2020ylv}
M.A. Antonioli et~al., Nucl. Instrum. Meth. A \textbf{962}, 163701 (2020)

\bibitem{Acker:2020qkv}
A.~Acker et~al., Nucl. Instrum. Meth. A \textbf{957}, 163423 (2020)

\bibitem{Carman:2020yma}
D.S. Carman et~al., Nucl. Instrum. Meth. A \textbf{960}, 163626 (2020)

\bibitem{Chatagnon:2020lwt}
P.~Chatagnon et~al., Nucl. Instrum. Meth. A \textbf{959}, 163441 (2020)

\bibitem{Segarra:2020txy}
E.P. Segarra et~al., Nucl. Instrum. Meth. A \textbf{978}, 164356 (2020)

\bibitem{Ungaro:2020hbs}
M.~Ungaro et~al., Nucl. Instrum. Meth. A \textbf{957}, 163420 (2020)

\bibitem{ftcal}
A.~Acker et~al., Nucl. Instrum. Meth. A \textbf{959}, 163475 (2020)

\bibitem{Mestayer:2020saf}
M.D. Mestayer et~al., Nucl. Instrum. Meth. A \textbf{959}, 163518 (2020)

\bibitem{Carman:2020fsv}
D.S. Carman et~al., Nucl. Instrum. Meth. A \textbf{960}, 163629 (2020)

\bibitem{coatjava}
V.~Ziegler et~al., Nucl. Instrum. Meth. A \textbf{959}, 163472 (2020)

\bibitem{ic}
R.~Niyazov et~al., \emph{{CLAS/DVCS Inner Calorimeter Calibration}}, \url{https://misportal.jlab.org/ul/Physics/Hall-B/clas/viewFile.cfm/2005-021.pdf?documentId=213}

\bibitem{HPS:2016rgp}
I.~Balossino et~al. (HPS), Nucl. Instrum. Meth. A \textbf{854}, 89 (2017)

\bibitem{Baltzell:2022rpd}
N.~Baltzell et~al. (2022), \texttt{2203.08324}

\bibitem{Xiong:2019umf}
W.~Xiong et~al., Nature \textbf{575}, 147 (2019)

\bibitem{doi:10.1142/9789812701978_0014}
A.~Gasparian, Proc. 11th Int. Conf. on Calorimetry in Particle Physics (eds Cecchi, C. et al.) pp. 109--115 (2005)

\bibitem{nps}
T.~Horn (NPS Collaboration), J. Phys. Conf. Ser. \textbf{587}, 012048 (2015)

\bibitem{Somov:2025eiq}
A.~Somov et~al. (2025), \texttt{2510.03500}

\bibitem{sgluex}
A.~Asaturyan et~al., Nucl. Instrum. Meth. A \textbf{1013}, 165683 (2021)

\bibitem{SAULI1997531}
F.~Sauli, Nucl. Instrum. Meth. A \textbf{386}, 531 (1997)

\bibitem{bonus_gem}
H.C. Fenker et~al., Nucl. Instrum. Meth. A \textbf{592}, 273 (2008)

\bibitem{eg6_gem}
R.~Dupr{\'e} et~al., Nucl. Instrum. Meth. A \textbf{898}, 90 (2018)

\bibitem{prad_gem}
K.~Gnanvo et~al., \url{https://wiki.jlab.org/pcrewiki/images/f/ff/KG_pRadReadinessReview_20160325.pdf}  (2016)

\bibitem{rd51_mw}
K.~Gnanvo, \url{https://indico.cern.ch/event/1110129/contributions/4714241/attachments/2386731/4079118/KG_RD51_MiniWeek20210208.pdf}  (2024)

\bibitem{COMPASS:2002}
C.~Altunbas et~al., Nucl. Instrum. Meth. A \textbf{490}, 177 (2002)

\bibitem{GNANVO:2016nim}
K.~Gnanvo et~al., Nucl. Instrum. Meth. A \textbf{808}, 83 (2016)

\bibitem{Bencivenni:2014exa}
G.~Bencivenni et~al., JINST \textbf{10}, P02008 (2015)

\bibitem{Bencivenni:2024jgp}
G.~Bencivenni et~al., Nucl. Instrum. Meth. A \textbf{1069}, 169725 (2024)

\bibitem{Boyarinov:2020yry}
S.~Boyarinov et~al., Nucl. Instrum. Meth. A \textbf{966}, 163698 (2020)

\bibitem{GRAPE_gen}
T.~Abe, Comput. Phys. Commun. \textbf{136}, 126 (2001)

\bibitem{Klimenko:incEG}
V.~Klimenko et~al., \emph{Inclusive electron generator}, \url{https://github.com/ValeriiKlimenko/IncEG/tree/master}

\bibitem{CLAS:2025zup}
V.~Klimenko et~al. (CLAS Collaboration), Phys. Rev. C \textbf{112}, 025201 (2025)

\bibitem{marcprl2}
M.~Vanderhaeghen et~al., Phys. Rev. D \textbf{60}, 094017 (1999)

\bibitem{Kroll:2019wug}
P.~Kroll, Eur. Phys. J. A \textbf{55}, 76 (2019)

\bibitem{clas12beam}
N.~Baltzell et~al., Nucl. Instrum. Meth. A \textbf{959}, 163421 (2020)

\bibitem{Heller:2021}
M.~Heller et~al., Phys. Rev. D \textbf{104}, 073007 (2021)

\bibitem{PhysRevD.100.034019}
D.~Winney et~al. (JPAC Collaboration), Phys. Rev. D \textbf{100}, 034019 (2019)

\bibitem{elSpectro_generator}
D.~Glazier, \emph{The elspectro generator}, \url{https://github.com/dglazier/elSpectro}

\bibitem{wolfschil}
K.~Schilling et~al., Nucl. Phys. B \textbf{61}, 381 (1973)

\bibitem{Aaij:2015tga}
R.~Aaij et~al. (LHCb Collaboration), Phys. Rev. Lett. \textbf{115}, 072001 (2015)

\bibitem{Wang:2015jsa}
Q.~Wang et~al., Phys. Rev. D \textbf{92}, 034022 (2015)

\bibitem{Karliner:2015voa}
M.~Karliner et~al., Phys. Lett. B \textbf{752}, 329 (2016)

\bibitem{PhotonPolarization}
H.~Olsen et~al., Phys. Rev. \textbf{114}, 887 (1959)

\bibitem{PhysRevD.88.014001}
C.~Mezrag et~al., Phys. Rev. D \textbf{88}, 014001 (2013)

\end{thebibliography}


\onecolumn
\section*{Full Author List}

\newcommand{\instANL}{1}
\newcommand{\instCPT}{2}
\newcommand{\instCFNS}{3}
\newcommand{\instCNU}{4}
\newcommand{\instDuke}{5}
\newcommand{\instGWU}{6}
\newcommand{\instGSI}{7}
\newcommand{\instGiessen}{8}
\newcommand{\instINFNFrascati}{9}
\newcommand{\instINFNBari}{10}
\newcommand{\instINFNCatania}{11}
\newcommand{\instINFNFerrara}{12}
\newcommand{\instINFNGenova}{13}
\newcommand{\instINFNRoma}{14}
\newcommand{\instINFNTorino}{15}
\newcommand{\instIJCLab}{16}
\newcommand{\instIRFU}{17}
\newcommand{\instJGU}{18}
\newcommand{\instKyungpook}{19}
\newcommand{\instMIT}{20}
\newcommand{\instMississippiState}{21}
\newcommand{\instNCBJ}{22}
\newcommand{\instNMSU}{23}
\newcommand{\instODU}{24}
\newcommand{\instShandong}{25}
\newcommand{\instJLab}{26}
\newcommand{\instUCRiverside}{27}
\newcommand{\instUConn}{28}
\newcommand{\instGlasgow}{29}
\newcommand{\instUNH}{30}
\newcommand{\instSouthCarolina}{31}
\newcommand{\instVirginia}{32}
\newcommand{\instYork}{33}
\newcommand{\instUniBrescia}{34}
\newcommand{\instUniGenova}{35}
\newcommand{\instUniFerrara}{36}
\newcommand{\instUniMessina}{37}
\newcommand{\instUniMilanoBicocca}{38}
\newcommand{\instUniRomaTorVergata}{39}
\newcommand{\instYerPhI}{40}

\begin{center}
\parbox{\textwidth}{
J.~S.~Alvarado$^{\instIJCLab}$,
N.~Baltzell$^{\instJLab}$,
M.~Bondi$^{\instINFNCatania}$,
P.~Chatagnon$^{\instIRFU}$,
R.~De~Vita$^{\instJLab,\instINFNGenova}$,
M.~Hoballah$^{\instIJCLab}$,
V.~Kubarovsky$^{\instJLab}$,
R.~Paremuzyan$^{\instJLab}$,
S.~Stepanyan$^{\instJLab}$,
P.~Achenbach$^{\instJGU}$
% ...
}
\end{center}

\begin{description}[labelsep=0.2em,align=right,labelwidth=0.7em,labelindent=0em,leftmargin=2em,noitemsep]

\item[$^{\instANL}$]Argonne National Laboratory, Argonne, Illinois 60439, USA
\item[$^{\instCPT}$] CPHT, CNRS, École Polytechnique, I.P. Paris, 91128 Palaiseau, France
\item[$^{\instCFNS}$] Center for Frontiers in Nuclear Science, Stony Brook University, Stony Brook, NY 11794, USA
\item[$^{\instCNU}$]Christopher Newport University, Newport News, Virginia 23606
\item[$^{\instDuke}$] Duke University, Durham, NC 27708, USA
\item[$^{\instGWU}$] George Washington University, Washington, DC 20052, USA
\item[$^{\instGSI}$] GSI Helmholtzzentrum für Schwerionenforschung GmbH, D-64291 Darmstadt, Germany
\item[$^{\instGiessen}$] II.~Physikalisches Institut, Universität Giessen, 35392 Giessen, Germany
\item[$^{\instINFNFrascati}$] INFN, Laboratori Nazionali di Frascati, 00044 Frascati (RM), Italy
\item[$^{\instINFNBari}$] INFN, Sezione di Bari, 70125 Bari, Italy
\item[$^{\instINFNCatania}$] INFN, Sezione di Catania, 95123 Catania, Italy
\item[$^{\instINFNFerrara}$] INFN, Sezione di Ferrara, 44122 Ferrara, Italy
\item[$^{\instINFNGenova}$] INFN, Sezione di Genova, 16146 Genova, Italy
\item[$^{\instINFNRoma}$] INFN, Sezione di Roma, 00185 Roma, Italy
\item[$^{\instINFNTorino}$] INFN, Sezione di Torino, 10125 Torino, Italy
\item[$^{\instIJCLab}$] IJCLab, Université Paris-Saclay, CNRS–IN2P3, 91405 Orsay, France
\item[$^{\instIRFU}$] IRFU, CEA, Université Paris-Saclay, 91191 Gif-sur-Yvette, France
\item[$^{\instJGU}$] Johannes Gutenberg University, 55128 Mainz, Germany
\item[$^{\instKyungpook}$] Kyungpook National University, Daegu 41566, Republic of Korea
\item[$^{\instMIT}$] Massachusetts Institute of Technology, Cambridge, MA 02139, USA
\item[$^{\instMississippiState}$] Mississippi State University, Mississippi State, MS 39762, USA
\item[$^{\instNCBJ}$] National Centre for Nuclear Research (NCBJ), Pasteura 7, 02-093 Warsaw, Poland
\item[$^{\instNMSU}$] New Mexico State University, Las Cruces, New Mexico 88003, USA
\item[$^{\instODU}$] Old Dominion University, Norfolk, VA 23529, USA
\item[$^{\instShandong}$] Shandong University, Qingdao, Shandong 266237, China
\item[$^{\instJLab}$] Thomas Jefferson National Accelerator Facility (JLab), Newport News, VA 23606, USA
\item[$^{\instUCRiverside}$] University of California, Riverside, CA 92521, USA
\item[$^{\instUConn}$] University of Connecticut, Storrs, Connecticut 06269, USA
\item[$^{\instGlasgow}$] University of Glasgow, Glasgow G12 8QQ, United Kingdom
\item[$^{\instUNH}$] University of New Hampshire, Durham, New Hampshire 03824, USA
\item[$^{\instSouthCarolina}$] University of South Carolina, Columbia, SC 29208, USA
\item[$^{\instVirginia}$] University of Virginia, Charlottesville, VA 22904, USA
\item[$^{\instYork}$] University of York, York, YO10 5DD, United Kingdom
\item[$^{\instUniBrescia}$] Università degli Studi di Brescia and INFN Sezione di Pavia, 25121 Brescia, Italy
\item[$^{\instUniGenova}$] Università degli Studi di Genova and INFN, Sezione di Genova, 16146 Genova, Italy
\item[$^{\instUniFerrara}$] Università di Ferrara and INFN, Sezione di Ferrara, 44100 Ferrara, Italy
\item[$^{\instUniMessina}$] Università di Messina and INFN, Sezione di Catania, 95123 Catania, Italy
\item[$^{\instUniMilanoBicocca}$] Università di Milano-Bicocca and INFN, Laboratori Nazionali del Sud, 20126 Milano, Italy
\item[$^{\instUniRomaTorVergata}$] Università di Roma Tor Vergata and INFN, Sezione di Roma Tor Vergata, 00133 Roma, Italy
\item[$^{\instYerPhI}$] Yerevan Physics Institute, 375036 Yerevan, Armenia

\end{description}

\end{document}